\documentclass[11pt]{article}
\usepackage{amsmath,amssymb,amsfonts}
\usepackage[english]{babel}
\usepackage{float}
\def\nn{\nonumber} \def\bd{\begin{document}} \def\ed{\end{document}}
\def\ds{\documentstyle}
\let\bm=\bibitem
\newcommand{\be}{\begin{equation}}
\newcommand{\ee}{\end{equation}}
\newcommand{\bea}{\setlength\arraycolsep{2pt} \begin{eqnarray}}
\newcommand{\eea}{\end{eqnarray}}
\newcommand{\hoch}[1]{$\, ^{#1}$}
\def\p{\partial}
\title{\large {\bf Field equations and Noether potentials
for higher-order theories of gravity with Lagrangians
involving $\Box^i R$, $\Box^i R_{\mu\nu}$ and
$\Box^i R_{\mu\nu\rho\sigma}$}}
\date{}

\author{Jun-Jin Peng$^{1,2}$\footnote{corresponding author:
pengjjph@163.com},
\quad Hua Li$^{1}$   \\\ \\
\small \sl $^1$School of Physics and Electronic Science,
\small \sl Guizhou Normal University,\\
\small Guiyang, Guizhou 550001, People's Republic of China; \\
\small \sl  $^{2}$Guizhou Provincial Key Laboratory of Radio Astronomy and
Data Processing, \\
\small \sl Guizhou Normal University, \\
\small Guiyang, Guizhou 550001, People's Republic of China
}

\begin{document}

\maketitle
\vspace{-5pt}

\begin{center}
\textbf{Abstract}
\end{center}

In this paper, we aim to perform a systematical investigation on
the field equations and Noether potentials for the higher-order
gravity theories endowed with Lagrangians depending on
the metric and the Riemann curvature tensor, together with
$i$th ($i=1,2,\cdot\cdot\cdot$) powers of the
Beltrami-d'Alembertian operator $\Box$ acting on the latter.
We start with a detailed derivation of the field equations
and the Noether potential corresponding to the Lagrangian
$\sqrt{-g}L_R(R,\Box R,\cdot\cdot\cdot,\Box^m R)$ through
the direct variation of the Lagrangian and a method based upon the
conserved current. Next the parallel analysis is extended
to a more generic Lagrangian
$\sqrt{-g}L_{\text{Ric}}(g^{\mu\nu}, R_{\mu\nu},\Box R_{\mu\nu},
\cdot\cdot\cdot,\Box^m R_{\mu\nu})$, as well as to the generalization
of the Lagrangian $\sqrt{-g}L_{\text{Ric}}$, which depends on
the metric $g^{\mu\nu}$, the Riemann
tensor $R_{\mu\nu\rho\sigma}$ and $\Box^i R_{\mu\nu\rho\sigma}$s.
Finally, all the results associated to the three types of Lagrangians
are extended to the Lagrangian relying on an arbitrary tensor
and the variables via $\Box^i$ acting on such a tensor.
In particular, we take into consideration of equations of
motion and Noether potentials for nonlocal gravity models.
For Lagrangians involving the variables
$\Box^i R$, $\Box^i R_{\mu\nu}$ and
$\Box^i R_{\mu\nu\rho\sigma}$, our investigation provides their
concrete Noether potentials and the field equations without
the derivative of the Lagrangian density with respect to the
metric. Besides, the Iyer-Wald potentials associated to these
Lagrangians are also presented.

%\noindent \textbf{PACS}: 04.20.-q, 04.50.Kd, 04.70.Bw

%%%%%%%%%%%%%%%%%%%%%%%%%%%%%%%%%%%%%%%%%%%%%%%%%%%%%%%%%%%%%%%%%%%%%%%%%
\voffset=-.90pt
\vspace{10pt}
\newpage

%%%%%%%%%%%%%%%%%%%%%%%%%%%%%%%%%%
\section{Introduction}\label{one}
%%%%%%%%%%%%%%%%%%%%%%%%%%%%%%%%%%

Quit recently, in the work \cite{JJP2306}, by means
of an off-shell conserved
current associated with an arbitrary smooth vector field, a
method to derive equations of motion and Noether potentials
for diffeomorphism invariant gravity theories
was put forward. The main idea of this method goes as follows.
For a general
diffeomorphism invariant Lagrangian
\be
\sqrt{-g}L=\sqrt{-g}L(g^{\mu\nu},R_{\alpha\beta\rho\sigma},
\nabla_\gamma{R}_{\alpha\beta\rho\sigma},
\nabla_{\gamma}\nabla_{\lambda}{R}_{\alpha\beta\rho\sigma},
\cdot\cdot\cdot)
\, ,\label{GenLag}
\ee
the variation of the Lagrangian
(\ref{GenLag}) with respect to all the variables
gives rise to
\be
\delta\big(\sqrt{-g}L\big)=\sqrt{-g}E_{\mu\nu} \delta g^{\mu\nu}
+\sqrt{-g}\nabla_\mu \Theta^\mu
\, . \label{Lagvari}
\ee
In Eq. (\ref{Lagvari}), without loss of generality, the expression
$E_{\mu\nu}$ for equations of motion can be written as the
following form
\be
E_{\mu\nu}=\frac{\partial L}{\partial
g^{\mu\nu}}-\frac{1}{2}Lg_{\mu\nu}+Y_{\mu\nu}
\, , \label{EEominY}
\ee
in which the second-rank symmetric tensor $Y_{\mu\nu}$
stands for all the contributions from the variation of
the Lagrangian with respect to the other variables but
the metric $g^{\mu\nu}$, such as $R_{\mu\nu\rho\sigma}$,
$\nabla_\alpha R_{\mu\nu\rho\sigma}$,
$\nabla_\alpha\nabla_\beta R_{\mu\nu\rho\sigma}$,
and so on. The surface term $\Theta^\mu$ within
Eq. (\ref{Lagvari}) embraces the sufficient information to
yield the field equations and the Noether potential related
to the Lagrangian (\ref{GenLag}). This indicates that it
is only required to handle $\Theta^\mu$ in order to obtain
these quantities. As a matter of fact, one
merely needs to compute such a surface term under the condition
that the variation operator $\delta$ in it is transformed into
the Lie derivative $\mathcal{L}_\zeta$ along an arbitrary smooth
vector $\zeta^\mu$. If doing this guarantees that the surface term
$\Theta^\mu(\delta\rightarrow\mathcal{L}_\zeta)$ can be
eventually decomposed into the following form
\be
\Theta^\mu(\delta\rightarrow\mathcal{L}_\zeta)
=2X^{\mu\nu}\zeta_\nu-\nabla_\nu K^{\mu\nu}
\, , \label{TheLiegen}
\ee
where $X^{\mu\nu}$ denotes some second-rank tensor
independent of the vector $\zeta^\mu$ and
$K^{\mu\nu}$ represents a second-rank anti-symmetric tensor,
one immediately obtains
the expression for field equations in an alternative form
\be
E_{\mu\nu}=X_{\mu\nu}-\frac{1}{2}Lg_{\mu\nu}
\, . \label{EoMgen}
\ee
In contrast with the expression (\ref{EEominY}) derived
straightforwardly out of the variation for the
Lagrangian (\ref{GenLag}), here the expression $E_{\mu\nu}$
completely originating from the surface term
is irrelevant to $\partial L/\partial g^{\mu\nu}$ since
the term $\big(\partial L/\partial
g^{\mu\nu}\big)\delta g^{\mu\nu}$
appearing in the variation of the Lagrangian is only
proportional to the variation of the metric
$\delta g^{\mu\nu}$ rather than its derivatives
so that it does not enter into the surface term $\Theta^\mu$.
Moreover, it has been proved that the two-form
$K^{\mu\nu}$ is just the desired Noether potential
corresponding to the vector $\zeta^\mu$.

In particular, within the situation where the surface
term $\Theta^\mu$ is decomposed as
\be
\Theta^\mu=\sum_i \Theta^\mu_{(i)}
\, , \label{Thetdeci}
\ee
each component $\Theta^\mu_{(i)}$ with the variation
operator $\delta$ substituted by the Lie derivative
$\mathcal{L}_\zeta$ is supposed to take a
similar structure displayed by Eq. (\ref{TheLiegen}),
namely,
\be
\Theta^\mu_{(i)}(\delta\rightarrow\mathcal{L}_\zeta)
=2\zeta_\nu{X}^{\mu\nu}_{(i)}
-\nabla_\nu K^{\mu\nu}_{(i)}
\, , \label{ThetidelLie}
\ee
where the rank-2 tensor $K^{\mu\nu}_{(i)}$ is anti-symmetric.
In such a case, the expression for field equations is expressed
as
\be
E^{\mu\nu}=\sum_i X^{\mu\nu}_{(i)}-\frac{1}{2}Lg^{\mu\nu}
\, , \label{EoMgeni}
\ee
and the Noether potential $K^{\mu\nu}$ has the form
\be
K^{\mu\nu}=\sum_i K^{\mu\nu}_{(i)}
\, . \label{NoePgeni}
\ee

On the basis of the two expressions (\ref{EEominY})
and (\ref{EoMgen}) for field equations, one is able to
get two identities. As a matter of fact,
the symmetry of the expression $E_{\mu\nu}$ for field
equations further determines that $X_{\mu\nu}$ is symmetric
as well, leading to an identity
\be
X_{[\mu\nu]}=\frac{1}{2}(X_{\mu\nu}-X_{\nu\mu})=0
\, . \label{IdenXmn}
\ee
In addition to this, from the comparison between
Eqs. (\ref{EEominY}) and (\ref{EoMgen}), one is able to
acquire the other identity
\be
\frac{\partial L}{\partial g^{\mu\nu}}
=X_{\mu\nu}-Y_{\mu\nu}
\, . \label{IdPLPgXY}
\ee
The above equation establishes the relation between
the second-rank tensor $\partial L/\partial g^{\mu\nu}$
and the derivatives of the Lagrangian density
with respect to all the other variables except for
the metric tensor.

In the present paper, following the method based on
conserved current proposed in the work \cite{JJP2306},
we delve into the field equations
and the Noether potentials for a series of diffeomorphism
invariant Lagrangians consisting of the higher-order
derivative terms $\Box^i R$, $\Box^i R_{\mu\nu}$ and
$\Box^i R_{\mu\nu\rho\sigma}$, where both $i$ and $\Box$
denote an arbitrary positive integer and
the conventional Beltrami-d'Alembertian operator,
respectively. As a generalization, we further concentrate
on applying this method to the Lagrangians that depend
upon a generic rank-$n$ tensor and the variables
generated by means of $i$th powers of the
Beltrami-d'Alembertian operator
$\Box$ acting on this tensor. The explicit expressions
for equations of motion and Nother potentials are obtained.
Apart from this, the similar analysis is extended to
derive field equations and Noether potentials for
a number of other types of Lagrangians. Some of them
can be incorporated into the nonlocal gravity theories
\cite{ALS97,BGKM12,Mods12,BCKM14,KKSrev23,CBnonloc22}.
Moreover, on the basis of the surface terms and the
Noether potentials, the Iyer-Wald potentials
\cite{LeeWald,IyWald,WalZo} associated to all the
involved Lagrangians are presented.

The remainder of this paper is structured as follows. In
Section \ref{two}, as a beginning of our investigation,
for simplicity, we consider the situation in which the
Lagrangian merely depends upon the Ricci scalar
$R$ and $\Box^i R$s, that is,
$\sqrt{-g}L_R(R,\Box R,\cdot\cdot\cdot,\Box^m R)$.
We acquire the equations
of motion and the Noether potential for such a Lagrangian.
In Section \ref{three}, we continue to take into account
the derivation for the field equations and the Noether
potential related to a more generic Lagrangian, which takes the
form $\sqrt{-g}L_{\text{Ric}}
(g^{\mu\nu},{R}_{\mu\nu},\Box{R}_{\mu\nu},
\cdot\cdot\cdot,\Box^m{R}_{\mu\nu})$. In
Section \ref{four}, we extend the analysis for both the
Lagrangians $\sqrt{-g}L_R$ and $\sqrt{-g}L_{\text{Ric}}$
to the Lagrangian $\sqrt{-g}L_{\text{Riem}}$ that is
dependent of the metric
$g^{\mu\nu}$, the Riemann tensor $R_{\mu\nu\rho\sigma}$,
and $\Box^i R_{\mu\nu\rho\sigma}$s. The field equations and the
Noether potential for this Lagrangian are derived. On the basis
of this, the Iyer-Wald potential built from the Noether one
and the surface term is presented as well. Within Section \ref{five},
for the sake of understanding all the previous results from a unified
perspective, we eventually generalize them to the Lagrangians
that rely on a general rank-$n$ tensor
$B_{\alpha_1\cdot\cdot\cdot\alpha_n}$ and
$\Box^i{B}_{\alpha_1\cdot\cdot\cdot\alpha_n}$s, where
the tensor $B_{\alpha_1\cdot\cdot\cdot\alpha_n}$ is supposed to
depend upon $g^{\mu\nu}$, $R_{\mu\nu\rho\sigma}$,
and $\Box^i R_{\mu\nu\rho\sigma}$s. As applications and extensions,
we take into consideration of the field equations and
the Noether potentials corresponding to several types of Lagrangians
that are made up of two funtionals. Our conclusions are
contained in Section \ref{six}. At the end, four appendixes are
given to provide some details on the derivation and a summary
of our main results.

%%%%%%%%%%%%%%%%%%%%%%%%%%%%%%%%%%%%%%%%%%%%%%%%%%%%%%%%%%%%%%%%%%%%%%%%
\section{Field equations and Noether potentials for the
Lagrangian density $L_R(R,\Box R,\cdot\cdot\cdot,\Box^m R)$}\label{two}
%%%%%%%%%%%%%%%%%%%%%%%%%%%%%%%%%%%%%%%%%%%%%%%%%%%%%%%%%%%%%%%%%%%%%%%%

%%%%%%%%%%%%%%%%%%%%%%%%%%%%%%%%%%%%%%%%%%%%%%%%%%%%%%%%
\subsection{The general formalism for field equations and
Noether potentials}\label{two1}
%%%%%%%%%%%%%%%%%%%%%%%%%%%%%%%%%%%%%%%%%%%%%%%%%%%%%%%%%

In the present section, we take into consideration of a higher-order
generalized theory of pure gravity with the Lagrangian
that only relies on the Ricci curvature scalar $R$, together with
its $(2i)$th-order $(i=1,2,\cdot\cdot\cdot,m)$ covariant
derivatives $\Box^iR$s, being of the form
\cite{HJSch90,Wan93,HOW96,CdMP16}
\be
\sqrt{-g}L_R=\sqrt{-g}
L_R(R,\Box R,\cdot\cdot\cdot,\Box^m R)
\, , \label{LagBoxR}
\ee
in which the Beltrami-d'Alembertian operator $\Box$
is defined in terms of both the inverse metric
$g^{\mu\nu}$ and the covariant derivative $\nabla_\mu$ as
$\Box=g^{\mu\nu}\nabla_\mu\nabla_\nu$. For the sake of
obtaining the field equations, as usual, we begin with
the variation with regard to all the variables
$R$ and $\Box^i R$s, giving rise to
\be
\delta\big(\sqrt{-g}L_R\big)
=\sqrt{-g}\left(\frac{1}{2}L_R
g^{\mu\nu}\delta g_{\mu\nu}+F_{(0)}\delta R
+\sum^{m}_{i=1}F_{(i)}\delta \Box^i R\right)
\, , \label{VaryLagR}
\ee
with the scalars $F_{(0)}$ and $F_{(i)}$s
$(i=1,\cdot\cdot\cdot,m)$ defined through
\be
F_{(0)}=\frac{\partial L_R}
{\partial R} \, , \qquad
F_{(i)}=\frac{\partial L_R}
{\partial \Box^i R}
\, . \label{Fidef}
\ee
Subsequently, we deal with the $F_{(i)}\delta\Box^i R$ term
to get rid of $\Box^i$ in it. In order to achieve this,
introducing scalars $\Phi_{(i,k)}$
$(k=1,\cdot\cdot\cdot,i+1)$ given by
\be
\Phi_{(i,k)}=\left(\Box^{k-1} F_{(i)}\right)
\delta\Box^{i-k+1} R
\, , \label{Phikdef}
\ee
where $\Phi_{(i,1)}=F_{(i)}\delta\Box^i{R}$
and $\Phi_{(i,i+1)}=\big(\Box^iF_{(i)}\big)\delta{R}$,
we figure out the relation between
$\Phi_{(i,k)}$ and $\Phi_{(i,k+1)}$ as
\be
\Phi_{(i,k)}=\Phi_{(i,k+1)}
+A^{\mu\nu}_{(i,k)} \delta g_{\mu\nu}
+B^\nu_{(i,k)}\delta \Gamma^\mu_{\mu\nu}
+\nabla_\mu C^\mu_{(i,k)}
\, . \label{PhikRel}
\ee
Within Eq. (\ref{PhikRel}), the three tensors
$A^{\mu\nu}_{(i,k)}$, $B^{\mu}_{(i,k)}$,
and $C^{\mu}_{(i,k)}$ are defined through
\bea
A^{\mu\nu}_{(i,k)}&=&
\left(\nabla^{(\mu}\Box^{k-1} F_{(i)}\right)
\left(\nabla^{\nu)}\Box^{i-k} R\right)
=A^{\nu\mu}_{(i,k)}\, , \nn \\
B^{\mu}_{(i,k)}&=&
\left(\Box^{k-1} F_{(i)}\right)
\left(\nabla^{\mu}\Box^{i-k} R\right) \, , \nn \\
C^{\mu}_{(i,k)}&=&
\left(\Box^{k-1} F_{(i)}\right)
\left(\delta\nabla^{\mu}\Box^{i-k} R\right)
-\left(\nabla^{\mu}\Box^{k-1} F_{(i)}\right)
\left(\delta\Box^{i-k} R\right)
\, , \label{ABCikdef}
\eea
respectively. It is easy to check that they satisfy
\be
C^{\mu}_{(i,k)}(\delta\rightarrow\nabla^\nu)
=\nabla^{\nu}B^{\mu}_{(i,k)}
-2A^{\mu\nu}_{(i,k)}
\, . \label{ABCrelat}
\ee
By means of summing both sides of Eq. (\ref{PhikRel})
over $k$ from 1 up to $i$, we further obtain
\be
\Phi_{(i,1)}=\Phi_{(i,i+1)}
+(\delta g_{\mu\nu})\sum^i_{k=1}A^{\mu\nu}_{(i,k)}
+\big(\delta \Gamma^\mu_{\mu\nu}\big)\sum^i_{k=1}B^\nu_{(i,k)}
+\sum^i_{k=1}\nabla_\mu C^\mu_{(i,k)}
\, . \label{SumPhik}
\ee
This equation establishes the relation between the scalars
$F_{(i)}\delta\Box^i{R}$ and $\big(\Box^iF_{(i)}\big)\delta{R}$.
Finally, with the help of the scalar
\be
F=F_{(0)}
+\sum^{m}_{i=1}\Box^i F_{(i)}
\, , \label{ScaFdef}
\ee
Equation (\ref{SumPhik}) renders Eq. (\ref{VaryLagR})
for the variation of the Lagrangian reformulated as
the form that is irrelevant to the terms
$F_{(i)}\delta\Box^i{R}$s, given by
\bea
\delta\left(\sqrt{-g}L_R\right)
&=& \frac{\sqrt{-g}}{2}\left[\sum^m_{i=1}\sum^i_{k=1}
\left(2A^{\mu\nu}_{(i,k)}-g^{\mu\nu}
\nabla_\sigma B^\sigma_{(i,k)}\right)
+L_R g^{\mu\nu}\right]\delta g_{\mu\nu} \nn \\
&&+\sqrt{-g}F\delta{R}
+\sqrt{-g}\sum^m_{i=1}\nabla_\mu\Theta_{R(i)}^\mu \nn \\
&=&\sqrt{-g}\left(E^R_{\mu\nu} \delta g^{\mu\nu}
+\nabla_\mu\Theta^\mu_R \right)
\, . \label{VaryLagR2}
\eea
Within Eq. (\ref{VaryLagR2}), the surface term $\Theta^\mu_R$
can be decomposed into
\be
\Theta^\mu_R=\Theta^\mu_{R(0)}
+\sum^m_{i=1}\Theta_{R(i)}^\mu
\, , \label{TThetR}
\ee
in which the $\Theta^\mu_{R(0)}$ term, coming from
the scalar $F\delta{R}$, taking the form
\be
\Theta^\mu_{R(0)}= 2Fg^{\rho[\mu}
\nabla^{\nu]}\delta g_{\rho\nu}
-2g^{\rho[\mu}
(\nabla^{\nu]} F)\delta g_{\rho\nu}
\, , \label{TheR0def}
\ee
while the $\Theta^\mu_{R(i)}$ term, incorporating all the
contributions from the divergence terms and the terms
proportional to the variation of the Levi-Civita connection
in Eq. (\ref{SumPhik}), is given by
\be
\Theta^\mu_{R(i)}= \sum^i_{k=1}
\left(C^{\mu}_{(i,k)}+\frac{1}{2}B^\mu_{(i,k)}
g^{\rho\sigma}\delta g_{\rho\sigma}\right)
\, . \label{TheRidef}
\ee
In addition to this, the expression for field equations
$E^R_{\mu\nu}$ is read off as
\be
E^{\mu\nu}_R=\frac{1}{2}\sum^m_{i=1}\sum^i_{k=1}
\left(g^{\mu\nu}\nabla_\sigma B^\sigma_{(i,k)}
-2A^{\mu\nu}_{(i,k)}\right)
-\frac{1}{2}L_R g^{\mu\nu}
+FR^{\mu\nu}-\nabla^\mu\nabla^\nu F
+g^{\mu\nu}\Box{F}
\, . \label{EoMforLagR0}
\ee
We substitute Eqs. (\ref{ABCikdef}) and (\ref{ScaFdef})
into Eq. (\ref{EoMforLagR0}) to write down its expression
in terms of the scalars ${F}_{(l)}$ and $R$ as
\bea
E^{\mu\nu}_R&=&\frac{1}{2}g^{\mu\nu}\sum^m_{i=1}\sum^i_{k=1}
\nabla_\lambda\left[\left(\Box^{k-1} F_{(i)}\right)
\nabla^{\lambda}\Box^{i-k} R\right]
-\sum^m_{i=1}\sum^i_{k=1}\left(\nabla^{(\mu}\Box^{k-1} F_{(i)}\right)
\left(\nabla^{\nu)}\Box^{i-k} R\right) \nn \\
&&+\sum^m_{l=0}\left(R^{\mu\nu}\Box^lF_{(l)}
-\nabla^\mu\nabla^\nu\Box^lF_{(l)}
+g^{\mu\nu}\Box^{l+1}{F}_{(l)}\right)
-\frac{1}{2}L_R g^{\mu\nu}
\, . \label{EoMforLagR}
\eea
Here $E^{\mu\nu}_R$ coincides with the expression for
equations of motion given by the work \cite{HJSch90}.
In the above equation, it can be proved that
\bea
\sum^i_{k=1}
\nabla_\lambda\big[\big(\Box^{k-1} F_{(i)}\big)
\nabla^{\lambda}\Box^{i-k} R\big]&=&
\sum^i_{k=1}
\nabla_\lambda\big[\big(\Box^{i-k} F_{(i)}\big)
\nabla^{\lambda}\Box^{k-1} R\big] \, , \nn \\
\sum^i_{k=1}\big(\nabla^{(\mu}\Box^{k-1} F_{(i)}\big)
\nabla^{\nu)}\Box^{i-k} R&=&
\sum^i_{k=1}\big(\nabla^{(\mu}\Box^{i-k} F_{(i)}\big)
\nabla^{\nu)}\Box^{k-1} R
\, . \label{IdesymikR}
\eea
In particular, when $m=1$, we obtain the expression for
field equations corresponding to the Lagrangian
$\sqrt{-g}L_R\big|_{m=1}=\sqrt{-g}L_R(R,\Box R)$,
being of the form
\bea
E^{\mu\nu}_R\big|_{m=1}
&=&\frac{1}{2}g^{\mu\nu}
\big(\nabla_\lambda{F}_{(1)}\big)
\nabla^{\lambda}R
+\frac{1}{2}g^{\mu\nu}F_{(1)}\Box{R}
-\frac{1}{2}g^{\mu\nu}L_R(R,\Box R)\nn \\
&&-\big(\nabla^{(\mu}F_{(1)}\big)\nabla^{\nu)}R
-\nabla^\mu\nabla^\nu F_{(0)}
-\nabla^\mu\nabla^\nu\Box F_{(1)}\nn \\
&&+g^{\mu\nu}\big(\Box{F}_{(0)}
+\Box^2{F}_{(1)}\big)
+R^{\mu\nu}\big(F_{(0)}+\Box{F}_{(1)}\big)
\, . \label{EoMLagRm1}
\eea

For the sake of providing a verification on the obtained
expression (\ref{EoMLagRm1}) for field equations,
in the remainder of this subsection, we shall
employ the method put forward in \cite{JJP2306} to derive
the field equations instead of the aforementioned procedure
in terms of the direct variation of the Lagrangian. By the way,
we will present the Noether potential for the Lagrangian
$\sqrt{-g}L_R$. According to this method, the surface term
$\Theta^\mu_{R}$ plays a crucial role in determining the
field equations and the Noether potentials. A prominent task
is to compute this term under the transformation
$\delta\rightarrow\mathcal{L}_\zeta$, where $\mathcal{L}_\zeta$
denotes the Lie derivative along an arbitrary smooth vector
$\zeta^\mu$. To do this, we start with the replacement of
the variation operator $\delta$ in $\Theta^\mu_{R(0)}$ with
the Lie derivative $\mathcal{L}_\zeta$, yielding
\be
\Theta^\mu_{R(0)}(\delta\rightarrow\mathcal{L}_\zeta)
= 2 \big(FR^{\mu\nu}-\nabla^\mu\nabla^\nu F
+g^{\mu\nu}\Box{F}\big)\zeta_\nu
-\nabla_\nu K^{\mu\nu}_{R(0)}
\, , \label{ThetR0Lie}
\ee
in which the second-rank anti-symmetric tensor
$K^{\mu\nu}_{R(0)}$ is given by
\be
K^{\mu\nu}_{R(0)}=2F\nabla^{[\mu}\zeta^{\nu]}
+4\zeta^{[\mu}\nabla^{\nu]} F
\, . \label{KmnR0def}
\ee
In the same way, we calculate
$\Theta^\mu_{R(i)}(\delta\rightarrow\mathcal{L}_\zeta)$
on the basis of Eq. (\ref{TheRidef}).
By the aid of Eq. (\ref{ABCrelat}), we have
\be
\Theta^\mu_{R(i)}(\delta\rightarrow\mathcal{L}_\zeta)
= 2\zeta_\nu\sum^i_{k=1}X^{\mu\nu}_{R(i,k)}
-\sum^i_{k=1}\nabla_\nu K^{\mu\nu}_{R(i,k)}
\, , \label{ThetRiLie}
\ee
with the second-rank symmetric tensor $X^{\mu\nu}_{R(i,k)}$
and the anti-symmetric one $K^{\mu\nu}_{R(i,k)}$ presented
respectively by
\bea
X^{\mu\nu}_{R(i,k)}&=& \frac{1}{2}g^{\mu\nu}\nabla_\sigma
B^\sigma_{(i,k)}-A^{\mu\nu}_{(i,k)} \, , \nn \\
K^{\mu\nu}_{R(i,k)}&=&2\zeta^{[\mu}B^{\nu]}_{(i,k)}
=2\left(\Box^{k-1} F_{(i)}\right)
\left(\zeta^{[\mu}\nabla^{\nu]}\Box^{i-k}R\right)
\, . \label{KmnRidef}
\eea
Interestingly, introducing a rank-3 tensor
$\tilde{B}^{\sigma\mu\nu}_{(i,k)}=g^{\sigma\mu}B^\nu_{(i,k)}$
to reexpress the term proportional to
the variation of the Levi-Civita connection on the right
hand side of Eq. (\ref{PhikRel}) as
$g_{\rho\sigma}\tilde{B}^{\sigma\mu\nu}_{(i,k)}
\delta \Gamma^\rho_{\mu\nu}$, one finds that the rank-2
tensor ${X}^{\mu\nu}_{R(i,k)}$ can be reexpressed as
\be
{X}^{\mu\nu}_{R(i,k)}=
\frac{1}{2}C^{\mu}_{(i,k)}(\delta\rightarrow\nabla^\nu)
+\frac{1}{2}\nabla_\lambda\left(
\tilde{B}^{(\mu\nu)\lambda}_{(i,k)}
-\tilde{B}^{\lambda(\mu\nu)}_{(i,k)}
+\tilde{B}^{[\mu|\lambda|\nu]}_{(i,k)}\right)
\, , \label{XRikdef2}
\ee
and the anti-symmetric tensor $K^{\mu\nu}_{R(i,k)}$
is transformed into
\be
K^{\mu\nu}_{R(i,k)}=\zeta_\lambda
\Big(\tilde{B}^{\lambda[\mu\nu]}_{(i,k)}
+\tilde{B}^{[\mu\nu]\lambda}_{(i,k)}
+\tilde{B}^{[\mu|\lambda|\nu]}_{(i,k)}\Big)
\, . \label{KmnRikdef}
\ee
Apart from Eq. (\ref{XRikdef2}) and (\ref{KmnRikdef}),
the vector $C^\mu_{(i,k)}(\delta\rightarrow\mathcal{L}_\zeta)$
is in connection with the rank-3 tensor
$\tilde{B}^{\sigma\mu\nu}_{(i,k)}$ in the following
manner
\be
C^\mu_{(i,k)}(\delta\rightarrow\mathcal{L}_\zeta)
=\zeta_\nu{C}^{\mu}_{(i,k)}(\delta\rightarrow\nabla^\nu)
-\tilde{B}^{\nu\mu\lambda}_{(i,k)}\nabla_\lambda\zeta_\nu
\, . \label{CikdelLieB}
\ee
What is more, it will be demonstrated in the next two
sections that it is completely allowed to extend
Eqs. (\ref{XRikdef2}), (\ref{KmnRikdef}) and (\ref{CikdelLieB})
to the Lagrangians that depends upon the variables
$\Box^i{R}_{\mu\nu}$s and $\Box^i{R}_{\mu\nu\rho\sigma}$s.
The most general extensions for them, which involve two
arbitrary tensors $\big(A^{\alpha_1\cdot\cdot\cdot\alpha_n},
B_{\alpha_1\cdot\cdot\cdot\alpha_n}\big)$
rather than the three pairs $\big(F_{(i)},{R}\big)$,
$\big({P}^{\mu\nu}_{(i)},{R}_{\mu\nu}\big)$,
and $\big({P}^{\mu\nu\rho\sigma}_{(i)},
{R}_{\mu\nu\rho\sigma}\big)$, will be
detailedly analyzed in Sec. \ref{five}.

On the basis of Eqs. (\ref{ThetR0Lie})
and (\ref{ThetRiLie}), the substitution of
$\delta$ in the surface term $\Theta^\mu_{R}$ by
the Lie derivative $\mathcal{L}_\zeta$ leads to
\bea
\Theta^\mu_{R}(\delta\rightarrow\mathcal{L}_\zeta)
&=&\Theta^\mu_{R(0)}(\delta\rightarrow\mathcal{L}_\zeta)
+\sum^m_{i=1}\Theta^\mu_{R(i)}(\delta\rightarrow\mathcal{L}_\zeta)\nn \\
&=&2\left(E^{\mu\nu}_R
+\frac{1}{2}L_R g^{\mu\nu}\right)\zeta_\nu
-\nabla_\nu K^{\mu\nu}_{R}
\, , \label{ThetRLie}
\eea
in which the anti-symmetric tensor $K^{\mu\nu}_{R}$
takes the form
\bea
K^{\mu\nu}_{R}&=&K^{\mu\nu}_{R(0)}
+\sum^m_{i=1}\sum^i_{k=1}K^{\mu\nu}_{R(i,k)} \nn \\
&=&2F\nabla^{[\mu}\zeta^{\nu]}
+4\zeta^{[\mu}\nabla^{\nu]}F
+2\sum^m_{i=1}\sum^i_{k=1}
\big(\Box^{k-1} F_{(i)}\big)
\zeta^{[\mu}\nabla^{\nu]}\Box^{i-k}R
\, . \label{KmnRdef}
\eea
From Eq. (\ref{ThetRLie}), an off-shell conserved current
$J^{\mu}_{R}$ reads \cite{TPad10,RievLL}
\be
J^{\mu}_{R}=\nabla_\nu{K}^{\mu\nu}_{R}
=2\zeta_\nu{E}^{\mu\nu}_{R}+\zeta^\mu{L}_{R}
-\Theta^\mu_{R}(\delta\rightarrow\mathcal{L}_\zeta)
\, . \label{ConCurLR}
\ee
According to Eqs. (\ref{TheLiegen}) and (\ref{EoMgen}),
the expression $E^{\mu\nu}_R$ for field equations
can be reproduced by Eq. (\ref{ThetRLie}),
while the Noether potentials
associated to the Lagrangian (\ref{LagBoxR}) are presented
by the second-rank anti-symmetric tensor $K^{\mu\nu}_{R}$.

%%%%%%%%%%%%%%%%%%%%%%%%%%%%%%%%%%%%%%%%%%%%%%%%%%%%%%%%
\subsection{Applications within three special cases of $L_R$:
$R^m\Box^nR$, $(\Box^iR)(\Box^jR)$ and
$R\Box^{i+j}R-(\Box^iR)(\Box^jR)$}\label{two2}
%%%%%%%%%%%%%%%%%%%%%%%%%%%%%%%%%%%%%%%%%%%%%%%%%%%%%%%%%

Within the present subsection, as some specifical examples
to demonstrate the above generic results,
we take into account the field equations and the Noether
potentials associated to three special cases of the
Lagrangian $\sqrt{-g}L_R$, which include the ones
$\sqrt{-g}R^m\Box^nR$, $\sqrt{-g}(\Box^iR)(\Box^jR)$ and
$\sqrt{-g}\big[R\Box^{i+j}R-(\Box^iR)(\Box^jR)\big]$.

Firstly, within the context of the Lagrangian
\be
\sqrt{-g}L_{R1}=\sqrt{-g}R^m\Box^nR
\, , \label{LagRmBoxRn}
\ee
makeing use of Eq. (\ref{EoMforLagR}), one is able to obtain
the expression for the field equations
\bea
E^{\mu\nu}_{R1}
&=&\frac{1}{2}g^{\mu\nu}\sum^n_{k=1}
\nabla_\lambda\big[\big(\Box^{k-1}R^m\big)
\nabla^{\lambda}\Box^{n-k} R\big]
-\sum^n_{k=1}\big(\nabla^{(\mu}\Box^{k-1}R^m\big)
\big(\nabla^{\nu)}\Box^{n-k}R\big) \nn \\
&&-\frac{1}{2}g^{\mu\nu}R^m\Box^nR
+g^{\mu\nu}\big[m\Box\big(R^{m-1}\Box^nR\big)
+\Box^{n+1}R^m\big]+R^{\mu\nu}\Box^nR^m\nn \\
&&+mR^{\mu\nu}R^{m-1}\Box^nR
-m\nabla^\mu\nabla^\nu\big(R^{m-1}\Box^nR\big)
-\nabla^\mu\nabla^\nu\Box^nR^m
\, , \label{EoMofLagRm}
\eea
together with the Noether potential $K^{\mu\nu}_{R1}$
derived out of the generic one (\ref{KmnRdef}),
read off as
\bea
K^{\mu\nu}_{R1}
&=&2mR^{m-1}\big(\Box^nR\big)\nabla^{[\mu}\zeta^{\nu]}
+2\big(\Box^nR^m\big)\nabla^{[\mu}\zeta^{\nu]}
+4m\zeta^{[\mu}\nabla^{\nu]}\big(R^{m-1}\Box^nR\big) \nn \\
&&+4\zeta^{[\mu}\nabla^{\nu]}\Box^nR^m
+2\sum^n_{k=1}
\big(\Box^{k-1}R^m\big)
\zeta^{[\mu}\nabla^{\nu]}\Box^{n-k}R
\, . \label{KmnR1def}
\eea
Particularly, when $m=0$, $E^{\mu\nu}_{R1}=0$,
attributed to the fact that the total divergence
term $\Box^nR=\nabla_\mu\big(\nabla^\mu\Box^{n-1}R\big)$
is non-dynamical. What is more, the surface term
$\Theta^\mu_{R1}$ for the Lagrangian $L_{R1}$ is given by
\bea
\Theta^\mu_{R1}&=&2\big(mR^{m-1}\Box^nR+\Box^nR^m\big)g^{\rho[\mu}
\nabla^{\nu]}\delta g_{\rho\nu}
-2mg^{\rho[\mu}
\Big[\nabla^{\nu]} \big(R^{m-1}\Box^nR\big)\Big]\delta g_{\rho\nu}
\nn \\
&&-2g^{\rho[\mu}\Big(\nabla^{\nu]} \Box^nR^{m}\Big)\delta g_{\rho\nu}
+\Theta^\mu_{R(n)}\big|_{F_{(n)}=R^m}
\, . \label{ThetR1}
\eea

Secondly, we take into consideration of the Lagrangian
\be
\sqrt{-g}L_{R2}=\sqrt{-g}\big(\Box^iR\big)\big(\Box^jR\big)
\, , \label{LagBoxRij}
\ee
According to Eq. (\ref{EoMforLagR}), the expression
for equations of motion corresponding to the Lagrangian
(\ref{LagBoxRij}) is read off as
\bea
E^{\mu\nu}_{R2}
&=&\frac{1}{2}g^{\mu\nu}\sum^i_{k=1}
\nabla_\lambda\big[\big(\Box^{j+k-1}R\big)
\nabla^{\lambda}\Box^{i-k} R\big]
+\frac{1}{2}g^{\mu\nu}\sum^j_{k=1}
\nabla_\lambda\big[\big(\Box^{i+k-1}R\big)
\nabla^{\lambda}\Box^{j-k} R\big]  \nn \\
&&-\sum^i_{k=1}\big(\nabla^{(\mu}\Box^{j+k-1}R\big)
\big(\nabla^{\nu)}\Box^{i-k}R\big)
-\sum^j_{k=1}\big(\nabla^{(\mu}\Box^{i+k-1}R\big)
\big(\nabla^{\nu)}\Box^{j-k}R\big)\nn \\
&&-\frac{1}{2}g^{\mu\nu}(\Box^iR)\Box^jR
+2R^{\mu\nu}\Box^{i+j}R
+2g^{\mu\nu}\Box^{i+j+1}R
-2\nabla^\mu\nabla^\nu\Box^{i+j}R
\, , \label{EoMofBoxRij}
\eea
while the Noether potential $K^{\mu\nu}_{R2}$ is given by
\bea
K^{\mu\nu}_{R2}
&=&8\zeta^{[\mu}\nabla^{\nu]}\Box^{i+j}R
+2\sum^i_{k=1}\big(\Box^{j+k-1}R\big)
\zeta^{[\mu}\nabla^{\nu]}\Box^{i-k}R
 \nn \\
&&+4\big(\Box^{i+j}R\big)\nabla^{[\mu}\zeta^{\nu]}
+2\sum^j_{k=1}\big(\Box^{i+k-1}R\big)
\zeta^{[\mu}\nabla^{\nu]}\Box^{j-k}R
\, . \label{KmnBoxRij}
\eea
Obviously, $E^{\mu\nu}_{R2}(i,j)=E^{\mu\nu}_{R2}(j,i)$
and $E^{\mu\nu}_{R1}(m=1)=E^{\mu\nu}_{R2}(i=0,j=n)$.
Besides, the surface term $\Theta^\mu_{R2}$
corresponding to the Lagrangian $L_{R2}$ has the form
\be
\Theta^\mu_{R2}=\Theta^\mu_{R(i)}\big|_{F_{(i)}=\Box^jR}
+\Theta^\mu_{R(j)}\big|_{F_{(j)}=\Box^iR}
\, . \label{ThetR2def}
\ee

Thirdly, let us pay attention to the Lagrangian
\be
\sqrt{-g}L_{R3}=\sqrt{-g}
\big[R\Box^{i+j}R-(\Box^iR)\Box^jR\big]
\, . \label{LagBoxRc3}
\ee
For simplicity, we consider the $j=1$ case of the Lagrangian
(\ref{LagBoxRc3}). In such a situation, after some
manipulations, the expression for the field equations of
the Lagrangian $\sqrt{-g}L_{R3}(j=1)$ is read off as
\bea
E^{\mu\nu}_{R3}\big|_{j=1}&=&E^{\mu\nu}_{R1}(m=1,n=i+1)
-E^{\mu\nu}_{R2}(j=1) \nn \\
&=&\frac{1}{2}g^{\mu\nu}\sum^i_{k=1}
\nabla_\lambda\big[\big(\nabla^{\lambda}\Box^{k-1}R\big)
\Box^{i-k+1} R-\big(\Box^{k}R\big)
\nabla^{\lambda}\Box^{i-k}R\big] \nn \\
&&-\sum^i_{k=1}
\big[\big(\nabla^{(\mu}\Box^{k-1}R\big)
\nabla^{\nu)}\Box^{i-k+1} R-\big(\nabla^{(\mu}\Box^{k}R\big)
\nabla^{\nu)}\Box^{i-k}R\big]
\, . \label{EoMofBoxRc31}
\eea
By making use of both the following identities
\bea
\sum^i_{k=1}
\nabla_\lambda\big[\big(\nabla^{\lambda}\Box^{k-1}R\big)
\Box^{i-k+1} R\big]&=&\sum^i_{k=1}
\nabla_\lambda\big[\big(\Box^{k}R\big)
\nabla^{\lambda}\Box^{i-k}R\big] \, , \nn \\
\sum^i_{k=1}
\big(\nabla^{(\mu}\Box^{k-1}R\big)
\nabla^{\nu)}\Box^{i-k+1} R&=&
\sum^i_{k=1}\big(\nabla^{(\mu}\Box^{k}R\big)
\nabla^{\nu)}\Box^{i-k}R
\, , \label{IdentR31ik}
\eea
one finds that $E^{\mu\nu}_{R3}\big|_{j=1}=0$, arising
from that
$L_{R3}(j=1)=\nabla_\mu\big[R\nabla^\mu\Box^iR
-\big(\nabla^\mu R\big)\big(\Box^iR\big)\big]$.
Furthermore, due to the fact that
\be
L_{R3}=\sum^i_{k=1}
\nabla_\mu\big[\big(\Box^{k-1}R\big)
\big(\nabla^{\mu}\Box^{i+j-k}R\big)
-\big(\nabla^{\mu}\Box^{k-1}R\big)
\big(\Box^{i+j-k}R\big)\big]
\, , \label{LR3divform}
\ee
we have the conclusion that the field equations for
the Lagrangian (\ref{LagBoxRc3}) vanishes identically,
that is,
\be
E^{\mu\nu}_{R3}=E^{\mu\nu}_{R1}(m=1,n=i+j)-E^{\mu\nu}_{R2}
=0
\, . \label{EoMofLagRc3}
\ee
From the above equation, $E^{\mu\nu}_{R2}$ can be reexpressed
as $E^{\mu\nu}_{R2}=E^{\mu\nu}_{R1}(m=1,n=i+j)$, which renders
$E^{\mu\nu}_{R2}$ simplified as
\bea
E^{\mu\nu}_{R2}&=&\frac{1}{2}g^{\mu\nu}\sum^{i+j}_{k=1}
\nabla_\lambda\left[\big(\Box^{k-1}R\big)
\nabla^{\lambda}\Box^{i+j-k} R\right]
-\sum^{i+j}_{k=1}\left(\nabla^{(\mu}\Box^{k-1}R\right)
\nabla^{\nu)}\Box^{i+j-k}R \nn \\
&&-\frac{1}{2}g^{\mu\nu}R\Box^{i+j}R
+2R^{\mu\nu}\Box^{i+j}R
+2g^{\mu\nu}\Box^{i+j+1}R
-2\nabla^\mu\nabla^\nu\Box^{i+j}R
\, . \label{EoMofBoxRij2}
\eea

%%%%%%%%%%%%%%%%%%%%%%%%%%%%%%%%%%%%%%%%%%%%%%%%%%%%%%%%%%%%%%%%%%%%%%%%
\section{Field equations and Noether potentials for the
Lagrangians relying on the variables $g^{\mu\nu}$ and
$\Box^i R_{\mu\nu}$s}\label{three}
%%%%%%%%%%%%%%%%%%%%%%%%%%%%%%%%%%%%%%%%%%%%%%%%%%%%%%%%%%%%%%%%%%%%%%%%

Within this section, by contrast with the situation for
the Lagrangian $\sqrt{-g}L_{R}$ in the previous section,
we will perform the same analysis to a more general
Lagrangian that is dependent of the inverse metric
$g^{\mu\nu}$ and the Ricci tensor $R_{\mu\nu}$,
together with $\Box^iR_{\mu\nu}$s
$(i=1,\cdot\cdot\cdot,m)$, taking the general form
\be
\sqrt{-g}L_{\text{Ric}}=\sqrt{-g}
L_{\text{Ric}}(g^{\mu\nu}, R_{\mu\nu},\Box R_{\mu\nu},
\cdot\cdot\cdot,\Box^m R_{\mu\nu})
\, . \label{LagBoxRic}
\ee

%%%%%%%%%%%%%%%%%%%%%%%%%%%%%%%%%%%%%%%%%%%%%%%%%%%%%%%%
\subsection{Equations of motion and
Noether potentials at general level}\label{three1}
%%%%%%%%%%%%%%%%%%%%%%%%%%%%%%%%%%%%%%%%%%%%%%%%%%%%%%%%%

With the help of the second-rank tensors
$P^{\mu\nu}_{(0)}$ and $P^{\mu\nu}_{(i)}$s
$(i=1,\cdot\cdot\cdot,m)$ defined by
\be
P^{\mu\nu}_{(0)}=\frac{\partial{L_{\text{Ric}}}}
{\partial{R}_{\mu\nu}} \, , \qquad
P^{\mu\nu}_{(i)}=\frac{\partial{L_{\text{Ric}}}}
{\partial\Box^i{R}_{\mu\nu}}
\, , \label{P0idef}
\ee
the variation of the Lagrangian (\ref{LagBoxRic}) with
respect to all the variables leads to
\be
\delta\big(\sqrt{-g}L_{\text{Ric}}\big)
=\sqrt{-g}\left[\left(
\frac{\partial{L_{\text{Ric}}}}{\partial{g}^{\mu\nu}}
-\frac{1}{2}L_{\text{Ric}}g_{\mu\nu}\right)
\delta g^{\mu\nu}+P^{\mu\nu}_{(0)}\delta{R}_{\mu\nu}
+\sum^{m}_{i=1}P^{\mu\nu}_{(i)}
\delta \Box^i{R}_{\mu\nu}\right]
\, . \label{VaryLagRic}
\ee
Let us deal with the terms $P^{\mu\nu}_{(i)}
\delta \Box^i{R}_{\mu\nu}$s in Eq. (\ref{VaryLagRic}).
For convenience to do this, we introduce scalars
$\Psi_{(i,k)}$ $(k=1,\cdot\cdot\cdot,i+1)$ defined through
\be
\Psi_{(i,k)}=\left(\Box^{k-1}P^{\mu\nu}_{(i)}\right)
\delta\Box^{i-k+1}{R}_{\mu\nu}
\, . \label{Psiikdef}
\ee
Obviously, $\Psi_{(i,1)}=P^{\mu\nu}_{(i)}
\delta\Box^i{R}_{\mu\nu}$ and
$\Psi_{(i,i+1)}=\big(\Box^iP^{\mu\nu}_{(i)}\big)
\delta{R}_{\mu\nu}$. By means of calculations on
$\Psi_{(i,k)}$, we relate it to $\Psi_{(i,k+1)}$
through
\be
\Psi_{(i,k)}=\Psi_{(i,k+1)}
+\nabla_\mu L^{\mu}_{(i,k)}
+g_{\rho\sigma}M^{\sigma\mu\nu}_{(i,k)}
\delta \Gamma^\rho_{\mu\nu}
+N^{\mu\nu}_{(i,k)} \delta g_{\mu\nu}
\, . \label{PsikRel}
\ee
Within Eq. (\ref{PsikRel}), the vector $L^{\mu}_{(i,k)}$ is given by
\be
L^{\mu}_{(i,k)}=
\left(\Box^{k-1} P^{\rho\sigma}_{(i)}\right)
\left(\delta\nabla^{\mu}\Box^{i-k} R_{\rho\sigma}\right)
-\left(\nabla^{\mu}\Box^{k-1} P^{\rho\sigma}_{(i)}\right)
\left(\delta\Box^{i-k} R_{\rho\sigma}\right)
\, , \label{Likdef}
\ee
the rank-3 tensor $M^{\sigma\mu\nu}_{(i,k)}$ takes the form
\bea
M^{\sigma\mu\nu}_{(i,k)}&=&
2\left(\nabla^{\mu}\Box^{k-1} P^{\nu\rho}_{(i)}\right)
\left(\Box^{i-k} R^{\sigma}_{\rho}\right)
-2\left(\Box^{k-1} P^{\rho\nu}_{(i)}\right)
\left(\nabla^{\mu}\Box^{i-k} R^{\sigma}_{\rho}\right) \nn \\
&&+g^{\sigma\mu}\left(\nabla^{\nu}\Box^{i-k}
R_{\alpha\beta}\right)
\left(\Box^{k-1} P^{\alpha\beta}_{(i)}\right)
\, , \label{Mikdef}
\eea
and the second-rank symmetric tensor $N^{\mu\nu}_{(i,k)}$
is read off as
\be
N^{\mu\nu}_{(i,k)}=
\left(\nabla^{(\mu}\Box^{i-k} R_{\rho\sigma}\right)
\left(\nabla^{\nu)}\Box^{k-1} P^{\rho\sigma}_{(i)}\right)
=N^{\nu\mu}_{(i,k)}
\, . \label{Nikdef}
\ee
Particularly, when $L_{\text{Ric}}=L_R$, it can be verified
that the substitution
of $P^{\mu\nu}_{(0)}=g^{\mu\nu}F_{(0)}$ and
$P^{\mu\nu}_{(i)}=g^{\mu\nu}F_{(i)}$ into Eq. (\ref{PsikRel})
yields Eq. (\ref{PhikRel}). From Eq. (\ref{PsikRel}), we further
obtain
\be
\Psi_{(i,1)}=\Psi_{(i,i+1)}
+\sum^i_{k=1}\nabla_\mu L^{\mu}_{(i,k)}
+\sum^i_{k=1}g_{\rho\sigma}M^{\sigma\mu\nu}_{(i,k)}
\delta\Gamma^\rho_{\mu\nu}
+\sum^i_{k=1}N^{\mu\nu}_{(i,k)}\delta{g}_{\mu\nu}
\, . \label{SumPsik0}
\ee
Apparently, the above equation enables us to remove
$\Box^i$ from the ingredient $\delta\Box^i{R}_{\mu\nu}$
in the scalar $P^{\mu\nu}_{(i)}\delta\Box^i{R}_{\mu\nu}$,
transforming it into the simpler
one $\big(\Box^i{P}^{\mu\nu}_{(i)}\big)\delta{R}_{\mu\nu}$.
As a consequence of Eq. (\ref{SumPsik0}), the scalar
$P^{\mu\nu}_{(i)}\delta\Box^i{R}_{\mu\nu}$ is explicitly
expressed as the desired form
\bea
P^{\mu\nu}_{(i)}\delta\Box^i{R}_{\mu\nu}
&=&\sum^i_{k=1}
\left[N^{\mu\nu}_{(i,k)}
-\frac{1}{2}\nabla_\lambda\Big(M^{(\mu\nu)\lambda}_{(i,k)}
+M^{(\mu|\lambda|\nu)}_{(i,k)}
-M^{\lambda(\mu\nu)}_{(i,k)}\Big)\right]\delta g_{\mu\nu}\nn \\
&&+\left(\Box^iP^{\mu\nu}_{(i)}\right)\delta{R}_{\mu\nu}
+\nabla_\mu\Theta_{\text{Ric}(i)}^\mu
\, , \label{SumPsik}
\eea
with $\Theta_{\text{Ric}(i)}^\mu$ given by
\be
\Theta_{\text{Ric}(i)}^\mu=
\sum^i_{k=1}\Theta_{\text{Ric}(i,k)}^\mu
\, , \label{TheRicidef}
\ee
in which $\Theta_{\text{Ric}(i,k)}^\mu$ is defined
through
\be
\Theta_{\text{Ric}(i,k)}^\mu
=L^{\mu}_{(i,k)}
+\frac{1}{2}\Big(M^{(\rho\sigma)\mu}_{(i,k)}
+M^{(\rho|\mu|\sigma)}_{(i,k)}
-M^{\mu(\rho\sigma)}_{(i,k)}\Big)
\delta{g}_{\rho\sigma}
\, . \label{TheRicik}
\ee

Furthermore, by means of substituting Eq. (\ref{SumPsik}) into
Eq. (\ref{VaryLagRic}), the variation of the Lagrangian
is reformulated as
\bea
\frac{\delta\big(\sqrt{-g}L_{\text{Ric}}\big)}{\sqrt{-g}}
&=& \sum^m_{i=1}\sum^i_{k=1}
\left[N^{\mu\nu}_{(i,k)}
-\frac{1}{2}\nabla_\lambda\Big(M^{(\mu\nu)\lambda}_{(i,k)}
+M^{(\mu|\lambda|\nu)}_{(i,k)}
-M^{\lambda(\mu\nu)}_{(i,k)}\Big)\right]
\delta g_{\mu\nu}\nn \\
&&+\left(\frac{\partial{L_{\text{Ric}}}}{\partial{g}^{\mu\nu}}
-\frac{1}{2}L_{\text{Ric}}g_{\mu\nu}\right)
\delta g^{\mu\nu}+P^{\mu\nu}\delta R_{\mu\nu}
+\sum^m_{i=1}\nabla_\mu\Theta_{\text{Ric}(i)}^\mu
\, , \label{VaryLagRic2}
\eea
where the second-rank symmetric tensor $P^{\mu\nu}$
is defined through
\be
P^{\mu\nu}=P^{\mu\nu}_{(0)}
+\sum^{m}_{i=1}\Box^i P^{\mu\nu}_{(i)}
\, . \label{TenPdef}
\ee
By utilizing
\be
P^{\mu\nu}\delta R_{\mu\nu}
=\mathcal{P}_{\mu\nu}\delta g^{\mu\nu}
+\nabla_\mu \Theta_{\text{Ric}(0)}^\mu
\, , \label{PmndelRic}
\ee
in which the symmetric tensor $\mathcal{P}_{\mu\nu}$
is given by
\be
\mathcal{P}_{\mu\nu}=
P^{\rho\sigma}R_{\mu\rho\nu\sigma}
-P^\sigma_{(\mu}{R}_{\nu)\sigma}
-\nabla_{(\mu}\nabla^\sigma{P}_{\nu)\sigma}
+\frac{1}{2}\Box{P}_{\mu\nu}
+\frac{1}{2}g_{\mu\nu}\nabla_\rho\nabla_\sigma
{P}^{\rho\sigma}
\, , \label{CalPmndef}
\ee
and the surface term $\Theta^\mu_{\text{Ric}(0)}$
is presented by
\be
\Theta^\mu_{\text{Ric}(0)}= P^{\rho[\mu}
\nabla^{\nu]}\delta g_{\rho\nu}
+g^{\rho[\mu}P^{\nu]\sigma}
\nabla_\sigma\delta g_{\rho\nu}
+(\delta g_{\rho\nu})\nabla^{[\mu}P^{\nu]\rho}
-(\delta g_{\rho\nu})g^{\rho[\mu}
\nabla_\sigma{P}^{\nu]\sigma}
\, , \label{TheRic0def}
\ee
the variation equation (\ref{VaryLagRic2}) is further expressed
as
\be
\delta\left(\sqrt{-g}L_{\text{Ric}}\right)
=\sqrt{-g}\left(E^{\text{Ric}}_{\mu\nu} \delta g^{\mu\nu}
+\nabla_\mu\Theta^\mu_{\text{Ric}}\right)
\, . \label{VaryLagRic3}
\ee
In the above equation, the surface term $\Theta^\mu_{\text{Ric}}$
takes the form
\be
\Theta^\mu_{\text{Ric}}=\Theta^\mu_{\text{Ric}(0)}
+\sum^m_{i=1}\Theta^\mu_{\text{Ric}(i)}
=\Theta^\mu_{\text{Ric}(0)}
+\sum^m_{i=1}\sum^i_{k=1}\Theta^\mu_{\text{Ric}(i,k)}
\, , \label{TThetRic}
\ee
while the expression for field equations
$E^{\text{Ric}}_{\mu\nu}$ is read off as
\bea
E^{\mu\nu}_{\text{Ric}}&=&
\frac{\partial{L_{\text{Ric}}}}{\partial{g}^{\rho\sigma}}
g^{\mu\rho}g^{\nu\sigma}
-\frac{1}{2}L_{\text{Ric}}g^{\mu\nu}
+\mathcal{P}^{\mu\nu}
-\sum^m_{i=1}\sum^i_{k=1}
N^{\mu\nu}_{(i,k)}\nn \\
&&+\frac{1}{2}\sum^m_{i=1}\sum^i_{k=1}
\nabla_\lambda\Big(M^{(\mu\nu)\lambda}_{(i,k)}
+M^{(\mu|\lambda|\nu)}_{(i,k)}
-M^{\lambda(\mu\nu)}_{(i,k)}\Big)
\, . \label{EoMforLagRic}
\eea
Here $E^{\mu\nu}_{\text{Ric}}$ is produced by following
the usual way to directly vary the Lagrangian.

Apparently, employing Eq. (\ref{EoMforLagRic}) to represent
the field equations for the Lagrangian (\ref{LagBoxRic})
involves the calculation on the derivative of the Lagrangian
density with respect to the inverse metric. To avoid this
like in the work \cite{JJP2306,Pady}, as well as to acquire
the Noether potential, we give an alternative
derivation of the field equations by following
the method based on the off-shell Noether current \cite{JJP2306}.
In light of this method, after getting surface terms via
the variation of the Lagrangian, it is merely demanded to
compute the surface terms with
the variation operator transformed into the Lie derivative
along an arbitrary smooth vector.

Substituting the variation operator
$\delta$ in the surface term $\Theta^\mu_{\text{Ric}(0)}$ by
the Lie derivative $\mathcal{L}_\zeta$, we have
\be
\Theta^\mu_{\text{Ric}(0)}(\delta\rightarrow\mathcal{L}_\zeta)
= 2 \big(\mathcal{P}^{\mu\nu}+P^{\sigma\mu}R^\nu_\sigma\big)\zeta_\nu
-\nabla_\nu K^{\mu\nu}_{\text{Ric}(0)}
\, , \label{ThetRic0Lie}
\ee
with the anti-symmetric tensor $K^{\mu\nu}_{\text{Ric}(0)}$
being of the form
\be
K^{\mu\nu}_{\text{Ric}(0)}=
P^{\sigma[\mu}\nabla_\sigma\zeta^{\nu]}
-P^{\sigma[\mu}\nabla^{\nu]}\zeta_\sigma
+2\zeta^{[\mu}\nabla_\sigma{P}^{\nu]\sigma}
-2\zeta_\sigma\nabla^{[\mu}{P}^{\nu]\sigma}
\, . \label{KmnRic0def}
\ee
In addition, calculations on
$\Theta^\mu_{\text{Ric}(i,k)}(\delta\rightarrow\mathcal{L}_\zeta)$
give rise to
\be
\Theta^\mu_{\text{Ric}(i,k)}(\delta\rightarrow\mathcal{L}_\zeta)
= 2 \zeta_\nu{X}^{\mu\nu}_{\text{Ric}(i,k)}
-\nabla_\nu K^{\mu\nu}_{\text{Ric}(i,k)}
\, . \label{ThetRiciLie}
\ee
Here, by employing the following identity
\be
L^\mu_{(i,k)}(\delta\rightarrow\mathcal{L}_\zeta)
=\zeta_\nu{L}^{\mu}_{(i,k)}(\delta\rightarrow\nabla^\nu)
-M^{\nu\mu\lambda}_{(i,k)}\nabla_\lambda\zeta_\nu
\, , \label{LikdelLie}
\ee
the second-rank tensor ${X}^{\mu\nu}_{\text{Ric}(i,k)}$
is read off as
\bea
{X}^{\mu\nu}_{\text{Ric}(i,k)}&=&\frac{1}{2}
\left(\Box^{k-1} P^{\rho\sigma}_{(i)}\right)
\left(\nabla^{\nu}\nabla^{\mu}\Box^{i-k} R_{\rho\sigma}\right)
-\frac{1}{2}\left(\nabla^{\mu}\Box^{k-1} P^{\rho\sigma}_{(i)}\right)
\left(\nabla^{\nu}\Box^{i-k} R_{\rho\sigma}\right) \nn \\
&&+\frac{1}{2}\nabla_\lambda\left(
M^{(\mu\nu)\lambda}_{(i,k)}
-M^{\lambda(\mu\nu)}_{(i,k)}
+M^{[\mu|\lambda|\nu]}_{(i,k)}\right)
\, , \label{XmnRicik}
\eea
and the second-rank anti-symmetric tensor
$K^{\mu\nu}_{\text{Ric}(i,k)}$ is given by
\be
K^{\mu\nu}_{\text{Ric}(i,k)}=
\zeta_\lambda\Big(M^{[\mu\nu]\lambda}_{(i,k)}
+M^{[\mu|\lambda|\nu]}_{(i,k)}
+M^{\lambda[\mu\nu]}_{(i,k)}\Big)
\, . \label{KmnRicik}
\ee
By using Eq. (\ref{Mikdef}), the tensor
$K^{\mu\nu}_{\text{Ric}(i,k)}$ takes the concrete form
\bea
K^{\mu\nu}_{\text{Ric}(i,k)}&=&
2\zeta^{[\mu}\Big(\nabla^{\nu]}\Box^{i-k} R_{\rho\sigma}\Big)
\Box^{k-1} P^{\rho\sigma}_{(i)}
-2\zeta_\lambda\Big(\nabla^{\lambda}\Box^{k-1}
P^{\alpha[\mu}_{(i)}\Big)
\Box^{i-k} R^{\nu]}_{\alpha} \nn \\
&&-2\zeta_\lambda
\Big(\nabla^{\lambda}\Box^{i-k} R^{[\mu}_{\alpha}\Big)
\Box^{k-1} P^{\nu]\alpha}_{(i)}
+2\zeta_\lambda
\Big(\Box^{i-k} R^{[\mu}_{\alpha}\Big)
\nabla^{\nu]}\Box^{k-1} P^{\lambda\alpha}_{(i)} \nn \\
&&+2\zeta_\lambda\Big(\Box^{k-1} P^{\alpha\lambda}_{(i)}\Big)
\nabla^{[\mu}\Box^{i-k} R^{\nu]}_{\alpha}
+2\zeta_\lambda\Big(\Box^{k-1} P^{\alpha[\mu}_{(i)}\Big)
\nabla^{\nu]}\Box^{i-k} R^{\lambda}_{\alpha} \nn \\
&&+2\zeta_\lambda\Big(\nabla^{[\mu}\Box^{k-1}
P^{\nu]\alpha}_{(i)}\Big)
\Box^{i-k} R^{\lambda}_{\alpha}
\, . \label{KmnRicik2}
\eea
Besides, for convenience to compute the field equations,
it is desirable to separate the tensor
${X}^{\mu\nu}_{\text{Ric}(i,k)}$ into a symmetric part
and an anti-symmetric one, namely,
\be
{X}^{\mu\nu}_{\text{Ric}(i,k)}={X}^{(\mu\nu)}_{\text{Ric}(i,k)}
+{X}^{[\mu\nu]}_{\text{Ric}(i,k)}
\, . \label{XmnRicik2}
\ee
Within Eq. (\ref{XmnRicik2}), the symmetric tensor
${X}^{(\mu\nu)}_{\text{Ric}(i,k)}$ is given by
\bea
{X}^{(\mu\nu)}_{\text{Ric}(i,k)}&=&
\frac{1}{2}g^{\mu\nu}\nabla_\lambda\Big[
\Big(\Box^{k-1} P^{\rho\sigma}_{(i)}\Big)
\nabla^{\lambda}\Box^{i-k}
R_{\rho\sigma}\Big]-\Big(\nabla^{(\mu}\Box^{k-1}
P^{|\rho\sigma|}_{(i)}\Big)
\nabla^{\nu)}\Box^{i-k} R_{\rho\sigma} \nn \\
&&+\nabla_\lambda\Big[\Big(\nabla^{(\mu}\Box^{k-1}
P^{|\lambda\alpha|}_{(i)}\Big)
\Box^{i-k} R^{\nu)}_{\alpha}\Big]
-\nabla_\lambda\Big[\Big(\Box^{k-1}
P^{\lambda\alpha}_{(i)}\Big)
\nabla^{(\mu}\Box^{i-k}
R^{\nu)}_{\alpha}\Big]\nn \\
&&-\nabla_\lambda\Big[\Big(\nabla^{(\mu}\Box^{k-1}
P^{\nu)\alpha}_{(i)}\Big)
\Box^{i-k} R^{\lambda}_{\alpha}\Big]
+\nabla_\lambda\Big[\Big(\Box^{k-1}
P^{\alpha(\mu}_{(i)}\Big)
\nabla^{\nu)}\Box^{i-k} R^{\lambda}_{\alpha}\Big]
\, , \label{SXmnRicik2}
\eea
and the anti-symmetric tensor
${X}^{[\mu\nu]}_{\text{Ric}(i,k)}$ is written as
\be
{X}^{[\mu\nu]}_{\text{Ric}(i,k)}=
-\nabla_\lambda\Big[\Big(\nabla^{\lambda}\Box^{k-1}
P^{\sigma[\mu}_{(i)}\Big)
\Box^{i-k} R^{\nu]}_{\sigma}\Big]
-\nabla_\lambda\Big[
\Big(\nabla^{\lambda}\Box^{i-k}
R^{[\mu}_{\sigma}\Big)\Box^{k-1}
P^{\nu]\sigma}_{(i)}\Big]
\, . \label{AXmnRicik2}
\ee
On the basis of Eqs. (\ref{ThetRic0Lie}) and (\ref{ThetRiciLie}),
we arrive at
\bea
\Theta^\mu_{\text{Ric}}(\delta\rightarrow\mathcal{L}_\zeta)
&=& 2 \zeta_\nu\left(\mathcal{P}^{\mu\nu}
+P^{\sigma\mu}R^\nu_\sigma
+\sum^m_{i=1}\sum^i_{k=1}
{X}^{\mu\nu}_{\text{Ric}(i,k)}\right) \nn \\
&&-\nabla_\nu \left(K^{\mu\nu}_{\text{Ric}(0)}
+\sum^m_{i=1}\sum^i_{k=1}
K^{\mu\nu}_{\text{Ric}(i,k)}\right)
\, . \label{TThetRicLie}
\eea
As expected, the surface term
$\Theta^\mu_{\text{Ric}}(\delta\rightarrow\mathcal{L}_\zeta)$
encodes the information for equations of motion together
with the Noether potential.

According to Eq. (\ref{EoMgen}), an alternative enhanced
expression for field equations corresponding to the Lagrangian
(\ref{LagBoxRic}) can be extracted from
Eq. (\ref{TThetRicLie}), which has the form
\be
E^{\mu\nu}_{\text{Ric}}=\mathcal{P}^{\mu\nu}
+P^{\sigma\mu}R^\nu_\sigma
-\frac{1}{2}L_{\text{Ric}}g^{\mu\nu}
+\sum^m_{i=1}\sum^i_{k=1}
{X}^{\mu\nu}_{\text{Ric}(i,k)}
\, . \label{EoMforLagRic2}
\ee
Due to the fact that
$E^{\mu\nu}_{\text{Ric}}=E^{\nu\mu}_{\text{Ric}}$,
one obtains an identity
$P^{\sigma[\mu}R^{\nu]}_\sigma
+\sum^m_{i=1}\sum^i_{k=1}
{X}^{[\mu\nu]}_{\text{Ric}(i,k)}=0$, namely,
\be
P^{\sigma[\mu}R^{\nu]}_\sigma
=\sum^m_{i=1}\sum^i_{k=1}
\nabla_\lambda\Big[\Big(\nabla^{\lambda}\Box^{k-1}
P^{\sigma[\mu}_{(i)}\Big)
\Box^{i-k} R^{\nu]}_{\sigma}+
\Big(\nabla^{\lambda}\Box^{i-k}
R^{[\mu}_{\sigma}\Big)\Box^{k-1}
P^{\nu]\sigma}_{(i)}\Big]
\, . \label{Ident1Ric}
\ee
By the aid of the following identity
\be
P^{\sigma\mu}_{(i)}\Box^iR^{\nu}_{\sigma}
-R^{\nu}_{\sigma}\Box^iP^{\sigma\mu}_{(i)}
=\sum^i_{k=1}\nabla_\lambda\Big[
\Big(\nabla^{\lambda}\Box^{i-k}
R^{\nu}_{\sigma}\Big)\Box^{k-1}
P^{\mu\sigma}_{(i)}
-\Big(\nabla^{\lambda}\Box^{k-1}
P^{\sigma\mu}_{(i)}\Big)
\Box^{i-k} R^{\nu}_{\sigma}\Big]
\, , \label{IdforId2Ric}
\ee
the identity (\ref{Ident1Ric}) turns into
\be
P^{\sigma[\mu}_{(0)}R^{\nu]}_\sigma
=-\sum^m_{i=1}P^{\sigma[\mu}_{(i)}
\Box^iR^{\nu]}_\sigma
\, . \label{Ident1Ric2}
\ee
Apart from this, the comparison between
Eqs. (\ref{EoMforLagRic}) and (\ref{EoMforLagRic2})
gives rise to another identity
\bea
\left(
\frac{\partial{L_{\text{Ric}}}}{\partial{g}^{\rho\sigma}}
\right)
g^{\mu\rho}g^{\nu\sigma}
&=&P^{\sigma(\mu}R^{\nu)}_\sigma
-\sum^m_{i=1}\sum^i_{k=1}
\nabla_\lambda\Big[\Big(\nabla^{\lambda}\Box^{k-1}
P^{\sigma(\mu}_{(i)}\Big)
\Box^{i-k} R^{\nu)}_{\sigma}\Big]\nn \\
&&+\sum^m_{i=1}\sum^i_{k=1}
\nabla_\lambda\Big[
\Big(\nabla^{\lambda}\Box^{i-k}
R^{(\mu}_{\sigma}\Big)\Box^{k-1}
P^{\nu)\sigma}_{(i)}\Big]
\, . \label{Ident2Ric}
\eea
Utilizing Eq. (\ref{IdforId2Ric}) to simplify
the identity (\ref{Ident2Ric}) leads to
\be
\frac{\partial{L_{\text{Ric}}}}{\partial{g}^{\mu\nu}}
=\frac{1}{2}\sum^m_{l=0}\left(
g_{\mu\rho}P^{\rho\sigma}_{(l)}\Box^l{R}_{\nu\sigma}
+g_{\nu\rho}P^{\rho\sigma}_{(l)}\Box^l{R}_{\mu\sigma}\right)
\, . \label{Ident2Ric2}
\ee
As a result of Eq. (\ref{Ident1Ric}),
$E^{\mu\nu}_{\text{Ric}}$ in Eq. (\ref{EoMforLagRic2})
is simplified as
\bea
E^{\mu\nu}_{\text{Ric}}&=&
P_{\rho\sigma}R^{\mu\rho\nu\sigma}
-\nabla^{(\mu}\nabla_\sigma{P}^{\nu)\sigma}
+\frac{1}{2}\Box{P}^{\mu\nu}
+\frac{1}{2}g^{\mu\nu}\nabla_\rho\nabla_\sigma
{P}^{\rho\sigma} \nn \\
&&-\frac{1}{2}L_{\text{Ric}}g^{\mu\nu}
+\sum^m_{i=1}\sum^i_{k=1}
{X}^{(\mu\nu)}_{\text{Ric}(i,k)}
\, . \label{EoMforLagRic3}
\eea
Here the ingredient $\big(P_{\rho\sigma}R^{\mu\rho\nu\sigma}
-\nabla^{(\mu}\nabla_\sigma{P}^{\nu)\sigma}\big)$ can be replaced
with the one $\big(R^{(\mu}_{\sigma}P^{\nu)\sigma}
-\nabla_\sigma\nabla^{(\mu}{P}^{\nu)\sigma}\big)$.
Along the way, from Eq. (\ref{TThetRicLie}), the Noether
potential $K^{\mu\nu}_{\text{Ric}}$ for the Lagrangian
(\ref{LagBoxRic}) is expressed as
\be
K^{\mu\nu}_{\text{Ric}}=
K^{\mu\nu}_{\text{Ric}(0)}
+\sum^m_{i=1}\sum^i_{k=1}
K^{\mu\nu}_{\text{Ric}(i,k)}
\, . \label{KmnRic}
\ee
The off-shell Noether current $J^{\mu}_{\text{Ric}}$ associated to
the Noether potential $K^{\mu\nu}_{\text{Ric}}$ is given by
\cite{TPad10,RievLL}
\be
J^{\mu}_{\text{Ric}}=\nabla_\nu{K}^{\mu\nu}_{\text{Ric}}
=2\zeta_\nu{E}^{\mu\nu}_{\text{Ric}}+\zeta^\mu{L}_{\text{Ric}}
-\Theta^\mu_{\text{Ric}}(\delta\rightarrow\mathcal{L}_\zeta)
\, . \label{ConCurLRic}
\ee

%%%%%%%%%%%%%%%%%%%%%%%%%%%%%%%%%%%%%%%%%%%%%%%%%%%%%%%%
\subsection{The application to the Lagrangian
$\sqrt{-g}L_R$}\label{three2}
%%%%%%%%%%%%%%%%%%%%%%%%%%%%%%%%%%%%%%%%%%%%%%%%%%%%%%%%%

In this subsection, for the sake of checking our results
in the previous subsection, we apply them to derive the
field equations and the Noether potential for the
Lagrangian $\sqrt{-g}L_R$ given by Eq. (\ref{LagBoxR}).

When the Lagrangian $\sqrt{-g}L_R$ is seen as a
functional depending on the variables $g^{\mu\nu}$,
$R_{\mu\nu}$ and $\Box^iR_{\mu\nu}$s, the tensors
$P^{\mu\nu}_{(0)}$, $P^{\mu\nu}_{(i)}$ and
${P}^{\mu\nu}$ are transformed into
\be
P^{\mu\nu}_{(0)}\big|_{L_R}=g^{\mu\nu}F_{(0)} \, ,
\quad
P^{\mu\nu}_{(i)}\big|_{L_R}=g^{\mu\nu}F_{(i)} \, ,
\quad
P^{\mu\nu}\big|_{L_R}=g^{\mu\nu}F
\, , \label{PiPforLR}
\ee
respectively. Consequently, the
${X}^{(\mu\nu)}_{\text{Ric}(i,k)}$
in Eq. (\ref{SXmnRicik2}) takes the value
\be
{X}^{(\mu\nu)}_{\text{Ric}(i,k)}\big|_{L_R}=
\frac{1}{2}g^{\mu\nu}\nabla_\lambda\big[
\big(\Box^{k-1} F_{(i)}\big)
\nabla^{\lambda}\Box^{i-k}
R\big]-\big(\nabla^{(\mu}\Box^{k-1}
F_{(i)}\big)
\nabla^{\nu)}\Box^{i-k}R
\, . \label{SXmnRicikLR}
\ee
Substituting $P^{\mu\nu}|_{L_R}$ and
${X}^{(\mu\nu)}_{\text{Ric}(i,k)}|_{L_R}$
into Eq. (\ref{EoMforLagRic3})
reproduces the expression $E^{\mu\nu}_{R}$ for
field equations in Eq. (\ref{EoMforLagR}).
What is more, the identity (\ref{Ident1Ric}) turns into
\be
FR^{[\mu\nu]}
=\sum^m_{i=1}\sum^i_{k=1}
\nabla_\lambda\big[\big(\nabla^{\lambda}\Box^{k-1}
F_{(i)}\big)\Box^{i-k} R^{[\mu\nu]}+
\big(\nabla^{\lambda}\Box^{i-k}
R^{[\mu\nu]}\big)\Box^{k-1}
F_{(i)}\big]=0
\, , \label{Ident1LR}
\ee
while the identity (\ref{Ident2Ric}) becomes
\bea
F_{(i)}\Box^i R^{\mu\nu}-R^{\mu\nu}\Box^iF_{(i)}
&=&\sum^i_{k=1}\nabla_\lambda
\big[\big(\Box^{k-1}F_{(i)}\big)
\nabla^{\lambda}\Box^{i-k}R^{\mu\nu}
-\big(\nabla^{\lambda}\Box^{k-1}
F_{(i)}\big)\Box^{i-k} R^{\mu\nu}\big]
\, . \quad\label{Ident2LR}
\eea
It can be proved that the above identity indeed holds.
At last, substituting Eq. (\ref{PiPforLR}) into
Eq. (\ref{KmnRic}), one obtains the Noether potential
$K^{\mu\nu}_{R}$ given by Eq. (\ref{KmnRdef}).

%%%%%%%%%%%%%%%%%%%%%%%%%%%%%%%%%%%%%%%%%%%%%%%%%%%%%%%%
\subsection{The application to the Lagrangian
$\sqrt{-g}R^{\mu\nu}\Box^nR_{\mu\nu}$}\label{three3}
%%%%%%%%%%%%%%%%%%%%%%%%%%%%%%%%%%%%%%%%%%%%%%%%%%%%%%%%%

We start with the Lagrangian
\be
\sqrt{-g}L_{\text{Ric1}}
=\sqrt{-g}R^{\mu\nu}\Box^nR_{\mu\nu}
\, . \label{LagRicBnRic}
\ee
The tensors $P^{\mu\nu}_{(0)}$, $P^{\mu\nu}_{(i)}$ and
${P}^{\mu\nu}$ corresponding to the Lagrangian
(\ref{LagRicBnRic}) are read off as
\be
P^{\mu\nu}_{(0)}\big|_{L_{\text{Ric1}}}
=\Box^nR^{\mu\nu} \, ,\quad
P^{\mu\nu}_{(n)}\big|_{L_{\text{Ric1}}}
=R^{\mu\nu} \, ,\quad
P^{\mu\nu}\big|_{L_{\text{Ric1}}}
=2\Box^nR^{\mu\nu}
\, , \label{PiPforLRic1}
\ee
respectively. By making use of them,
calculations in terms of Eq. (\ref{EoMforLagRic2})
yield the field equations for the Lagrangian
$\sqrt{-g}L_{\text{Ric1}}$, given by
\bea
E^{\mu\nu}_{\text{Ric1}}&=&
2R^{(\mu}_\sigma\Box^nR^{\nu)\sigma}
-2\nabla_\sigma\nabla^{(\mu}\Box^nR^{\nu)\sigma}
+\Box^{n+1}R^{\mu\nu}
+\frac{1}{2}g^{\mu\nu}(2\nabla^\rho\nabla_\sigma
-R^{\rho}_{\sigma})\Box^nR^{\sigma}_\rho \nn \\
&&+2\sum^n_{k=1}
\nabla^\lambda\Big[\Big(\nabla^{(\mu}\Box^{k-1}
R^{\sigma}_{\lambda}\Big)
\Box^{n-k} R^{\nu)}_{\sigma}\Big]
-2\sum^n_{k=1}\nabla^\lambda\Big[\Big(\Box^{k-1}
R^{\sigma}_{\lambda}\Big)
\nabla^{(\mu}\Box^{n-k}R^{\nu)}_{\sigma}\Big] \nn \\
&&+\frac{1}{2}g^{\mu\nu}\sum^n_{k=1}\nabla^\lambda\Big[
\Big(\Box^{k-1} R^{\rho}_{\sigma}\Big)
\nabla_{\lambda}\Box^{n-k}R^{\sigma}_{\rho}\Big]
-\sum^n_{k=1}\Big(\nabla^{(\mu}\Box^{k-1}
R_{\rho\sigma}\Big)
\nabla^{\nu)}\Box^{n-k}R^{\rho\sigma}
\, . \label{EomLagRic1c}
\eea
Particularly, in the simplest $n=0$ case,
$E^{\mu\nu}_{\text{Ric1}}$ has the form
\be
E^{\mu\nu}_{\text{Ric1}}\big|_{n=0}=
2R_{\rho\sigma}{R}^{\mu\rho\nu\sigma}
-\nabla^\mu\nabla^{\nu}R
+\Box{R}^{\mu\nu}
+\frac{1}{2}g^{\mu\nu}\Box{R}
-\frac{1}{2}g^{\mu\nu}R^{\rho}_{\sigma}R^{\sigma}_\rho
\, , \label{EomLaRic1n0}
\ee
and in the $n=1$ case, $E^{\mu\nu}_{\text{Ric1}}$ is
written as
\bea
E^{\mu\nu}_{\text{Ric1}}\big|_{n=1}&=&
2R^{(\mu}_\sigma\Box{R}^{\nu)\sigma}
-2\nabla_\sigma\nabla^{(\mu}\Box{R}^{\nu)\sigma}
+2\nabla^\rho\Big(R^{\sigma(\mu}
\nabla^{\nu)}R_{\rho\sigma}\Big)
\nn \\
&&-2\nabla_\rho\Big(R^{\rho\sigma}
\nabla^{(\mu}R^{\nu)}_{\sigma}\Big)
-\Big(\nabla^{(\mu}R_{\rho\sigma}\Big)
\nabla^{\nu)}R^{\rho\sigma}
+\Box^{2}R^{\mu\nu}  \nn \\
&&+g^{\mu\nu}\nabla_\rho\nabla_\sigma
\Box{R}^{\rho\sigma}
+\frac{1}{2}g^{\mu\nu}
\big(\nabla_\alpha{R}_{\rho\sigma}\big)
\nabla^\alpha{R}^{\rho\sigma}
\, . \label{EomLaRic1n1}
\eea
Furthermore, substituting Eq. (\ref{PiPforLRic1}) into
Eq. (\ref{KmnRic}) to compute the Noether potential
$K^{\mu\nu}_{\text{Ric1}}$ corresponding to the
Lagrangian (\ref{LagRicBnRic}), we have
\bea
K^{\mu\nu}_{\text{Ric1}}&=&
2\Big(\Box^nR^{\sigma[\mu}\Big)\nabla_\sigma\zeta^{\nu]}
-2\Big(\Box^nR^{[\mu}_{\sigma}\Big)\nabla^{\nu]}\zeta^\sigma
+4\zeta^{[\mu}\nabla_\sigma\Box^n{R}^{\nu]\sigma}
-4\zeta^\sigma\nabla^{[\mu}\Box^n{R}^{\nu]}_{\sigma}\nn \\
&&+2\sum^n_{k=1}\zeta^{[\mu}\Big(\nabla^{\nu]}\Box^{n-k}
R^{\rho}_{\sigma}\Big)
\Box^{k-1} R^{\sigma}_\rho
+4\zeta^\lambda\sum^n_{k=1}
\Big(\Box^{n-k} R^{[\mu}_\sigma\Big)
\nabla^{\nu]}\Box^{k-1} R^{\sigma}_{\lambda} \nn \\
&&+4\zeta^\lambda\sum^n_{k=1}
\Big(\Box^{n-k} R^{\sigma[\mu}\Big)
\nabla_{\lambda}\Box^{k-1} R^{\nu]}_{\sigma}
+4\zeta^\lambda\sum^n_{k=1}\Big(\Box^{n-k}
R^{\sigma}_{\lambda}\Big)
\nabla^{[\mu}\Box^{k-1} R^{\nu]}_{\sigma}
\, . \label{KmnRicc1}
\eea
As a special case in which $n=1$, the Noether potential
$K^{\mu\nu}_{\text{Ric1}}$ reduces to the one
$K^{\mu\nu}_{\text{Ric1}}\big|_{n=1}$ associated to
the Lagrangian $\sqrt{-g}R^{\mu\nu}\Box{R}_{\mu\nu}$,
given by
\bea
K^{\mu\nu}_{\text{Ric1}}\big|_{n=1}&=&
2\Big(\Box{R}^{\sigma[\mu}\Big)\nabla_\sigma\zeta^{\nu]}
-2\Big(\Box{R}^{[\mu}_{\sigma}\Big)\nabla^{\nu]}\zeta^\sigma
+2R^{\sigma}_\rho\zeta^{[\mu}\nabla^{\nu]}
R^{\rho}_{\sigma}
+4\zeta^{[\mu}\nabla_\sigma\Box{R}^{\nu]\sigma}\nn \\
&&-4\zeta^\rho\nabla^{[\mu}\Box{R}^{\nu]}_{\rho}
+4\zeta^\rho{R}^{[\mu}_\sigma
\nabla^{\nu]}R^{\sigma}_{\rho}
+4\zeta^\rho{R}^{[\mu}_\sigma
\nabla_{\rho}R^{\nu]\sigma}
+4\zeta^\rho{R}^{\sigma}_{\rho}
\nabla^{[\mu}R^{\nu]}_{\sigma}
\, . \label{KmnRicc1n1}
\eea
Furthermore, the surface term $\Theta^\mu_{\text{Ric1}}$
for the Lagrangian $L_{\text{Ric1}}$ is read off as
\bea
\Theta^\mu_{\text{Ric1}}&=& 2\Big(\Box^nR^{\rho[\mu}\Big)
\nabla^{\nu]}\delta g_{\rho\nu}
+2g^{\rho[\mu}\Big(\Box^nR^{\nu]\sigma}\Big)
\nabla_\sigma\delta g_{\rho\nu}
+2(\delta g_{\rho\nu})\nabla^{[\mu}\Box^nR^{\nu]\rho} \nn \\
&&-2(\delta g_{\rho\nu})g^{\rho[\mu}
\nabla_\sigma\Box^n{R}^{\nu]\sigma}
+\Theta^\mu_{\text{Ric}(n)}\big|_{P^{\mu\nu}_{(n)}=R^{\mu\nu}}
\, . \label{TheRic1def}
\eea

Within the framework for the Lagrangian
(\ref{LagRicBnRic}), the identity (\ref{Ident1Ric})
takes the form
\be
R^{[\mu}_\sigma\Box^nR^{\nu]\sigma}
=-\sum^n_{k=1}
\nabla_\lambda\Big[\Big(\nabla^{\lambda}\Box^{k-1}
R^{[\mu}_\sigma\Big)\Box^{n-k} R^{\nu]\sigma}\Big]
=-R^{[\nu}_\sigma\Box^nR^{\mu]\sigma}
\, . \label{Id1Ricc1}
\ee
Apparently, Eq. (\ref{Id1Ricc1})
holds identically. Apart from this, the identity
(\ref{Ident2Ric}) is specific to
\be
\frac{\partial{L_{\text{Ric1}}}}{\partial{g}^{\mu\nu}}
=R_{\sigma\mu}\Box^nR^\sigma_{\nu}
+R_{\sigma\nu}\Box^nR^\sigma_{\mu}
\, . \label{Id2Ricc1}
\ee
As a matter of fact, the above equality can be reproduced
via making a straightforward computation for the derivative
of $L_{\text{Ric1}}$ with respect to the inverse metric
${g}^{\mu\nu}$.

%%%%%%%%%%%%%%%%%%%%%%%%%%%%%%%%%%%%%%%%%%%%%%%%%%%%%%%%%%%%%%%%%%%%%%%%
\section{Equations of motion and Noether potentials for the
Lagrangians with the variables $\Box^i R_{\mu\nu\rho\sigma}$s}\label{four}
%%%%%%%%%%%%%%%%%%%%%%%%%%%%%%%%%%%%%%%%%%%%%%%%%%%%%%%%%%%%%%%%%%%%%%%%

By analogy with the previous section, we pay attention to
the field equations and the Noether potential associated with
the general Lagrangian depending upon the inverse metric
${g}^{\mu\nu}$ and the Riemann tensor $R_{\mu\nu\rho\sigma}$,
together with $\Box^i R_{\mu\nu\rho\sigma}$s obtained via
$i$th ($i=1,2,\cdot\cdot\cdot,m$) powers of the
Beltrami-d'Alembertian operator $\Box$ acting on the latter,
presented by the following form
\be
\sqrt{-g}L_{\text{Riem}}=\sqrt{-g}
L_{\text{Riem}}(g^{\mu\nu}, R_{\mu\nu\rho\sigma},
\Box R_{\mu\nu\rho\sigma},
\cdot\cdot\cdot,\Box^m R_{\mu\nu\rho\sigma})
\, , \label{LagBoxRiem}
\ee
which is supposed to satisfy the fundamental requirements
for invariance under diffeomorphisms and
includes the Lagrangians $\sqrt{-g}L_{R}$
and $\sqrt{-g}L_{\text{Ric}}$ as its two special cases.
This implies that the results obtained in the present
section are applicable to both the Lagrangians.

%%%%%%%%%%%%%%%%%%%%%%%%%%%%%%%%%%%%%%%%%%%%%%%%%%%%%%%%
\subsection{The generic outcomes for field equations
and Noether potentials}\label{four1}
%%%%%%%%%%%%%%%%%%%%%%%%%%%%%%%%%%%%%%%%%%%%%%%%%%%%%%%%%

When the Lagrangian (\ref{LagBoxRiem}) is varied with
respect to all the variables ${g}^{\mu\nu}$,
$R_{\mu\nu\rho\sigma}$ and $\Box^i R_{\mu\nu\rho\sigma}$s
$(i=1,\cdot\cdot\cdot,m)$, the result is read off as
\bea
\delta\big(\sqrt{-g}L_{\text{Riem}}\big)
&=&\sqrt{-g}\left[\left(
\frac{\partial{L_{\text{Riem}}}}{\partial{g}^{\mu\nu}}
-\frac{1}{2}L_{\text{Riem}}g_{\mu\nu}\right)
\delta g^{\mu\nu}
+P^{\mu\nu\rho\sigma}_{(0)}
\delta{R}_{\mu\nu\rho\sigma}\right. \nn \\
&&\left.+\sum^{m}_{i=1}P^{\mu\nu\rho\sigma}_{(i)}
\delta \Box^i{R}_{\mu\nu\rho\sigma}\right]
\, , \label{VaryLagRiem}
\eea
in which all the fourth-rank tensors
$P^{\mu\nu\rho\sigma}_{(0)}$ and
$P^{\mu\nu\rho\sigma}_{(i)}$s
are defined through
\be
P^{\mu\nu\rho\sigma}_{(0)}=\frac{\partial{L_{\text{Riem}}}}
{\partial{R}_{\mu\nu\rho\sigma}} \, , \qquad
P^{\mu\nu\rho\sigma}_{(i)}=\frac{\partial{L_{\text{Riem}}}}
{\partial\Box^i{R}_{\mu\nu\rho\sigma}}
\, , \label{P40idef}
\ee
respectively. Here all the tensors $P^{\mu\nu\rho\sigma}_{(l)}$s
$(l=0,\cdot\cdot\cdot,m)$ exhibit the algebraic symmetries
\be
P^{\mu\nu\rho\sigma}_{(l)}=-P^{\nu\mu\rho\sigma}_{(l)}
=-P^{\mu\nu\sigma\rho}_{(l)}
=P^{\rho\sigma\mu\nu}_{(l)}
\, . \label{P4iAlSym}
\ee
As before, with the purpose to provide convenience to extract
terms proportional to $\delta{R}_{\mu\nu\rho\sigma}$
out of all the ones $P^{\mu\nu\rho\sigma}_{(i)}\delta\Box^i
{R}_{\mu\nu\rho\sigma}$s $(i=1,\cdot\cdot\cdot,m)$
in Eq. (\ref{VaryLagRiem}), our first task is to introduce
scalars $\Upsilon_{(i,k)}$ $(k=1,\cdot\cdot\cdot,i+1)$
given by
\be
\Upsilon_{(i,k)}
=\left(\Box^{k-1}P^{\mu\nu\rho\sigma}_{(i)}\right)
\delta\Box^{i-k+1}{R}_{\mu\nu\rho\sigma}
\, , \label{Upsiikdef}
\ee
in addition to three tensors $U^{\mu}_{(i,k)}$,
$V^{\mu\nu}_{(i,k)}$, and $W^{\lambda\mu\nu}_{(i,k)}$.
Specifically, the vector $U^{\mu}_{(i,k)}$ is expressed
in terms of $\delta\nabla^{\mu}\Box^{i-k} R_{\alpha\beta\rho\sigma}$
and $\delta\Box^{i-k} R_{\alpha\beta\rho\sigma}$ as
\be
U^{\mu}_{(i,k)}=
\left(\Box^{k-1} P^{\alpha\beta\rho\sigma}_{(i)}\right)
\left(\delta\nabla^{\mu}\Box^{i-k} R_{\alpha\beta\rho\sigma}\right)
-\left(\nabla^{\mu}\Box^{k-1}
P^{\alpha\beta\rho\sigma}_{(i)}\right)
\left(\delta\Box^{i-k} R_{\alpha\beta\rho\sigma}\right)
\, , \label{Uikdef}
\ee
the second-rank symmetric tensor $V^{\mu\nu}_{(i,k)}$
is read off as
\be
V^{\mu\nu}_{(i,k)}=
\left(\nabla^{(\mu}\Box^{i-k} R_{\alpha\beta\rho\sigma}\right)
\left(\nabla^{\nu)}\Box^{k-1} P^{\alpha\beta\rho\sigma}_{(i)}\right)
\, , \label{Vikdef}
\ee
and the third-rank tensor $W^{\lambda\mu\nu}_{(i,k)}$
has the form
\bea
W^{\lambda\mu\nu}_{(i,k)}&=&H^{\lambda\mu\nu}_{\text{Riem}(i,k)}
+g^{\lambda\mu}\left(\nabla^{\nu}\Box^{i-k}
R_{\alpha\beta\rho\sigma}\right)
\left(\Box^{k-1} P^{\alpha\beta\rho\sigma}_{(i)}\right)
\, , \nn \\
H^{\lambda\mu\nu}_{\text{Riem}(i,k)}&=&
4\left(\nabla^{\mu}\Box^{k-1} P^{\nu\tau\rho\sigma}_{(i)}\right)
\Box^{i-k} R^{\lambda}_{~\tau\rho\sigma}
-4\left(\Box^{k-1} P^{\nu\tau\rho\sigma}_{(i)}\right)
\nabla^{\mu}\Box^{i-k}R^{\lambda}_{~\tau\rho\sigma}
\, . \label{Wikdef}
\eea
According to the definitions for the four tensors
$U^{\mu}_{(i,k)}$, $V^{\mu\nu}_{(i,k)}$,
$W^{\lambda\mu\nu}_{(i,k)}$ and $H^{\lambda\mu\nu}_{\text{Riem}(i,k)}$,
we find that they fulfill a useful relation
\bea
U^{\mu}_{(i,k)}(\delta\rightarrow\nabla^\nu)
&=&\nabla^{\nu}\left[\left(\nabla^{\mu}\Box^{i-k}
R_{\alpha\beta\rho\sigma}\right)
\left(\Box^{k-1}P^{\alpha\beta\rho\sigma}_{(i)}\right)\right]
-2V^{\mu\nu}_{(i,k)} \nn \\
&=&\nabla_\lambda{W}^{\nu\lambda\mu}_{(i,k)}
-\nabla_\lambda{H}^{\nu\lambda\mu}_{\text{Riem}(i,k)}
-2V^{\mu\nu}_{(i,k)}
\, . \label{UVWrelat}
\eea
With all the quantities defined by equations from (\ref{Upsiikdef})
to (\ref{Wikdef}), complex calculations indicate that the
scalar $\Upsilon_{(i,k)}$ can be associated to the one
$\Upsilon_{(i,k+1)}$ in the following manner
\be
\Upsilon_{(i,k)}=\Upsilon_{(i,k+1)}
+\nabla_\mu U^{\mu}_{(i,k)}
+V^{\mu\nu}_{(i,k)} \delta g_{\mu\nu}
+g_{\gamma\lambda}W^{\lambda\mu\nu}_{(i,k)}
\delta \Gamma^\gamma_{\mu\nu}
\, . \label{UpsikRel}
\ee
Here we point out that a generalization of Eq. (\ref{UpsikRel})
with respect to two arbitrary tensors instead of both the fourth-rank
ones $P^{\mu\nu\rho\sigma}_{(i)}$ and ${R}_{\mu\nu\rho\sigma}$
will be given by Eq. (\ref{OmegikRel}) in the next section.
As before, on the basis of Eq. (\ref{UpsikRel}), we further
arrive at
\be
\Upsilon_{(i,1)}=\Upsilon_{(i,i+1)}
+\sum^i_{k=1}\nabla_\mu U^{\mu}_{(i,k)}
+\sum^i_{k=1}V^{\mu\nu}_{(i,k)} \delta g_{\mu\nu}
+g_{\gamma\lambda}\sum^i_{k=1}W^{\lambda\mu\nu}_{(i,k)}
\delta \Gamma^\gamma_{\mu\nu}
\, . \label{SumUpsik0}
\ee
Here $\Upsilon_{(i,1)}=P^{\mu\nu\rho\sigma}_{(i)}
\delta\Box^i{R}_{\mu\nu\rho\sigma}$ and
$\Upsilon_{(i,i+1)}=\big(\Box^iP^{\mu\nu\rho\sigma}_{(i)}\big)
\delta{R}_{\mu\nu\rho\sigma}$ according to the
definition (\ref{Upsiikdef}) for the scalar $\Upsilon_{(i,k)}$.
As a consequence of Eq. (\ref{SumUpsik0}), we find that
the contraction between the tensors $P^{\mu\nu\rho\sigma}_{(i)}$
and $\delta\Box^i{R}_{\mu\nu\rho\sigma}$
is able to be expressed as
\bea
P^{\mu\nu\rho\sigma}_{(i)}\delta\Box^i{R}_{\mu\nu\rho\sigma}
&=&\sum^i_{k=1}
\left[V^{\mu\nu}_{(i,k)}
-\frac{1}{2}\nabla_\lambda\Big(W^{(\mu\nu)\lambda}_{(i,k)}
+W^{(\mu|\lambda|\nu)}_{(i,k)}
-W^{\lambda(\mu\nu)}_{(i,k)}\Big)\right]\delta g_{\mu\nu}\nn \\
&&+\big(\Box^iP^{\mu\nu\rho\sigma}_{(i)}\big)
\delta{R}_{\mu\nu\rho\sigma}
+\nabla_\mu\Theta_{\text{Riem}(i)}^\mu
\, . \label{SumUpsik}
\eea
Within Eq. (\ref{SumUpsik}), by introducing
a vector $\Theta_{\text{Riem}(i,k)}^\mu$ defined
in terms of both the tensors $U^{\mu}_{(i,k)}$ and
$W^{\mu\rho\sigma}_{(i,k)}$ as
\be
\Theta_{\text{Riem}(i,k)}^\mu
=U^{\mu}_{(i,k)}
+\frac{1}{2}\Big(W^{(\rho\sigma)\mu}_{(i,k)}
+W^{(\rho|\mu|\sigma)}_{(i,k)}
-W^{\mu(\rho\sigma)}_{(i,k)}\Big)
\delta{g}_{\rho\sigma}
\, , \label{TheRiemik}
\ee
the surface term $\Theta_{\text{Riem}(i)}^\mu$ has the
form
\be
\Theta_{\text{Riem}(i)}^\mu=
\sum^i_{k=1}\Theta_{\text{Riem}(i,k)}^\mu
\, . \label{TheRiemidef}
\ee
With the help of a fourth-rank tensor $P^{\mu\nu\rho\sigma}$
defined through
\be
P^{\mu\nu\rho\sigma}=P^{\mu\nu\rho\sigma}_{(0)}
+\sum^{m}_{i=1}\Box^i P^{\mu\nu\rho\sigma}_{(i)}
\, , \label{TenP4def}
\ee
substituting Eq. (\ref{SumUpsik}) into Eq. (\ref{VaryLagRiem})
renders the variation of the Lagrangian to be further
expressed as the linear combination of divergence terms
together with terms proportional to the variations of
the metric and the Riemann tensor, namely,
\bea
\delta\big(\sqrt{-g}L_{\text{Riem}}\big)
&=& \sqrt{-g}\sum^m_{i=1}\sum^i_{k=1}
\left[V^{\mu\nu}_{(i,k)}
-\frac{1}{2}\nabla_\lambda\Big(W^{(\mu\nu)\lambda}_{(i,k)}
+W^{(\mu|\lambda|\nu)}_{(i,k)}
-W^{\lambda(\mu\nu)}_{(i,k)}\Big)\right]
\delta g_{\mu\nu}\nn \\
&&+\sqrt{-g}
\left(\frac{\partial{L_{\text{Riem}}}}{\partial{g}^{\mu\nu}}
-\frac{1}{2}L_{\text{Riem}}g_{\mu\nu}\right)
\delta g^{\mu\nu}
+\sqrt{-g}P^{\mu\nu\rho\sigma}\delta R_{\mu\nu\rho\sigma}
\nn \\
&&+\sqrt{-g}\sum^m_{i=1}\nabla_\mu\Theta_{\text{Riem}(i)}^\mu
\, . \label{VaryLagRiem2}
\eea
Here the scalar $P^{\mu\nu\rho\sigma}\delta R_{\mu\nu\rho\sigma}$
can be written as the linear combination for a term proportional
to the variation of the metric and the divergence of a
surface term $\Theta^\mu_{\text{Riem}(0)}$,
presented by
\be
\Theta^\mu_{\text{Riem}(0)}=2P^{\mu\nu\rho\sigma}
\nabla_\sigma\delta g_{\rho\nu}
-2(\delta g_{\nu\rho})
\nabla_\sigma{P}^{\mu\nu\rho\sigma}
\, . \label{TheRiem0def}
\ee
Specifically, it takes the following form
\cite{JJP2306,Pady,PWG23}
\be
P^{\mu\nu\rho\sigma}\delta R_{\mu\nu\rho\sigma}
=\big(P^{\mu\tau\rho\sigma}
R^\nu_{~\tau\rho\sigma}+2\nabla_\rho\nabla_\sigma
P^{\rho\mu\nu\sigma}\big)\delta g_{\mu\nu}
+\nabla_\mu \Theta_{\text{Riem}(0)}^\mu
\, , \label{P4mndelRic}
\ee
in which the ati-symmetric component of the tensor $P^{\mu\tau\rho\sigma}
R^\nu_{~\tau\rho\sigma}$ satisfies identically
\be
P^{[\mu|\tau\rho\sigma|}
R^{\nu]}_{~~\tau\rho\sigma}=-2\nabla_\rho\nabla_\sigma
P^{\rho[\mu\nu]\sigma}
\, . \label{PRiemAnSym}
\ee
As a consequence of the substitution of Eq. (\ref{P4mndelRic})
into Eq. (\ref{VaryLagRiem2}), the variation for the Lagrangian
(\ref{LagBoxRiem}) is ultimately written as the following
conventional form
\be
\delta\left(\sqrt{-g}L_{\text{Riem}}\right)
=\sqrt{-g}\left(E^{\text{Riem}}_{\mu\nu} \delta g^{\mu\nu}
+\nabla_\mu\Theta^\mu_{\text{Riem}}\right)
\, . \label{VaryLagRiem3}
\ee
In the above equation, the expression $E^{\text{Riem}}_{\mu\nu}$
for field equations is read off as
\bea
E^{\mu\nu}_{\text{Riem}}&=&
\frac{\partial{L_{\text{Riem}}}}{\partial{g}^{\rho\sigma}}
g^{\mu\rho}g^{\nu\sigma}
-\frac{1}{2}L_{\text{Riem}}g^{\mu\nu}
-P^{(\mu|\lambda\rho\sigma|}
R^{\nu)}_{~\lambda\rho\sigma}-2\nabla_\rho\nabla_\sigma
P^{\rho(\mu\nu)\sigma}\nn \\
&&
-\sum^m_{i=1}\sum^i_{k=1}
V^{\mu\nu}_{(i,k)}
+\frac{1}{2}\sum^m_{i=1}\sum^i_{k=1}
\nabla_\lambda\Big(W^{(\mu\nu)\lambda}_{(i,k)}
+W^{(\mu|\lambda|\nu)}_{(i,k)}
-W^{\lambda(\mu\nu)}_{(i,k)}\Big)
\, , \label{EoMforLagRiem}
\eea
and the surface term $\Theta^\mu_{\text{Riem}}$
takes the form
\bea
\Theta^\mu_{\text{Riem}}&=&\Theta^\mu_{\text{Riem}(0)}
+\sum^m_{i=1}\Theta^\mu_{\text{Riem}(i)} \nn \\
&=&2P^{\mu\nu\rho\sigma}
\nabla_\sigma\delta g_{\rho\nu}
-2(\delta g_{\nu\rho})
\nabla_\sigma{P}^{\mu\nu\rho\sigma}
+\sum^m_{i=1}\sum^i_{k=1}\Theta^\mu_{\text{Riem}(i,k)}
\, . \label{TThetRiem}
\eea

Like before, in what follows, we shall follow the method in terms
of the conserved current to derive an economic and simple
form for the field equations depending on the Riemann
tensor and its covariant derivatives but in the absence
of the term consisting of the derivative for the Lagrangian
density with regard to the metric. Meanwhile,
the Noether potential corresponding to any smooth vector
$\zeta^\mu$ will be obtained. As what has been shown before,
a key characteristic of this method is to calculate the surface
term under the transformation for the variation operator into
the Lie derivative along an arbitrary smooth vector. According to
this, we begin with performing computations on the surface terms
$\Theta^\mu_{\text{Riem}(0)}(\delta\rightarrow\mathcal{L}_\zeta)$
and
$\Theta^\mu_{\text{Riem}(i,k)}(\delta\rightarrow\mathcal{L}_\zeta)$.
The first one is presented by
\be
\Theta^\mu_{\text{Riem}(0)}(\delta\rightarrow\mathcal{L}_\zeta)
= 2 \big(P^{\mu\lambda\rho\sigma}R^{\nu}_{~\lambda\rho\sigma}
-2\nabla_\rho\nabla_\sigma{P}^{\rho\mu\nu\sigma}\big)\zeta_\nu
-\nabla_\nu K^{\mu\nu}_{\text{Riem}(0)}
\, , \label{ThetRiem0Lie}
\ee
in which the anti-symmetric tensor $K^{\mu\nu}_{\text{Riem}(0)}$
has the form \cite{JJP2306}
\be
K^{\mu\nu}_{\text{Riem}(0)}=
2P^{\mu\nu\rho\sigma}\nabla_{\rho}\zeta_{\sigma}
+4\zeta_\rho\nabla_\sigma P^{\mu\nu\rho\sigma}
-6P^{\mu[\nu\rho\sigma]}\nabla_\rho\zeta_\sigma
\, . \label{KmnRiem0def}
\ee
And the second quantity is read off as
\be
\Theta^\mu_{\text{Riem}(i,k)}(\delta\rightarrow\mathcal{L}_\zeta)
= 2 \zeta_\nu{X}^{\mu\nu}_{\text{Riem}(i,k)}
-\nabla_\nu K^{\mu\nu}_{\text{Riem}(i,k)}
\, . \label{ThetRiemikLie}
\ee
By means of the following equation
\be
U^\mu_{(i,k)}(\delta\rightarrow\mathcal{L}_\zeta)
=\zeta_\nu{U}^{\mu}_{(i,k)}(\delta\rightarrow\nabla^\nu)
-W^{\nu\mu\lambda}_{(i,k)}\nabla_\lambda\zeta_\nu
\, , \label{UikdelLie}
\ee
the second-rank tensor ${X}^{\mu\nu}_{\text{Riem}(i,k)}$
in Eq. (\ref{ThetRiemikLie}) is given by
\bea
{X}^{\mu\nu}_{\text{Riem}(i,k)}&=&
\frac{1}{2}U^{\mu}_{(i,k)}(\delta\rightarrow\nabla^\nu)
+\frac{1}{2}\nabla_\lambda\Big(
W^{(\mu\nu)\lambda}_{(i,k)}
-W^{\lambda(\mu\nu)}_{(i,k)}
+W^{[\mu|\lambda|\nu]}_{(i,k)}\Big) \nn \\
&=&\frac{1}{2}\nabla_\lambda\Big(
W^{(\mu\nu)\lambda}_{(i,k)}
+W^{(\mu|\lambda|\nu)}_{(i,k)}
-W^{\lambda(\mu\nu)}_{(i,k)}\Big)
-V^{\mu\nu}_{(i,k)}
-\frac{1}{2}\nabla_\lambda{H}^{\nu\lambda\mu}_{\text{Riem}(i,k)}
\, . \label{XmnRiemik}
\eea
In order to obtain the concrete expression for the tensor
${X}^{\mu\nu}_{\text{Riem}(i,k)}$ expressed in terms of
the tensors $P^{\alpha\beta\rho\sigma}_{(i)}$s, we find
that it is of great
convenience to split it up into the form
\be
{X}^{\mu\nu}_{\text{Riem}(i,k)}={X}^{(\mu\nu)}_{\text{Riem}(i,k)}
+{X}^{[\mu\nu]}_{\text{Riem}(i,k)}
\, , \label{XmnRiemik2}
\ee
in which the symmetric component
${X}^{(\mu\nu)}_{\text{Riem}(i,k)}$ is given by
\bea
{X}^{(\mu\nu)}_{\text{Riem}(i,k)}&=&
\frac{1}{2}g^{\mu\nu}\nabla_\lambda\Big[
\Big(\nabla^{\lambda}\Box^{i-k}
R_{\alpha\beta\rho\sigma}\Big)
\Box^{k-1} P^{\alpha\beta\rho\sigma}_{(i)}\Big]
-\Big(\nabla^{(\mu}\Box^{i-k}
R_{\alpha\beta\rho\sigma}\Big)
\nabla^{\nu)}\Box^{k-1}
P^{\alpha\beta\rho\sigma}_{(i)} \nn \\
&&+2\nabla_\lambda\Big[\Big(\Box^{i-k}
R^{(\mu}_{~~\tau\rho\sigma}\Big)
\nabla^{\nu)}\Box^{k-1}
P^{\lambda\tau\rho\sigma}_{(i)}\Big]
-2\nabla_\lambda\Big[\Big(\nabla^{(\mu}\Box^{i-k}
R^{\nu)}_{~~\tau\rho\sigma}\Big)
\Box^{k-1}P^{\lambda\tau\rho\sigma}_{(i)}\Big]\nn \\
&&+2\nabla^\lambda\Big[\Big(\nabla^{(\mu}\Box^{i-k}
R_{\lambda\tau\rho\sigma}\Big)
\Box^{k-1}P^{\nu)\tau\rho\sigma}_{(i)}\Big]
-2\nabla^\lambda\Big[\Big(\Box^{i-k}
R_{\lambda\tau\rho\sigma}\Big)
\nabla^{(\mu}\Box^{k-1}
P^{\nu)\tau\rho\sigma}_{(i)}\Big]
\, ,  \nn \\
\label{SXmnRiemik2}
\eea
while the anti-symmetric one
${X}^{[\mu\nu]}_{\text{Riem}(i,k)}$ can be directly read off
from Eq. (\ref{XmnRiemik}), having the following form
\bea
{X}^{[\mu\nu]}_{\text{Riem}(i,k)}&=&
\frac{1}{2}\nabla_\lambda{H}^{[\mu|\lambda|\nu]}_{\text{Riem}(i,k)}\nn\\
&=&2\nabla^\lambda\Big[
\Big(\Box^{i-k} R^{[\mu}_{~~\tau\rho\sigma}\Big)
\Big(\nabla_{\lambda}\Box^{k-1}
P^{\nu]\tau\rho\sigma}_{(i)}\Big)\Big] \nn \\
&&-2\nabla^\lambda\Big[
\Big(\nabla_{\lambda}\Box^{i-k}
R^{[\mu}_{~~\tau\rho\sigma}\Big)\Big(\Box^{k-1}
P^{\nu]\tau\rho\sigma}_{(i)}\Big)\Big]
\, . \label{AXmnRiemik2}
\eea
Besides, within Eq. (\ref{ThetRiemikLie}), the second-rank
anti-symmetric tensor
$K^{\mu\nu}_{\text{Riem}(i,k)}$ is given by
\be
K^{\mu\nu}_{\text{Riem}(i,k)}=
\zeta_\lambda\Big(W^{[\mu\nu]\lambda}_{(i,k)}
+W^{[\mu|\lambda|\nu]}_{(i,k)}
+W^{\lambda[\mu\nu]}_{(i,k)}\Big)
\, . \label{KmnRiemik}
\ee
By substituting $W^{\lambda\mu\nu}_{(i,k)}$ given by
Eq. (\ref{Wikdef}) into Eq. (\ref{KmnRiemik}), the tensor
$K^{\mu\nu}_{\text{Riem}(i,k)}$ is specifically written as
\bea
K^{\mu\nu}_{\text{Riem}(i,k)}&=&
2\zeta^{[\mu}\Big(\nabla^{\nu]}\Box^{i-k}
R_{\alpha\beta\rho\sigma}\Big)
\Box^{k-1} P^{\alpha\beta\rho\sigma}_{(i)}
+4\zeta_\lambda\Big(\nabla^{[\mu}\Box^{i-k}
R^{\nu]}_{~~\tau\rho\sigma}\Big)
\Box^{k-1} P^{\lambda\tau\rho\sigma}_{(i)}
\nn \\
&&+4\zeta_\lambda\Big(\Box^{i-k}
R^{[\mu}_{~~\tau\rho\sigma}\Big)\nabla^{\nu]}
\Box^{k-1} P^{\lambda\tau\rho\sigma}_{(i)}
-4\zeta^\lambda\Big(\nabla^{[\mu}\Box^{i-k}
R_{\lambda\tau\rho\sigma}\Big)
\Box^{k-1} P^{\nu]\tau\rho\sigma}_{(i)} \nn \\
&&+4\zeta^\lambda\Big(\Box^{i-k}
R_{\lambda\tau\rho\sigma}\Big)
\nabla^{[\mu}\Box^{k-1}
P^{\nu]\tau\rho\sigma}_{(i)}
-4\zeta^\lambda\Big(\nabla_{\lambda}\Box^{i-k}
R^{[\mu}_{~~\tau\rho\sigma}\Big)
\Box^{k-1} P^{\nu]\tau\rho\sigma}_{(i)}\nn \\
&&+4\zeta^\lambda\Big(\Box^{i-k}
R^{[\mu}_{~~\tau\rho\sigma}\Big)
\nabla_{\lambda}\Box^{k-1} P^{\nu]\tau\rho\sigma}_{(i)}
\, . \label{KmnRiemik2}
\eea

By the aid of Eqs. (\ref{ThetRiem0Lie}) and (\ref{ThetRiemikLie}),
from Eq. (\ref{TThetRiem}) we obtain
\be
\Theta^\mu_{\text{Riem}}(\delta\rightarrow\mathcal{L}_\zeta)
=2 \zeta_\nu{X}^{\mu\nu}_{\text{Riem}}
-\nabla_\nu{K}^{\mu\nu}_{\text{Riem}}
\, . \label{TThetRiemLie}
\ee
This indicates that the surface term
$\Theta^\mu_{\text{Riem}}$ under
$\delta\rightarrow\mathcal{L}_\zeta$ is guaranteed to be
similarly expressed as the form (\ref{TheLiegen}).
As a consequence, equation (\ref{TThetRiemLie}) harbours
the sufficient ingredients to produce the Noether
potential and field equations. In Eq. (\ref{TThetRiemLie}),
the rank-2 tensor $X^{\mu\nu}_{\text{Riem}}$ is presented by
\be
X^{\mu\nu}_{\text{Riem}}=P^{\mu\lambda\rho\sigma}
R^{\nu}_{~\lambda\rho\sigma}
-2\nabla_\rho\nabla_\sigma{P}^{\rho\mu\nu\sigma}
+\sum^m_{i=1}\sum^i_{k=1}
{X}^{\mu\nu}_{\text{Riem}(i,k)}
\, , \label{XRiemdef}
\ee
and the anti-symmetric tensor $K^{\mu\nu}_{\text{Riem}}$,
standing for the Noether potential associated to the Lagrangian
(\ref{LagBoxRiem}) due to Eq. (\ref{TheLiegen}), is expressed as
\be
K^{\mu\nu}_{\text{Riem}}=
K^{\mu\nu}_{\text{Riem}(0)}
+\sum^m_{i=1}\sum^i_{k=1}
K^{\mu\nu}_{\text{Riem}(i,k)}
\, , \label{KmnRiem}
\ee
which corresponds to the conserved current
\be
J^{\mu}_{\text{Riem}}=\nabla_\nu{K}^{\mu\nu}_{\text{Riem}}
=2\zeta_\nu{E}^{\mu\nu}_{\text{Riem}}+\zeta^\mu{L}_{\text{Riem}}
-\Theta^\mu_{\text{Riem}}(\delta\rightarrow\mathcal{L}_\zeta)
\, . \label{ConCurLRiim}
\ee
By making use of the Noether potential $K^{\mu\nu}_{\text{Riem}}$,
together with the surface term $\Theta^\mu_{\text{Riem}}$, we are
able to further define the well-known Iyer-Wald potential
corresponding to a Killing vector $\xi^\mu$, being of the form
\cite{LeeWald,IyWald,WalZo}
\be
Q^{\mu\nu}_{\text{Riem}}=\frac{1}{\sqrt{-g}}
\delta\left(\sqrt{-g}
{K}^{\mu\nu}_{\text{Riem}}(\zeta\rightarrow\xi)\right)
-\xi^{[\mu}\Theta^{\nu]}_{\text{Riem}}
\, . \label{IWpotofLagRiem}
\ee
Here the Iyer-Wald potential $Q^{\mu\nu}_{\text{Riem}}$ can be
adopted to define conserved charges for gravity theories described
by the Lagrangian (\ref{LagBoxRiem}), such as the entropy, the mass
and the angular momentum.
Apart from the Noether potential,
according to Eq. (\ref{EoMgen}), an alternative economic
and simple formulation for equations of motion corresponding
to the Lagrangian (\ref{LagBoxRiem}) can be extracted
out of Eq. (\ref{TThetRiemLie}), which is read off as
\bea
E^{\mu\nu}_{\text{Riem}}&=&X^{\mu\nu}_{\text{Riem}}
-\frac{1}{2}L_{\text{Riem}}g^{\mu\nu} \nn \\
&=&P^{\mu\lambda\rho\sigma}
R^{\nu}_{~\lambda\rho\sigma}
-2\nabla_\rho\nabla_\sigma{P}^{\rho\mu\nu\sigma}
-\frac{1}{2}L_{\text{Riem}}g^{\mu\nu}
+\sum^m_{i=1}\sum^i_{k=1}
{X}^{\mu\nu}_{\text{Riem}(i,k)}
\, . \label{EoMforLagRiem2}
\eea
By contrast with Eq. (\ref{EoMforLagRiem}), here the expression
$E^{\mu\nu}_{\text{Riem}}$ for equations of motion does not
incorporate the term composed of the derivative of the Lagrangian
density with respect to
the metric. As a matter of fact, ${E}^{\mu\nu}_{\text{Riem}}$ in
the absence of such a term possesses the advantage to render it much
easier to prove that the field equations are divergence-free, namely,
$\nabla_\mu{E}^{\mu\nu}_{\text{Riem}}=0$. This will be
demonstrated within Appendix \ref{appendC}.

Finally, we consider two identities in connection with the field equations.
Since the second-rank tensor $E^{\mu\nu}_{\text{Riem}}$ is symmetric,
one obtains an identity
${X}^{[\mu\nu]}_{\text{Riem}}=0$, or specifically,
\be
P^{[\mu|\lambda\rho\sigma|}
R^{\nu]}_{~~\lambda\rho\sigma}
= -\frac{1}{2}\sum^m_{i=1}\sum^i_{k=1}
{X}^{[\mu\nu]}_{\text{Riem}(i,k)}
=-\frac{1}{4}\sum^m_{i=1}\sum^i_{k=1}
\nabla_\lambda{H}^{[\mu|\lambda|\nu]}_{\text{Riem}(i,k)}
\, . \label{Ident1Riem}
\ee
By making use of the following identity
\bea
P^{\mu\tau\rho\sigma}_{(i)}\Box^iR^{\nu}_{~\tau\rho\sigma}
&=&R^{\nu}_{~\tau\rho\sigma}\Box^i
P^{\mu\tau\rho\sigma}_{(i)}+
\sum^i_{k=1}\nabla_\lambda\Big[
\Big(\nabla^{\lambda}\Box^{i-k}
R^{\nu}_{~\tau\rho\sigma}\Big)\Box^{k-1}
P^{\mu\tau\rho\sigma}_{(i)}\Big] \nn \\
&&-\sum^i_{k=1}\nabla_\lambda\Big[
\Big(\nabla^{\lambda}\Box^{k-1}
P^{\mu\tau\rho\sigma}_{(i)}\Big)
\Box^{i-k}R^{\nu}_{~\tau\rho\sigma}\Big]
\,  \label{IdforId2Riem}
\eea
to express the sum of
$\nabla_\lambda{H}^{[\mu|\lambda|\nu]}_{\text{Riem}(i,k)}$
as
\be
\sum^i_{k=1}\nabla_\lambda{H}^{[\mu|\lambda|\nu]}_{\text{Riem}(i,k)}=
4R^{[\mu}_{~~\tau\rho\sigma}\Box^iP^{\nu]\tau\rho\sigma}_{(i)}
-4\Big(\Box^iR^{[\mu}_{~~\tau\rho\sigma}\Big)P^{\nu]\tau\rho\sigma}_{(i)}
\, , \label{SumXPiemik}
\ee
the identity (\ref{Ident1Riem}) is transformed into
a much simpler form
\be
P^{[\mu|\lambda\rho\sigma|}_{(0)}
R^{\nu]}_{~~\lambda\rho\sigma}
= -\sum^m_{i=1}P^{[\mu|\lambda\rho\sigma|}_{(i)}
\Box^iR^{\nu]}_{~~\lambda\rho\sigma}
\, . \label{Ident1Riem2}
\ee
Apart from the above identity, by the aid of
the relation (\ref{UVWrelat}) among the
$U^{\mu}_{(i,k)}$, $V^{\mu\nu}_{(i,k)}$ and
${W}^{\lambda\mu\nu}_{(i,k)}$ tensors,
the straightforward
comparison between Eqs. (\ref{EoMforLagRiem})
and (\ref{EoMforLagRiem2}) gives rise to the second one
\bea
\left(
\frac{\partial{L_{\text{Riem}}}}{\partial{g}^{\rho\sigma}}
\right)
g^{\mu\rho}g^{\nu\sigma}
&=&2P^{(\mu|\tau\rho\sigma|}
R^{\nu)}_{~~\tau\rho\sigma}
+2\sum^m_{i=1}\sum^i_{k=1}\nabla^\lambda\Big[
\Big(\nabla_\lambda\Box^{i-k}
R^{(\mu}_{~~\tau\rho\sigma}\Big)
\Box^{k-1}P^{\nu)\tau\rho\sigma}_{(i)}\Big] \nn \\
&&-2\sum^m_{i=1}\sum^i_{k=1}\nabla^\lambda\Big[\Big(\Box^{i-k}
R^{(\mu}_{~~\tau\rho\sigma}\Big)
\nabla_\lambda\Box^{k-1}P^{\nu)\tau\rho\sigma}_{(i)}\Big]
\, . \label{Ident2Riem}
\eea
As a matter of fact, by means of Eq. (\ref{IdforId2Riem}),
the identity (\ref{Ident2Riem}) is simplified as
\be
\frac{\partial{L_{\text{Riem}}}}{\partial{g}^{\mu\nu}}
=\sum^m_{l=0}\left(
g_{\mu\alpha}P^{\alpha\beta\rho\sigma}_{(l)}\Box^l
{R}_{\nu\beta\rho\sigma}
+g_{\nu\alpha}P^{\alpha\beta\rho\sigma}_{(l)}\Box^l
{R}_{\mu\beta\rho\sigma}\right)
\, . \label{Ident2Riem2}
\ee
Particularly, when $m=0$, the Lagrangian $L_{\text{Riem}}$
takes the form $L_{\text{Riem}}(g^{\mu\nu}, R_{\mu\nu\rho\sigma})$,
and the identity (\ref{Ident2Riem2}) accordingly turns into
the one given by the work \cite{Pady}.
Thanks to the above identity, it is feasible to verify that
the expression ${E}^{\mu\nu}_{\text{Riem}}$ for field equations
is divergence-free via direct computations on its divergence.
This will be illustrated within Appendix \ref{appendC}.

%%%%%%%%%%%%%%%%%%%%%%%%%%%%%%%%%%%%%%%%%%%%%%%%%%%%%%%%
\subsection{The re-derivation of the results related to
the Lagrangian $\sqrt{-g}L_{\text{Ric}}$}\label{four2}
%%%%%%%%%%%%%%%%%%%%%%%%%%%%%%%%%%%%%%%%%%%%%%%%%%%%%%%%%

Within the present subsection, in an attempt to check the results
related to the Lagrangian $\sqrt{-g}L_{\text{Riem}}$,
we utilize them to re-derive the corresponding
ones for the Lagrangian $\sqrt{-g}L_{\text{Ric}}$.
When $L_{\text{Riem}}=L_{\text{Ric}}$, all the fourth-rank
tensors $P^{\mu\nu\rho\sigma}_{(0)}$,
$P^{\mu\nu\rho\sigma}_{(i)}$s and
$P^{\mu\nu\rho\sigma}$ take the following forms
\bea
\bar{P}^{\mu\nu\rho\sigma}_{(0)}
&=&g^{[\mu|[\rho}P^{\sigma]|\nu]}_{(0)}
=g^{[\rho|[\mu}P^{\nu]|\sigma]}_{(0)}
=P_{(0)}^{[\mu|[\rho}g^{\sigma]|\nu]}
=P_{(0)}^{[\rho|[\mu}g^{\nu]|\sigma]} \, ,\nn \\
\bar{P}^{\mu\nu\rho\sigma}_{(i)}
&=&g^{[\mu|[\rho}P^{\sigma]|\nu]}_{(i)}
=g^{[\rho|[\mu}P^{\nu]|\sigma]}_{(i)}
=P_{(i)}^{[\mu|[\rho}g^{\sigma]|\nu]}
=P_{(i)}^{[\rho|[\mu}g^{\nu]|\sigma]}\, , \nn \\
\bar{P}^{\mu\nu\rho\sigma}
&=&g^{[\mu|[\rho}P^{\sigma]|\nu]}
=g^{[\rho|[\mu}P^{\nu]|\sigma]}
=P^{[\mu|[\rho}g^{\sigma]|\nu]}
=P^{[\rho|[\mu}g^{\nu]|\sigma]}
\, , \label{P4iP4forLRic}
\eea
respectively. Substituting Eq. (\ref{P4iP4forLRic}) into
$U^{\mu}_{(i,k)}$, $V^{\mu\nu}_{(i,k)}$ and
$W^{\lambda\mu\nu}_{(i,k)}$, given by Eqs. (\ref{Uikdef}),
(\ref{Vikdef}), and (\ref{Wikdef}), respectively,
we have
\bea
U^{\mu}_{(i,k)}\big|_{P\rightarrow\bar{P}}
&=&L^{\mu}_{(i,k)}
-\left(\Box^{k-1} P^{\rho\sigma}_{(i)}\right)
\left(\nabla^{\mu}\Box^{i-k}R_{\alpha\rho\beta\sigma}\right)
\delta{g}^{\alpha\beta} \nn \\
&&+\left(\nabla^{\mu}\Box^{k-1} P^{\rho\sigma}_{(i)}\right)
\left(\Box^{i-k}R_{\alpha\rho\beta\sigma}\right)
\delta{g}^{\alpha\beta} \, , \nn \\
V^{\mu\nu}_{(i,k)}\big|_{P\rightarrow\bar{P}}
&=&N^{\mu\nu}_{(i,k)}
\, , \label{UVLNrel}
\eea
together with
\bea
W^{\lambda\mu\nu}_{(i,k)}\big|_{P\rightarrow\bar{P}}&=&
M^{\lambda\mu\nu}_{(i,k)}+
2\left(\nabla^{\mu}\Box^{k-1} P_{(i)\rho\sigma}\right)
\left(\Box^{i-k} R^{\lambda\rho\nu\sigma}\right)
 \nn \\
&&-2\left(\Box^{k-1} P_{(i)\rho\sigma}\right)
\left(\nabla^{\mu}\Box^{i-k} R^{\lambda\rho\nu\sigma}\right)
\, . \label{WMikrel}
\eea
Furthermore, as what will be shown in Appendix \ref{appendA},
the substitution of Eqs. (\ref{UVLNrel})
and (\ref{WMikrel}) into Eq. (\ref{UpsikRel})
yields the relation (\ref{PsikRel}) between
$\Psi_{(i,k)}$ and $\Psi_{(i,k+1)}$ corresponding to
the Lagrangian $\sqrt{-g}L_{\text{Ric}}$. Apart from this,
we obtain
\be
\Theta^{\mu}_{\text{Ric}(0)}=
\Theta^{\mu}_{\text{Riem}(0)}\big|_{P\rightarrow\bar{P}}
 \, , \qquad
\Theta^{\mu}_{\text{Ric}(i,k)}=
\Theta^{\mu}_{\text{Riem}(i,k)}\big|_{P\rightarrow\bar{P}}
\, , \label{TheRiemRicRel}
\ee
which leads to $\Theta^{\mu}_{\text{Ric}}
=\Theta^{\mu}_{\text{Riem}}\big|_{P\rightarrow\bar{P}}$,
as well as the following relations
\be
X^{(\mu\nu)}_{\text{Ric}(i,k)}=
X^{(\mu\nu)}_{\text{Riem}(i,k)}\big|_{P\rightarrow\bar{P}}
 \, , \qquad
X^{[\mu\nu]}_{\text{Ric}(i,k)}=
X^{[\mu\nu]}_{\text{Riem}(i,k)}\big|_{P\rightarrow\bar{P}}
\, . \label{XRiemRicRel}
\ee
As a consequence of Eqs. (\ref{TheRiemRicRel}) and
(\ref{XRiemRicRel}), the expression $E^{\mu\nu}_{\text{Riem}}$
for field equations given by Eq. (\ref{EoMforLagRiem2})
turns into the one $E^{\mu\nu}_{\text{Ric}}$ in
Eq. (\ref{EoMforLagRic2}) when $L_{\text{Riem}}=L_{\text{Ric}}$.

By the aid of the equation $K^{\mu\nu}_{\text{Ric}(i,k)}=
K^{\mu\nu}_{\text{Riem}(i,k)}\big|_{P\rightarrow\bar{P}}$
derived out of the following one
\be
\Big(W^{[\mu\nu]\lambda}_{(i,k)}
+W^{[\mu|\lambda|\nu]}_{(i,k)}
+W^{\lambda[\mu\nu]}_{(i,k)}\Big)
\Big|_{P\rightarrow\bar{P}}
=M^{[\mu\nu]\lambda}_{(i,k)}
+M^{[\mu|\lambda|\nu]}_{(i,k)}
+M^{\lambda[\mu\nu]}_{(i,k)}
\, , \label{WLantisRel}
\ee
together with the equation $K^{\mu\nu}_{\text{Ric}(0)}=
K^{\mu\nu}_{\text{Riem}(0)}\big|_{P\rightarrow\bar{P}}$,
Eqs. (\ref{KmnRic}) and (\ref{KmnRiem}) enable us
to arrive at
\be
K^{\mu\nu}_{\text{Ric}}=K^{\mu\nu}_{\text{Riem}}
\big|_{P\rightarrow\bar{P}}
\, . \label{NoePRiemRicrel}
\ee
This reproduces the Noether potential $K^{\mu\nu}_{\text{Ric}}$
in the framework of the Lagrangian $\sqrt{-g}L_{\text{Riem}}$.
What is more, the identity (\ref{Ident1Ric2}) associated
with the Lagrangian $\sqrt{-g}L_{\text{Ric}}$ can be straightforwardly
derived out of the one (\ref{Ident1Riem2}) via the substitution
of Eq. (\ref{P4iP4forLRic}) into the latter. Besides,
by utilizing
\be
\frac{\partial{L_{\text{Riem}}}}{\partial{g}^{\mu\nu}}
\Big|_{P\rightarrow\bar{P}}
=\frac{\partial{L_{\text{Ric}}}}{\partial{g}^{\mu\nu}}
+P^{\rho\sigma}_{(0)}{R}_{\mu\rho\nu\sigma}
+\sum^m_{i=1}P^{\rho\sigma}_{(i)}\Box^i
{R}_{\mu\rho\nu\sigma}
\, , \label{PLagRiemPgRic}
\ee
the identity (\ref{Ident2Riem2}) becomes the one in
Eq. (\ref{Ident2Ric2}).

At the end of this subsection, we point out that the
equations of motion and the Noether potential for the
Lagrangian $\sqrt{-g}L_R$ can be also directly derived
out of the corresponding ones for
$\sqrt{-g}L_{\text{Riem}}$. Actually, in the situation
where $L_{\text{Riem}}=L_R$,
within Eqs. (\ref{KmnRiem}) and (\ref{EoMforLagRiem2}),
by performing the following replacements
\be
P^{\mu\nu\rho\sigma}_{(0)}
\rightarrow{g}^{\mu[\rho}g^{\sigma]\nu}F_{(0)} \, ,
\quad
P^{\mu\nu\rho\sigma}_{(i)}
\rightarrow{g}^{\mu[\rho}g^{\sigma]\nu}F_{(i)} \, ,
\quad
{P}^{\mu\nu\rho\sigma}
\rightarrow{F}g^{\mu[\rho}g^{\sigma]\nu}
\, , \label{P4iP4forLR}
\ee
one is able to obtain the Noether potential $K^{\mu\nu}_{R}$
in Eq. (\ref{KmnRdef}) and the expression $E^{\mu\nu}_R$ for
field equations in Eq. (\ref{EoMforLagR}), respectively.
In addition, within such a case, the identity (\ref{Ident2Riem2})
becomes
\be
\frac{\partial{L_{\text{Riem}}}}{\partial{g}^{\mu\nu}}
\Big|_{L_{\text{Riem}}=L_R}
=2\sum^m_{l=0}F_{(l)}\Box^l{R}_{\mu\nu}
\, , \label{IdentRiemtoRiS}
\ee
which can be directly verified via the computation on the
derivative of the Lagrangian $L_R$ with respect to the
inverse metric, that is,
\be
\frac{\partial{L_R}}{\partial{g}^{\mu\nu}}
=\sum^m_{l=0}\frac{\partial{L_R}}{\partial\Box^lR}
\cdot\frac{\partial{\Box^lR}}{\partial{g}^{\mu\nu}}
=2\sum^m_{l=0}F_{(l)}\Box^l{R}_{\mu\nu}
\, . \label{IdentRiemtoRiS2}
\ee

%%%%%%%%%%%%%%%%%%%%%%%%%%%%%%%%%%%%%%%%%%%%%%%%%%%%%%%%
\subsection{The application to the Lagrangian
$\sqrt{-g}R^{\mu\nu\rho\sigma}\Box^n{R}_{\mu\nu\rho\sigma}$}
\label{four3}
%%%%%%%%%%%%%%%%%%%%%%%%%%%%%%%%%%%%%%%%%%%%%%%%%%%%%%%%%

As another example, let us consider the Lagrangian
\be
\sqrt{-g}L_{\text{Riem1}}=
\sqrt{-g}R^{\mu\nu\rho\sigma}\Box^n{R}_{\mu\nu\rho\sigma}
\, . \label{LagRiemBnRiem}
\ee
In the context of the above Lagrangian, the fourth-rank
tensors $P^{\mu\nu\rho\sigma}_{(0)}$,
$P^{\mu\nu\rho\sigma}_{(n)}$ and
$P^{\mu\nu\rho\sigma}$ are given respectively by
\be
P^{\mu\nu\rho\sigma}_{(0)}\big|_{L_{\text{Riem1}}}
=\Box^nR^{\mu\nu\rho\sigma} \, ,
\quad
P^{\mu\nu\rho\sigma}_{(n)}\big|_{L_{\text{Riem1}}}
=R^{\mu\nu\rho\sigma} \, , \quad
P^{\mu\nu\rho\sigma}\big|_{L_{\text{Riem1}}}
=2\Box^nR^{\mu\nu\rho\sigma}
\, . \label{P4iP4forLRiem1}
\ee
Then substituting Eq. (\ref{P4iP4forLRiem1}) into
Eqs. (\ref{KmnRiem}) and (\ref{EoMforLagRiem2}) yields
the Noether potential $K^{\mu\nu}_{\text{Riem1}}$  and
the expression $E^{\mu\nu}_{\text{Riem1}}$ for
field equations associated to the Lagrangian
(\ref{LagRiemBnRiem}), respectively. Concretely,
the former has the form
\bea
K^{\mu\nu}_{\text{Riem1}}&=&
8\zeta^\lambda\sum^n_{k=1}\Big(\nabla^{[\mu}\Box^{n-k}
R^{\nu]\tau\rho\sigma}\Big)
\Box^{k-1} R_{\lambda\tau\rho\sigma}
-8\zeta^\lambda\sum^n_{k=1}\Big(\nabla^{[\mu}\Box^{n-k}
R_{\lambda\tau\rho\sigma}\Big)
\Box^{k-1} R^{\nu]\tau\rho\sigma}\nn \\
&&+4\big(\Box^nR^{\mu\nu\rho\sigma}\big)
\nabla_\rho\zeta_\sigma
+2\sum^n_{k=1}\zeta^{[\mu}\Big(\nabla^{\nu]}\Box^{n-k}
R_{\alpha\beta\rho\sigma}\Big)
\Big(\Box^{k-1} R^{\alpha\beta\rho\sigma}\Big) \nn \\
&&+8\zeta_\rho\nabla_\sigma\Box^nR^{\mu\nu\rho\sigma}
+8\zeta^\lambda\sum^n_{k=1}\Big(\Box^{n-k}
R^{[\mu}_{~~\tau\rho\sigma}\Big)
\nabla_{\lambda}\Box^{k-1} R^{\nu]\tau\rho\sigma}
\, , \label{KmnRieBnRiem}
\eea
and the latter is read off as
\bea
E^{\mu\nu}_{\text{Riem1}}
&=&2\big(\Box^nR^{(\mu}_{~~\tau\rho\sigma}\big)
R^{\nu)\tau\rho\sigma}
+\frac{1}{2}g^{\mu\nu}\sum^n_{k=1}\nabla_\lambda\Big[
\Big(\nabla^{\lambda}\Box^{n-k}
R_{\alpha\beta\rho\sigma}\Big)
\Box^{k-1} R^{\alpha\beta\rho\sigma}\Big] \nn \\
&&-\frac{1}{2}g^{\mu\nu}R^{\alpha\beta\rho\sigma}
\Box^n{R}_{\alpha\beta\rho\sigma}
-\sum^n_{k=1}\Big(\nabla^{(\mu}\Box^{n-k}
R_{\alpha\beta\rho\sigma}\Big)
\Big(\nabla^{\nu)}\Box^{k-1}
R^{\alpha\beta\rho\sigma}\Big) \nn \\
&&-4\nabla_\rho\nabla_\sigma\Box^n{R}^{\rho(\mu\nu)\sigma}
+4\sum^n_{k=1}\nabla^\lambda\Big[\Big(\nabla^{(\mu}\Box^{n-k}
R_{\lambda\tau\rho\sigma}\Big)
\Big(\Box^{k-1}R^{\nu)\tau\rho\sigma}\Big)\Big] \nn \\
&&-4\sum^n_{k=1}\nabla^\lambda\Big[\Big(\Box^{n-k}
R_{\lambda\tau\rho\sigma}\Big)
\Big(\nabla^{(\mu}\Box^{k-1}
R^{\nu)\tau\rho\sigma}\Big)\Big]
\, . \label{EoMLagRemBnRem}
\eea
In addition, the surface term $\Theta^\mu_{\text{Riem1}}$
for the Lagrangian $L_{\text{Riem1}}$ takes the following form
\bea
\Theta^\mu_{\text{Riem1}}&=& 4\big(\Box^nR^{\mu\nu\rho\sigma}\big)
\nabla_\sigma\delta g_{\rho\nu}
-4(\delta g_{\nu\rho})
\nabla_\sigma\Box^nR^{\mu\nu\rho\sigma}
+\Theta^\mu_{\text{Riem}(n)}
\big|_{P^{\mu\nu\rho\sigma}_{(n)}=R^{\mu\nu\rho\sigma}}
\, . \label{TheRiem1def}
\eea

In particular, when $n=0$, Eq. (\ref{EoMLagRemBnRem}) gives
the expression of field equations for the Lagrangian
$\sqrt{-g}R^{\alpha\beta\rho\sigma}{R}_{\alpha\beta\rho\sigma}$,
being of the form
\bea
E^{\mu\nu}_{\text{Riem1}}\big|_{n=0}&=&
2R^{\mu\tau\rho\sigma}R^{\nu}_{~\tau\rho\sigma}
+4R_{\rho\sigma}{R}^{\mu\rho\nu\sigma}
-\frac{1}{2}g^{\mu\nu}
R^{\alpha\beta\rho\sigma}{R}_{\alpha\beta\rho\sigma} \nn \\
&&-4R^{\mu}_\sigma{R}^{\nu\sigma}
+4\Box{R}^{\mu\nu}-2\nabla^\mu\nabla^\nu{R}
\, , \label{EomRiemsquar}
\eea
and the Noether potential for this Lagrangian is
\be
K^{\mu\nu}_{\text{Riem1}}\big|_{n=0}=
4R^{\mu\nu\rho\sigma}\nabla_\rho\zeta_\sigma
-16\zeta^\sigma\nabla^{[\mu}{R}^{\nu]}_{\sigma}
\, . \label{NPRiemsquar}
\ee

What is more, it is easy to confirm that the identity
(\ref{Ident1Riem2}) holds true for the Lagrangian
(\ref{LagRiemBnRiem}). Besides, for such a Lagrangian,
the identity (\ref{Ident2Riem2}) turns into
\be
\frac{\partial{L_{\text{Riem1}}}}{\partial{g}^{\mu\nu}}
=2g_{\mu\alpha}R^{\alpha\beta\rho\sigma}\Box^n
{R}_{\nu\beta\rho\sigma}
+2g_{\nu\alpha}R^{\alpha\beta\rho\sigma}\Box^n
{R}_{\mu\beta\rho\sigma}
\, . \label{Ident2RieBnRiem}
\ee
This equality can be also obtained by a direct
computation on the partial derivative of
$L_{\text{Riem1}}$ with respect to the inverse metric.

It is worth pointing out that Eq. (\ref{EoMLagRemBnRem})
is able to be used to determine the equations of motion for
the Lagrangian $\sqrt{-g}L_{\text{Riem2}}=
\sqrt{-g}\big(\Box^iR^{\mu\nu\rho\sigma}\big)
\Box^j{R}_{\mu\nu\rho\sigma}$. Actually, due to the fact
that the scalar $L_{\text{Riem2}}$ can be expressed as
$L_{\text{Riem2}}=R^{\mu\nu\rho\sigma}
\Box^{i+j}{R}_{\mu\nu\rho\sigma}+\nabla_\mu (\bullet)^\mu$,
where the divergence term does not contribute to the
field equations, the equations of motion for
the Lagrangian density $L_{\text{Riem2}}$ are given by
$E^{\mu\nu}_{\text{Riem1}}\big|_{n=i+j}=0$.

%%%%%%%%%%%%%%%%%%%%%%%%%%%%%%%%%%%%%%%%%%%%%%%%%%%%%%%%
\section{The relation between two generic scalars
$A^{\alpha_1\cdot\cdot\cdot\alpha_n}
(\delta\Box^{i}B_{\alpha_1\cdot\cdot\cdot\alpha_n})$ and
$(\Box^{i}A^{\alpha_1\cdot\cdot\cdot\alpha_n})
\delta{B}_{\alpha_1\cdot\cdot\cdot\alpha_n}$
and its application in deriving field equations
and Noether potentials}
\label{five}
%%%%%%%%%%%%%%%%%%%%%%%%%%%%%%%%%%%%%%%%%%%%%%%%%%%%

In the previous three sections, the relation between
$\Phi_{(i,1)}$ and $\Phi_{(i,i+1)}$, the one
between $\Psi_{(i,1)}$ and $\Psi_{(i,i+1)}$, and the one
between $\Upsilon_{(i,1)}$ and $\Upsilon_{(i,i+1)}$,
given respectively by Eqs. (\ref{SumPhik}), (\ref{SumPsik0})
and (\ref{SumUpsik0}), have played a crucial role in
deriving the expression for equations of motion, as well as
the Noether potentials. Within the present section, inspired
with these three crucial relations, we are going to perform a
detailed demonstration that they can actually be
generalized to the situation for a general scalar
$A^{\alpha_1\cdot\cdot\cdot\alpha_n}
(\delta\Box^{i}B_{\alpha_1\cdot\cdot\cdot\alpha_n})$,
where $A^{\alpha_1\cdot\cdot\cdot\alpha_n}$ and
$B_{\alpha_1\cdot\cdot\cdot\alpha_n}$ stand for
two arbitrary rank-$n$ tensors (both of them are
allowed to be independent of the metric tensor).
After figuring out the relation between the scalar
$A^{\alpha_1\cdot\cdot\cdot\alpha_n}
(\delta\Box^{i}B_{\alpha_1\cdot\cdot\cdot\alpha_n})$ and
the one $(\Box^{i}A^{\alpha_1\cdot\cdot\cdot\alpha_n})
\delta{B}_{\alpha_1\cdot\cdot\cdot\alpha_n}$, as well as
the concrete expression for the surface term with the variation
operator substituted by the Lie derivative along an
arbitrary vector, we will carry out both of them for
the derivation of the field equations and Noether potentials
associated to Lagrangians armed with diffeomorphism invariance,
which depend upon $g^{\mu\nu}$, $R_{\mu\nu\rho\sigma}$,
$\Box^i{R}_{\mu\nu\rho\sigma}$s, together with the variables
through $\Box^i$ acting on a generic tensor. All the results
related to the Lagrangians $\sqrt{-g}L_{R}$,
$\sqrt{-g}L_{\text{Ric}}$ and $\sqrt{-g}L_{\text{Riem}}$
will be reproduced from a unified perspective.
Particularly, we shall pay attention to a type of Lagrangian
that can be extended to nonlocal gravity theories.

%%%%%%%%%%%%%%%%%%%%%%%%%%%%%%%%%%%%%%%%%%%%%%%%%%%%
\subsection{General formalism} \label{Five1}
%%%%%%%%%%%%%%%%%%%%%%%%%%%%%%%%%%%%%%%%%%%

To proceed, in a similar fashion, we begin with introducing a scalar
$\Omega_{(i,k)}$ defined in terms of the contraction of
the rank-$n$ contravariant tensor
$\Box^{k-1}A^{\alpha_1\cdot\cdot\cdot\alpha_n}$
with the variation of the rank-$n$ covariant tensor
$\Box^{i-k+1}B_{\alpha_1\cdot\cdot\cdot\alpha_n}$
as
\be
\Omega_{(i,k)}=
\left(\Box^{k-1}A^{\alpha_1\cdot\cdot\cdot\alpha_n}\right)
\left(\delta\Box^{i-k+1}B_{\alpha_1\cdot\cdot\cdot\alpha_n}\right)
\, , \label{Omegikdef}
\ee
where the integer $k$ is allowed to run from 1 up to $i+1$,
together with three tensors $S^{\mu}_{(i,k)}$,
$T^{\mu\nu}_{(i,k)}$, and $Z^{\sigma\mu\nu}_{(i,k)}$.
Specifically, the vector $S^{\mu}_{(i,k)}$ takes the following
form
\be
S^{\mu}_{(i,k)}=
\left(\Box^{k-1}A^{\alpha_1\cdot\cdot\cdot\alpha_n}\right)
\left(\delta\nabla^{\mu}\Box^{i-k}
B_{\alpha_1\cdot\cdot\cdot\alpha_n}\right)
-\left(\nabla^{\mu}\Box^{k-1}
A^{\alpha_1\cdot\cdot\cdot\alpha_n}\right)
\left(\delta\Box^{i-k}B_{\alpha_1\cdot\cdot\cdot\alpha_n}\right)
\, , \label{Sikdef}
\ee
the second-rank symmetric tensor $T^{\mu\nu}_{(i,k)}$
is expressed as
\be
T^{\mu\nu}_{(i,k)}=
\Big(\nabla^{(\mu}\Box^{k-1}
A^{|\alpha_1\cdot\cdot\cdot\alpha_n|}\Big)
\Big(\nabla^{\nu)}\Box^{i-k}
B_{\alpha_1\cdot\cdot\cdot\alpha_n}\Big)
\, , \label{Tikdef}
\ee
and the third-rank tensor $Z^{\sigma\mu\nu}_{(i,k)}$
is presented by
\be
Z^{\sigma\mu\nu}_{(i,k)}=
H^{\sigma\mu\nu}_{(i,k)}
+g^{\sigma\mu}\left(\nabla^{\nu}\Box^{i-k}
B_{\alpha_1\cdot\cdot\cdot\alpha_n}\right)
\left(\Box^{k-1}A^{\alpha_1\cdot\cdot\cdot\alpha_n}\right)
\, , \label{Zikdef}
\ee
with $H^{\sigma\mu\nu}_{(i,k)}$ being of the form
\bea
H^{\sigma\mu\nu}_{(i,k)}&=&g^{\rho\sigma}
\sum^n_{j=1}\left(\nabla^{\mu}\Box^{k-1}
A^{\alpha_1\cdot\cdot\cdot\alpha_{j-1}\nu
\alpha_{j+1}\cdot\cdot\cdot\alpha_n}\right)
\left(\Box^{i-k}
B_{\alpha_1\cdot\cdot\cdot\alpha_{j-1}\rho
\alpha_{j+1}\cdot\cdot\cdot\alpha_n}\right) \nn \\
&&-g^{\rho\sigma}
\sum^n_{j=1}\left(\Box^{k-1}
A^{\alpha_1\cdot\cdot\cdot\alpha_{j-1}\nu
\alpha_{j+1}\cdot\cdot\cdot\alpha_n}\right)
\nabla^{\mu}\Box^{i-k}
B_{\alpha_1\cdot\cdot\cdot\alpha_{j-1}\rho
\alpha_{j+1}\cdot\cdot\cdot\alpha_n}
\, . \label{Hikdef}
\eea
It can be proved that the sum for the divergence of the
rank-3 tensor ${H}^{\mu\lambda\nu}_{(i,k)}$ over $k$ from
1 to $i$ satisfies identically
\bea
\sum^i_{k=1}\nabla_\lambda{H}^{\mu\lambda\nu}_{(i,k)}
&=&g^{\mu\lambda}\sum^n_{j=1}
B_{\alpha_1\cdot\cdot\cdot\alpha_{j-1}\lambda
\alpha_{j+1}\cdot\cdot\cdot\alpha_n}
\Box^{i}A^{\alpha_1\cdot\cdot\cdot\alpha_{j-1}\nu
\alpha_{j+1}\cdot\cdot\cdot\alpha_n} \nn \\
&&-g^{\mu\lambda}
\sum^n_{j=1}A^{\alpha_1\cdot\cdot\cdot\alpha_{j-1}\nu
\alpha_{j+1}\cdot\cdot\cdot\alpha_n}\Box^{i}
B_{\alpha_1\cdot\cdot\cdot\alpha_{j-1}\lambda
\alpha_{j+1}\cdot\cdot\cdot\alpha_n}
\, . \label{SumHikdef}
\eea
In light of the definitions for the three tensors
$S^{\mu}_{(i,k)}$, $T^{\mu\nu}_{(i,k)}$ and
$Z^{\lambda\mu\nu}_{(i,k)}$, substituting the variation
operator $\delta$ in the vector $S^{\mu}_{(i,k)}$
by the covariant derivative $\nabla^\nu$,
we obtain a useful relation given by
\bea
S^{\mu}_{(i,k)}(\delta\rightarrow\nabla^\nu)
&=&\big(\Box^{k-1}A^{\alpha_1\cdot\cdot\cdot\alpha_n}\big)
\nabla^{\nu}\nabla^{\mu}\Box^{i-k}
B_{\alpha_1\cdot\cdot\cdot\alpha_n}
-\big(\nabla^{\mu}\Box^{k-1}
A^{\alpha_1\cdot\cdot\cdot\alpha_n}\big)
\nabla^{\nu}\Box^{i-k}
B_{\alpha_1\cdot\cdot\cdot\alpha_n} \nn\\
&=&\nabla_\lambda{Z}^{\nu\lambda\mu}_{(i,k)}
-2T^{\mu\nu}_{(i,k)}
-\nabla_\lambda{H}^{\nu\lambda\mu}_{(i,k)}
\, . \label{STZrelat}
\eea
In addition to Eq. (\ref{STZrelat}), if the variation
operator in $S^{\mu}_{(i,k)}$ is replaced with the Lie
derivative $\mathcal{L}_\zeta$ along an arbitrary
smooth vector $\zeta^\mu$ instead of the covariant
derivative $\nabla^\nu$, we have another significant identity
\bea
S^\mu_{(i,k)}(\delta\rightarrow\mathcal{L}_\zeta)
&=&\zeta_\nu{S}^{\mu}_{(i,k)}(\delta\rightarrow\nabla^\nu)
-Z^{\nu\mu\lambda}_{(i,k)}\nabla_\lambda\zeta_\nu\nn \\
&=&2\zeta_\nu\left[\frac{1}{2}\nabla_\lambda\Big(
Z^{\nu\mu\lambda}_{(i,k)}+{Z}^{\nu\lambda\mu}_{(i,k)}
-{H}^{\nu\lambda\mu}_{(i,k)}\Big)
-T^{\mu\nu}_{(i,k)}\right] \nn \\
&&-\nabla_\nu\Big(\zeta_\lambda{Z}^{\lambda\mu\nu}_{(i,k)}\Big)
\, . \label{SikdelLie}
\eea
This identity is of great importance for the simplification
for the calculations of the surface term below.

Next, with the three tensors
$S^{\mu}_{(i,k)}$, $T^{\mu\nu}_{(i,k)}$ and
$Z^{\lambda\mu\nu}_{(i,k)}$ in hand, we compute the
scalar $\Omega_{(i,k)}$ and then establish its connection
to $\Omega_{(i,k+1)}$, taking the form
\be
\Omega_{(i,k)}=\Omega_{(i,k+1)}
+\nabla_\mu S^{\mu}_{(i,k)}
+T^{\mu\nu}_{(i,k)} \delta g_{\mu\nu}
+g_{\rho\sigma}Z^{\sigma\mu\nu}_{(i,k)}
\delta \Gamma^\rho_{\mu\nu}
\, . \label{OmegikRel}
\ee
Starting from Eq. (\ref{OmegikRel}), we further
arrive at the relation between $\Omega_{(i,1)}$ and
$\Omega_{(i,i+1)}$,
\be
\Omega_{(i,1)}=\Omega_{(i,i+1)}
+\sum^i_{k=1}\nabla_\mu S^{\mu}_{(i,k)}
+\sum^i_{k=1}T^{\mu\nu}_{(i,k)} \delta g_{\mu\nu}
+g_{\rho\sigma}\sum^i_{k=1}Z^{\sigma\mu\nu}_{(i,k)}
\delta \Gamma^\rho_{\mu\nu}
\, . \label{SumOmegik0}
\ee
By substituting the three tensors
$S^{\mu}_{(i,k)}$, $T^{\mu\nu}_{(i,k)}$ and
$Z^{\lambda\mu\nu}_{(i,k)}$ into Eq. (\ref{SumOmegik0}),
we find that both the scalars
$\Omega_{(i,1)}=A^{\alpha_1\cdot\cdot\cdot\alpha_n}
(\delta\Box^{i}B_{\alpha_1\cdot\cdot\cdot\alpha_n})$ and
$\Omega_{(i,i+1)}=(\Box^{i}A^{\alpha_1\cdot\cdot\cdot\alpha_n})
\delta{B}_{\alpha_1\cdot\cdot\cdot\alpha_n}$ are associated
with each other in the following manner
\bea
A^{\alpha_1\cdot\cdot\cdot\alpha_n}
(\delta\Box^{i}B_{\alpha_1\cdot\cdot\cdot\alpha_n})
&=&\sum^i_{k=1}
\left[T^{\mu\nu}_{(i,k)}
-\frac{1}{2}\nabla_\lambda\Big(Z^{(\mu\nu)\lambda}_{(i,k)}
+Z^{(\mu|\lambda|\nu)}_{(i,k)}
-Z^{\lambda(\mu\nu)}_{(i,k)}\Big)\right]\delta g_{\mu\nu}\nn \\
&&+\left(\Box^{i}A^{\alpha_1\cdot\cdot\cdot\alpha_n}\right)
\delta{B}_{\alpha_1\cdot\cdot\cdot\alpha_n}
+\sum^i_{k=1}\nabla_\mu\Theta_{(i,k)}^\mu
\, . \label{SumOmegik}
\eea
This equation assists us to peel off the operator $\Box^i$
in the variation term
$\delta\Box^{i}B_{\alpha_1\cdot\cdot\cdot\alpha_n}$ so that
we only need to deal with terms proportional to
$\delta{B}_{\alpha_1\cdot\cdot\cdot\alpha_n}$ during the
process of deriving the field equations.
Within Eq. (\ref{SumOmegik}), the vector $\Theta_{(i,k)}^\mu$
in the divergence term is defined in terms of both the tensors
$S^{\mu}_{(i,k)}$ and $Z^{\mu\rho\sigma}_{(i,k)}$ as
\be
\Theta_{(i,k)}^\mu
=S^{\mu}_{(i,k)}
+\frac{1}{2}\Big(Z^{(\rho\sigma)\mu}_{(i,k)}
+Z^{(\rho|\mu|\sigma)}_{(i,k)}
-Z^{\mu(\rho\sigma)}_{(i,k)}\Big)
\delta{g}_{\rho\sigma}
\, . \label{Thegenik}
\ee
Here the quantity $\Theta_{(i,k)}^\mu$ is of great importance
for the derivation of the field equations and the Noether
potentials in the framework of the method based upon the
conserved current, according to which there is a basic
requirement to compute $\Theta_{(i,k)}^\mu$ with the
substitution of $\delta$ by
the Lie derivative along any smooth vector.
For convenience, we substitute Eqs. (\ref{Sikdef}) and
(\ref{Zikdef}) into Eq. (\ref{Thegenik}) to reexpress
$\Theta_{(i,k)}^\mu$ in terms of the third-rank tensor
$H^{\sigma\mu\nu}_{(i,k)}$ as
\bea
\Theta_{(i,k)}^\mu&=&
\left(\Box^{k-1}A^{\alpha_1\cdot\cdot\cdot\alpha_n}\right)
\left(\delta\nabla^{\mu}\Box^{i-k}
B_{\alpha_1\cdot\cdot\cdot\alpha_n}\right)
-\left(\nabla^{\mu}\Box^{k-1}
A^{\alpha_1\cdot\cdot\cdot\alpha_n}\right)
\left(\delta\Box^{i-k}B_{\alpha_1\cdot\cdot\cdot\alpha_n}\right)
\nn \\
&&+\frac{1}{2}g^{\rho\sigma}
\left(\Box^{k-1}A^{\alpha_1\cdot\cdot\cdot\alpha_n}\right)
\left(\nabla^{\mu}\Box^{i-k}
B_{\alpha_1\cdot\cdot\cdot\alpha_n}\right)
\delta{g}_{\rho\sigma} \nn \\
&&+\frac{1}{2}\Big(H^{\rho\sigma\mu}_{(i,k)}
+H^{\rho\mu\sigma}_{(i,k)}
-H^{\mu\rho\sigma}_{(i,k)}\Big)
\delta{g}_{\rho\sigma}
\, . \label{Thegenik2}
\eea

We move on to compute $\Theta_{(i,k)}^\mu$ under the condition that
the variation operator $\delta$ is transformed into the
Lie derivative $\mathcal{L}_\zeta$. By making use of
Eq. (\ref{SikdelLie}), we have
\be
\Theta^\mu_{(i,k)}(\delta\rightarrow\mathcal{L}_\zeta)
=2\zeta_\nu{X}^{\mu\nu}_{(i,k)}
-\nabla_\nu{K}^{\mu\nu}_{(i,k)}
\, . \label{TheikdelLie}
\ee
Within Eq. (\ref{TheikdelLie}), the tensor
${X}^{\mu\nu}_{(i,k)}$ is given by
\bea
{X}^{\mu\nu}_{(i,k)}&=&
\frac{1}{2}S^{\mu}_{(i,k)}(\delta\rightarrow\nabla^\nu)
+\frac{1}{2}\nabla_\lambda\Big(
Z^{(\mu\nu)\lambda}_{(i,k)}
-Z^{\lambda(\mu\nu)}_{(i,k)}
+Z^{[\mu|\lambda|\nu]}_{(i,k)}\Big)
\, , \label{Xmngenik0}
\eea
which is equivalently expressed as
\be
{X}^{\mu\nu}_{(i,k)}
=\frac{1}{2}\nabla_\lambda\Big(
Z^{(\mu\nu)\lambda}_{(i,k)}
-Z^{\lambda(\mu\nu)}_{(i,k)}
+Z^{(\mu|\lambda|\nu)}_{(i,k)}\Big)
-T^{\mu\nu}_{(i,k)}
-\frac{1}{2}\nabla_\lambda{H}^{\nu\lambda\mu}_{(i,k)}
\, , \label{Xmngenik}
\ee
and the second-rank anti-symmetric tensor
${K}^{\mu\nu}_{(i,k)}={K}^{[\mu\nu]}_{(i,k)}$ is
read off as
\be
K^{\mu\nu}_{(i,k)}=
\zeta_\lambda\Big(Z^{[\mu\nu]\lambda}_{(i,k)}
+Z^{[\mu|\lambda|\nu]}_{(i,k)}
+Z^{\lambda[\mu\nu]}_{(i,k)}\Big)
\, . \label{Kmngenik}
\ee
By reformulating $K^{\mu\nu}_{(i,k)}$ in terms of the rank-3
tensor $H^{\lambda\mu\nu}_{(i,k)}$, one finds that
\bea
K^{\mu\nu}_{(i,k)}&=&
2\zeta^{[\mu}\left(\nabla^{\nu]}\Box^{i-k}
B_{\alpha_1\cdot\cdot\cdot\alpha_n}\right)
\Big(\Box^{k-1}A^{\alpha_1\cdot\cdot\cdot\alpha_n}\Big) \nn \\
&&+\zeta_\lambda\left(H^{[\mu\nu]\lambda}_{(i,k)}
+H^{[\mu|\lambda|\nu]}_{(i,k)}
+H^{\lambda[\mu\nu]}_{(i,k)}\right)
\, . \label{Kmngenik2}
\eea
Here the anti-symmetric tensor ${K}^{\mu\nu}_{(i,k)}$
only consists of terms proportional to the vector
$\zeta^\mu$, without any term comprising its derivatives.
In particular, if the variation of the Lagrangian
includes $A^{\alpha_1\cdot\cdot\cdot\alpha_n}
(\delta\Box^{i}B_{\alpha_1\cdot\cdot\cdot\alpha_n})$
as one of its ingredients,
the sum of ${K}^{\mu\nu}_{(i,k)}$ over $k$ from 1 to $i$
is responsible for all the contributions to the Noether
potential out of the difference between the scalar
$A^{\alpha_1\cdot\cdot\cdot\alpha_n}
(\delta\Box^{i}B_{\alpha_1\cdot\cdot\cdot\alpha_n})$
and the one $(\Box^{i}A^{\alpha_1\cdot\cdot\cdot\alpha_n})
\delta{B}_{\alpha_1\cdot\cdot\cdot\alpha_n}$.
From Eq. (\ref{TheikdelLie}), one is able to define a
conserved current associated to an arbitrary vector
$\zeta^\mu$ as
\be
{J}^{\mu}_{(i,k)}=
2\zeta_\nu{X}^{\mu\nu}_{(i,k)}
-\Theta^\mu_{(i,k)}(\delta\rightarrow\mathcal{L}_\zeta)
\, , \label{ConCurJik}
\ee
attributed to the fact that
$\nabla_\mu\nabla_\nu{K}^{\mu\nu}_{(i,k)}=0$.
With the help of Eqs. (\ref{SumHikdef}) and (\ref{Xmngenik}),
Eq. (\ref{SumOmegik}) can be reformulated as a
more practical form
\bea
A^{\alpha_1\cdot\cdot\cdot\alpha_n}
\delta\Box^{i}B_{\alpha_1\cdot\cdot\cdot\alpha_n}
&=&\left(\Box^{i}A^{\alpha_1\cdot\cdot\cdot\alpha_n}\right)
\delta{B}_{\alpha_1\cdot\cdot\cdot\alpha_n}
-\sum^i_{k=1}\left[\left(X^{\mu\nu}_{(i,k)}
+\frac{1}{2}\nabla_\lambda{H}^{\nu\lambda\mu}_{(i,k)}\right)
\delta g_{\mu\nu}
-\nabla_\mu\Theta_{(i,k)}^\mu \right]\nn \\
&=&\left(\Box^{i}A^{\alpha_1\cdot\cdot\cdot\alpha_n}\right)
\delta{B}_{\alpha_1\cdot\cdot\cdot\alpha_n}
-\sum^i_{k=1}X^{\mu\nu}_{(i,k)}\delta g_{\mu\nu}
+\sum^i_{k=1}\nabla_\mu\Theta_{(i,k)}^\mu \nn \\
&&-\frac{1}{2}g^{\nu\lambda}\big(\delta g_{\mu\nu}\big)
\sum^n_{j=1}B_{\alpha_1\cdot\cdot\cdot\alpha_{j-1}\lambda
\alpha_{j+1}\cdot\cdot\cdot\alpha_n}
\Box^{i}A^{\alpha_1\cdot\cdot\cdot\alpha_{j-1}\mu
\alpha_{j+1}\cdot\cdot\cdot\alpha_n} \nn \\
&&+\frac{1}{2}g^{\nu\lambda}\big(\delta g_{\mu\nu}\big)
\sum^n_{j=1}A^{\alpha_1\cdot\cdot\cdot\alpha_{j-1}\mu
\alpha_{j+1}\cdot\cdot\cdot\alpha_n}\Box^{i}
B_{\alpha_1\cdot\cdot\cdot\alpha_{j-1}\lambda
\alpha_{j+1}\cdot\cdot\cdot\alpha_n}
\, . \label{SumOmegik2}
\eea
For a direct application of Eq. (\ref{SumOmegik2}) see
Eq. (\ref{QijBoxjRiem}) in Appendix \ref{appendB}.

Furthermore, since the second-rank tensor ${X}^{\mu\nu}_{(i,k)}$
plays a significant role in determining the field equations,
we pay much more attention to analysing its properties. If this
tensor is decomposed as
\be
{X}^{\mu\nu}_{(i,k)}={X}^{(\mu\nu)}_{(i,k)}
+{X}^{[\mu\nu]}_{(i,k)}
\, , \label{Xmngenik2}
\ee
by the aid of Eq. (\ref{Xmngenik0}) or (\ref{Xmngenik}) ,
after some manipulations,
we find that the symmetric component
${X}^{(\mu\nu)}_{(i,k)}$ can be put into the following form
\bea
{X}^{(\mu\nu)}_{(i,k)}&=&
\frac{1}{2}\nabla_\lambda\Big(
Z^{(\mu\nu)\lambda}_{(i,k)}
-Z^{\lambda(\mu\nu)}_{(i,k)}
+Z^{(\mu|\lambda|\nu)}_{(i,k)}-H^{(\mu|\lambda|\nu)}_{(i,k)}\Big)
-T^{\mu\nu}_{(i,k)}\nn \\
&=&\frac{1}{2}\nabla_\lambda
H^{(\mu\nu)\lambda}_{(i,k)}
+\frac{1}{2}g^{\mu\nu}\nabla_\lambda
\left[\left(\Box^{k-1}A^{\alpha_1\cdot\cdot\cdot\alpha_n}\right)
\nabla^{\lambda}\Box^{i-k}
B_{\alpha_1\cdot\cdot\cdot\alpha_n}\right] \nn \\
&&-\frac{1}{2}\nabla_\lambda{H}^{\lambda(\mu\nu)}_{(i,k)}
-\Big(\nabla^{(\mu}\Box^{k-1}
A^{|\alpha_1\cdot\cdot\cdot\alpha_n|}\Big)
\nabla^{\nu)}\Box^{i-k}
B_{\alpha_1\cdot\cdot\cdot\alpha_n}
\, , \label{Xmnikgensym}
\eea
while the anti-symmetric component ${X}^{[\mu\nu]}_{(i,k)}$
is written as
\bea
{X}^{[\mu\nu]}_{(i,k)}
&=&\frac{1}{4}S^{\mu}_{(i,k)}(\delta\rightarrow\nabla^\nu)
-\frac{1}{4}S^{\nu}_{(i,k)}(\delta\rightarrow\nabla^\mu)
+\frac{1}{2}\nabla_\lambda
Z^{[\mu|\lambda|\nu]}_{(i,k)} \nn \\
&=&\frac{1}{2}\nabla_\lambda{H}^{[\mu|\lambda|\nu]}_{(i,k)}
\, . \label{Xmnikgenansym}
\eea
The above identity can be alternatively derived out of
Eq. (\ref{Xmngenik}).
When the scalar $A^{\alpha_1\cdot\cdot\cdot\alpha_n}
(\delta\Box^{i}B_{\alpha_1\cdot\cdot\cdot\alpha_n})$ enters into
the variation of the Lagrangian,
according to Eqs. (\ref{EoMgen}) and (\ref{IdenXmn}), the
symmetric tensor $\sum^i_{k=1}{X}^{(\mu\nu)}_{(i,k)}$ actually
accounts for all the contributions to the field equations
from the difference between both the terms
$A^{\alpha_1\cdot\cdot\cdot\alpha_n}
(\delta\Box^{i}B_{\alpha_1\cdot\cdot\cdot\alpha_n})$ and
$(\Box^{i}A^{\alpha_1\cdot\cdot\cdot\alpha_n})
\delta{B}_{\alpha_1\cdot\cdot\cdot\alpha_n}$.
By employing Eq. (\ref{SumHikdef}), the sum of the
anti-symmetric tensor ${X}^{[\mu\nu]}_{(i,k)}$ over $k$
from 1 to $i$ gives rise to
\bea
\sum^i_{k=1}{X}^{[\mu\nu]}_{(i,k)}
&=&\frac{1}{2}\sum^n_{j=1}
B_{\alpha_1\cdot\cdot\cdot\alpha_{j-1}\lambda
\alpha_{j+1}\cdot\cdot\cdot\alpha_n}g^{\lambda[\mu}
\Big(\Box^{i}A^{|\alpha_1\cdot\cdot\cdot\alpha_{j-1}|\nu]
\alpha_{j+1}\cdot\cdot\cdot\alpha_n}\Big) \nn \\
&&-\frac{1}{2}
\sum^n_{j=1}g^{\lambda[\mu}
A^{|\alpha_1\cdot\cdot\cdot\alpha_{j-1}|\nu]
\alpha_{j+1}\cdot\cdot\cdot\alpha_n}\big(\Box^{i}
B_{\alpha_1\cdot\cdot\cdot\alpha_{j-1}\lambda
\alpha_{j+1}\cdot\cdot\cdot\alpha_n}\big)
\, . \label{SumXikdef}
\eea
In particular, when
\be
A^{\alpha_1\cdot\cdot\cdot\alpha_{j-1}\mu
\alpha_{j+1}\cdot\cdot\cdot\alpha_n}
B_{\alpha_1\cdot\cdot\cdot\alpha_{j-1}\nu
\alpha_{j+1}\cdot\cdot\cdot\alpha_n}
=A^{\mu\alpha_1\cdot\cdot\cdot\alpha_{n-1}}
B_{\nu\alpha_1\cdot\cdot\cdot\alpha_{n-1}}
\, \label{ABcontra}
\ee
holds for $j$ running from 1 up to $n$, Eq. (\ref{SumXikdef})
becomes
\be
\sum^i_{k=1}{X}^{[\mu\nu]}_{(i,k)}
=\frac{n}{2}
B^{[\mu}_{~~\alpha_1\cdot\cdot\cdot\alpha_{n-1}}
\Box^{i}A^{\nu]\alpha_1\cdot\cdot\cdot\alpha_{n-1}}
-\frac{n}{2}\Big(\Box^{i}
B^{[\mu}_{~~\alpha_1\cdot\cdot\cdot\alpha_{n-1}}\Big)
A^{\nu]\alpha_1\cdot\cdot\cdot\alpha_{n-1}}
\, . \label{SumXikSpeC}
\ee
For instance, when both the tensors
$\big(A^{\alpha_1\cdot\cdot\cdot\alpha_{n}},
B_{\alpha_1\cdot\cdot\cdot\alpha_{n}})
\rightarrow\big({P}^{\mu\nu\rho\sigma}_{(i)},
{R}_{\mu\nu\rho\sigma}\big)$, equation (\ref{SumXikSpeC})
becomes Eq. (\ref{SumXPiemik}).

Let us deal with the divergences for ${X}^{(\mu\nu)}_{(i,k)}$
and ${X}^{[\mu\nu]}_{(i,k)}$. Both of them are related to each
other through
\bea
2\nabla_\nu{X}^{(\mu\nu)}_{(i,k)}&=&
\left(\Box^{k-1}
{A}^{\alpha_1\cdot\cdot\cdot\alpha_n}\right)
\nabla^{\mu}\Box^{i-k+1}B_{\alpha_1\cdot\cdot\cdot\alpha_n}
-\left(\Box^{k}
{A}^{\alpha_1\cdot\cdot\cdot\alpha_n}\right)
\nabla^{\mu}\Box^{i-k}B_{\alpha_1\cdot\cdot\cdot\alpha_n}\nn\\
&&+2\left(\nabla_\nu\Box^{k-1}
A^{\alpha_1\cdot\cdot\cdot\alpha_n}\right)
\nabla^{[\mu}\nabla^{\nu]}\Box^{i-k}
B_{\alpha_1\cdot\cdot\cdot\alpha_n}
+R^\mu_{~\nu\rho\sigma}
{H}^{\rho\nu\sigma}_{(i,k)}\nn \\
&&+2\left(\nabla^{[\mu}\nabla^{\nu]}
\Box^{k-1}A^{\alpha_1\cdot\cdot\cdot\alpha_n}\right)
\nabla_{\nu}\Box^{i-k}
B_{\alpha_1\cdot\cdot\cdot\alpha_n}
+2\nabla_\nu{X}^{[\mu\nu]}_{(i,k)}
\, . \label{DiveXsymasy}
\eea
After making use of the following identity
\bea
R^\mu_{~\nu\rho\sigma}
{H}^{\rho\nu\sigma}_{(i,k)}&=&
-2\left(\nabla_\nu\Box^{k-1}
A^{\alpha_1\cdot\cdot\cdot\alpha_n}\right)
\left(\nabla^{[\mu}\nabla^{\nu]}\Box^{i-k}
B_{\alpha_1\cdot\cdot\cdot\alpha_n}\right)\nn \\
&&-2\left(\nabla^{[\mu}\nabla^{\nu]}
\Box^{k-1}A^{\alpha_1\cdot\cdot\cdot\alpha_n}\right)
\left(\nabla_{\nu}\Box^{i-k}
B_{\alpha_1\cdot\cdot\cdot\alpha_n}\right)
\,  \label{IdeRiemHik}
\eea
to eliminate the third-rank tensor ${H}^{\sigma\mu\nu}_{(i,k)}$
in Eq. (\ref{DiveXsymasy}),
the sum of Eq. (\ref{DiveXsymasy}) over $k$ from 1 to $i$
gives rise to an identity
\be
\sum^i_{k=1}\nabla_\nu{X}^{\nu\mu}_{(i,k)}=
\frac{1}{2}A^{\alpha_1\cdot\cdot\cdot\alpha_n}
\nabla^{\mu}\Box^{i}B_{\alpha_1\cdot\cdot\cdot\alpha_n}
-\frac{1}{2}(\Box^{i}A^{\alpha_1\cdot\cdot\cdot\alpha_n})
\nabla^{\mu}B_{\alpha_1\cdot\cdot\cdot\alpha_n}
\, . \label{IdeRiemHik2}
\ee
The above identity plays an important role in proving
that the field equations
are divergence-free via straightforward calculations
on the expression for equations of motion (see Appendix
\ref{appendC} for its significant applications in the proofs
for $\nabla_\mu{E}^{\mu\nu}_{\text{Riem}}=0$ and the generalizations
of such a Bianchi-type identity).

As examples to check all the above results, letting
both the tensors $\big(A^{\alpha_1\cdot\cdot\cdot\alpha_n},
B_{\alpha_1\cdot\cdot\cdot\alpha_n}\big)$ take the values
$\big(F_{(i)},R)$, $\big(P^{\mu\nu}_{(i)},R_{\mu\nu}\big)$
and $\big(P^{\mu\nu\rho\sigma}_{(i)},R_{\mu\nu\rho\sigma}\big)$,
respectively, one reproduces all the corresponding results
associated respectively to the Lagrangians
$\sqrt{-g}L_{R}$, $\sqrt{-g}L_{\text{Ric}}$ and
$\sqrt{-g}L_{\text{Riem}}$ appearing in the previous
three sections.

Finally, on the basis of the aforementioned results in this
section, we focus on their applications in the derivation for
the field equations and the Noether potentials of Lagrangians
with diffeomorphism invariance. Without loss of generality,
we take into account the situation in which the quantity
$(\Box^{i}A^{\alpha_1\cdot\cdot\cdot\alpha_n})
\delta{B}_{\alpha_1\cdot\cdot\cdot\alpha_n}$
is able to be expressed as
\be
(\Box^{i}A^{\alpha_1\cdot\cdot\cdot\alpha_n})
\delta{B}_{\alpha_1\cdot\cdot\cdot\alpha_n}=
-\bar{E}^{\mu\nu}_{B(i)}\delta{g}_{\mu\nu}+
\nabla_\mu \Theta^\mu_{B(i)}
\, . \label{BoxiAdelB}
\ee
Here $\bar{E}^{\mu\nu}_{B(i)}=\bar{E}^{\nu\mu}_{B(i)}$ is symmetric.
Under such a situation, according to Eq. (\ref{SumOmegik}),
the quantity $A^{\alpha_1\cdot\cdot\cdot\alpha_n}
(\delta\Box^{i}B_{\alpha_1\cdot\cdot\cdot\alpha_n})$
can be put into the form
\be
A^{\alpha_1\cdot\cdot\cdot\alpha_n}
(\delta\Box^{i}B_{\alpha_1\cdot\cdot\cdot\alpha_n})
=-\bar{E}^{\mu\nu}_{(i)}\delta g_{\mu\nu}
+\nabla_\mu\Theta_{(i)}^\mu
\, . \label{ABoxidel}
\ee
in which the symmetric tensor
$\bar{E}^{\mu\nu}_{(i)}=\bar{E}^{(\mu\nu)}_{(i)}$ is given by
\bea
\bar{E}^{\mu\nu}_{(i)}&=&\bar{E}^{\mu\nu}_{B(i)}
+\sum^i_{k=1}\left[\frac{1}{2}\nabla_\lambda
\Big(Z^{(\mu\nu)\lambda}_{(i,k)}
+Z^{(\mu|\lambda|\nu)}_{(i,k)}
-Z^{\lambda(\mu\nu)}_{(i,k)}\Big)
-T^{\mu\nu}_{(i,k)}\right] \nn \\
&=&\bar{E}^{\mu\nu}_{B(i)}
+\sum^i_{k=1}{X}^{\mu\nu}_{(i,k)}
+\frac{1}{2}\sum^i_{k=1}\nabla_\lambda
{H}^{\nu\lambda\mu}_{(i,k)}
\, , \label{EmniAB}
\eea
while the surface term $\Theta_{(i)}^\mu$ is decomposed into
\be
\Theta_{(i)}^\mu=\Theta_{B(i)}^\mu
+\sum^i_{k=1}\Theta_{(i,k)}^\mu
\, . \label{ThetaABi}
\ee
What is more, after the variation operator $\delta$ is substituted
by the Lie derivative $\mathcal{L}_\zeta$ along an arbitrary vector
$\zeta^\mu$, it is assumed that $\Theta_{(i)}^\mu$ can be written as
\be
\Theta^\mu_{B(i)}(\delta\rightarrow\mathcal{L}_\zeta)
=2\zeta_\nu{X}^{\mu\nu}_{B(i)}
-\nabla_\nu{K}^{\mu\nu}_{B(i)}
\, , \label{TheBidelLie}
\ee
where the second-rank tensor ${K}^{\mu\nu}_{B(i)}$
is required to be anti-symmetric with respect to
$(\mu,\nu)$. In order to illustrate
Eq. (\ref{TheBidelLie}), an example will be given in
Appendix \ref{appendB}.
From Eqs. (\ref{TheikdelLie}), (\ref{ThetaABi}) and
(\ref{TheBidelLie}), then we arrive at
\be
\Theta^\mu_{(i)}(\delta\rightarrow\mathcal{L}_\zeta)
=2\zeta_\nu\left({X}^{\mu\nu}_{B(i)}
+\sum^i_{k=1}{X}^{\mu\nu}_{(i,k)}\right)
-\nabla_\nu\left({K}^{\mu\nu}_{B(i)}
+\sum^i_{k=1}{K}^{\mu\nu}_{(i,k)}\right)
\, . \label{TheABiLie}
\ee
Equation (\ref{TheABiLie}) is our desired outcome.
It paves the way for determining the field equations
and the Noether potentials associated to the Lagrangians
involving the variables
$\Box^i{B}_{\alpha_1\cdot\cdot\cdot\alpha_n}$s,
where ${B}_{\alpha_1\cdot\cdot\cdot\alpha_n}$ denotes
an arbitrary tensor, which can be specific to the Ricci
scalar $R$, or the Ricci tensor $R_{\mu\nu}$,
or the Riemann tensor $R_{\mu\nu\rho\sigma}$, or the
tensor depending upon the metric and
$\Box^jR_{\mu\nu\rho\sigma}$s. Some concrete applications of
Eq. (\ref{TheABiLie}) will be presented in the
remainder of this section.

%%%%%%%%%%%%%%%%%%%%%%%%%%%%%%%%%%%%%%%%%%%%%%%%%%%%
\subsection{Field equations and Noether potentials
associated to the Lagrangian $\sqrt{-g}L_{B}
\big({g}^{\mu\nu},B_{\alpha_1\cdot\cdot\cdot\alpha_n},
\Box{B}_{\alpha_1\cdot\cdot\cdot\alpha_n},
\cdot\cdot\cdot,\Box^{m}
B_{\alpha_1\cdot\cdot\cdot\alpha_n}\big)$}\label{Five2}
%%%%%%%%%%%%%%%%%%%%%%%%%%%%%%%%%%%%%%%%%%%

In this subsection, by making use of
Eq. (\ref{SumOmegik2}), which displays the
relation between the scalars
$A^{\alpha_1\cdot\cdot\cdot\alpha_n}
(\delta\Box^{i}B_{\alpha_1\cdot\cdot\cdot\alpha_n})$
and $\big(\Box^{i}{A}^{\alpha_1\cdot\cdot\cdot\alpha_n}\big)
(\delta{B}_{\alpha_1\cdot\cdot\cdot\alpha_n})$,
as well as Eq. (\ref{TheABiLie}), we delve into the field
equations and the Noether potentials corresponding to
the Lagrangian whose variables incorporate
$B_{\alpha_1\cdot\cdot\cdot\alpha_n}$ and
$\Box^{i}B_{\alpha_1\cdot\cdot\cdot\alpha_n}$s.
Without loss of generality, here the
tensor $B_{\alpha_1\cdot\cdot\cdot\alpha_n}$  is supposed
to be dependent of the metric ${g}^{\mu\nu}$ and the
Riemann tensor $R_{\mu\nu\rho\sigma}$, together with
$\Box^j{R}_{\mu\nu\rho\sigma}$s.

As a beginning of our investigation, we concentrate on a
simple situation
in which the scalar $A^{\alpha_1\cdot\cdot\cdot\alpha_n}
(\delta\Box^{i}B_{\alpha_1\cdot\cdot\cdot\alpha_n})$ completely
results from the variation of the Lagrangian
admitting diffeomorphism invariance,
\be
\sqrt{-g}L_{(i)}=\sqrt{-g}L_{(i)}
\big({g}^{\mu\nu},\Box^{i}B_{\alpha_1\cdot\cdot\cdot\alpha_n}\big)
\, , \label{LagBoxiB}
\ee
with the rank-$n$ tensor $B_{\alpha_1\cdot\cdot\cdot\alpha_n}$
that is constrained to depend upon both the inverse metric
$g^{\mu\nu}$ and the Riemann curvature tensor
$R_{\alpha\beta\rho\sigma}$ for simplicity.
The variation of Eq. (\ref{LagBoxiB}) is read off as
\bea
\delta\Big(\sqrt{-g}L_{(i)}\Big)
&=&\sqrt{-g}\left[\left(\frac{\partial{L}_{(i)}}{\partial{g}^{\mu\nu}}
-\frac{1}{2}L_{(i)}g_{\mu\nu}\right)\delta{g}^{\mu\nu}
+A^{\alpha_1\cdot\cdot\cdot\alpha_n}
\big(\delta\Box^{i}B_{\alpha_1\cdot\cdot\cdot\alpha_n}\big)\right] \nn \\
&=&\sqrt{-g}\left[\left(\frac{\partial{L}_{(i)}}{\partial{g}^{\mu\nu}}
-\frac{1}{2}L_{(i)}g_{\mu\nu}
+g_{\mu\rho}g_{\nu\sigma}\bar{E}^{\rho\sigma}_{(i)}\right)
\delta{g}^{\mu\nu}
+\nabla_\mu\tilde{\Theta}_{(i)}^\mu\right]
\, . \label{VariLiAB}
\eea
Within Eq. (\ref{VariLiAB}) together with all the quantities
associated to the Lagrangian (\ref{LagBoxiB}) below, note that
the tensor $A^{\alpha_1\cdot\cdot\cdot\alpha_n}$
is specific to
\be
A^{\alpha_1\cdot\cdot\cdot\alpha_n}\rightarrow
\frac{\partial{L}_{(i)}}{
\partial\Box^{i}{B}_{\alpha_1\cdot\cdot\cdot\alpha_n}}
\, . \label{AforLagBi}
\ee
The rank-2 symmetric tensor $\bar{E}^{\mu\nu}_{(i)}$
in Eq. (\ref{VariLiAB}) is presented by Eq. (\ref{EmniAB})
with the tensor $\bar{E}^{\mu\nu}_{B(i)}$ in it substituted by
the one $\tilde{E}^{\mu\nu}_{B(i)}$ appearing in
Eq. (\ref{TildEBidef}), and the surface
term $\tilde{\Theta}_{(i)}^\mu$ takes the form
\be
\tilde{\Theta}_{(i)}^\mu=\tilde{\Theta}_{B(i)}^\mu
+\sum^i_{k=1}\Theta_{(i,k)}^\mu
\, , \label{ThetaLagi}
\ee
with $\tilde{\Theta}_{B(i)}^\mu$ given by Eq. (\ref{TilThetAB}).
Due to Eq. (\ref{EoMgen}), the expression for equations
of motion associated to the Lagrangian $\sqrt{-g}L_{(i)}$
is given by
\be
{E}^{\mu\nu}_{(i)}=\tilde{X}^{\mu\nu}_{B(i)}
+\sum^i_{k=1}{X}^{\mu\nu}_{(i,k)}
-\frac{1}{2}L_{(i)}{g}^{\mu\nu}
\, , \label{EomABLi}
\ee
or presented by
\bea
{E}^{\mu\nu}_{(i)}&=&
\sum^i_{k=1}\left[\frac{1}{2}\nabla_\lambda
\Big(Z^{(\mu\nu)\lambda}_{(i,k)}
+Z^{(\mu|\lambda|\nu)}_{(i,k)}
-Z^{\lambda(\mu\nu)}_{(i,k)}\Big)
-T^{\mu\nu}_{(i,k)}\right] \nn \\
&&+\tilde{E}^{\mu\nu}_{B(i)}
+{g}^{\mu\rho}{g}^{\nu\sigma}
\frac{\partial{L}_{(i)}}{\partial{g}^{\rho\sigma}}
-\frac{1}{2}L_{(i)}{g}^{\mu\nu}
\, . \label{EomABLi2}
\eea
Within Eq. (\ref{EomABLi}), the second-rank tensor
$\tilde{X}^{\mu\nu}_{B(i)}$ is given by Eq. (\ref{TildXBidef}).
Then the comparison between Eqs. (\ref{EomABLi}) and
(\ref{EomABLi2}) results in that the derivative of $L_{(i)}$
with respect to the metric has to be constrained by
\be
\frac{\partial{L}_{(i)}}{\partial{g}^{\mu\nu}}
={g}_{\mu\rho}{g}_{\nu\sigma}
\left(\tilde{X}^{\rho\sigma}_{B(i)}-\tilde{E}^{\rho\sigma}_{B(i)}
-\frac{1}{2}\sum^i_{k=1}\nabla_\lambda{H}^{\sigma\lambda\rho}_{(i,k)}
\right)
\, . \label{XBiEBirel}
\ee
Apart from this, the symmetry of ${E}^{\mu\nu}_{(i)}$ leads
to that the anti-symmetric tensor $\tilde{X}^{[\mu\nu]}_{B(i)}$ has to
satisfy the following identity
\be
\tilde{X}^{[\mu\nu]}_{B(i)}=
-\frac{1}{2}\sum^i_{k=1}
\nabla_\lambda{H}^{[\mu|\lambda|\nu]}_{(i,k)}
=-\sum^i_{k=1}{X}^{[\mu\nu]}_{(i,k)}
\, . \label{XBiansym}
\ee
From Eq. (\ref{TheABiLie}), by the aid of the anti-symmetric
tensor $\tilde{K}^{\mu\nu}_{B(i)}$ in Eq. (\ref{KmnABidef}),
the Noether potential
${K}^{\mu\nu}_{(i)}$ for the Lagrangian $\sqrt{-g}L_{(i)}$
has the form
\be
{K}^{\mu\nu}_{(i)}=\tilde{K}^{\mu\nu}_{B(i)}
+\sum^i_{k=1}{K}^{\mu\nu}_{(i,k)}
\, . \label{NoePABiLie}
\ee
On the basis of the Nother potential ${K}^{\mu\nu}_{(i)}$,
one is able to further investigate the conserved quantities
of various gravity theories endowed with the Lagrangian (\ref{LagBoxiB}),
such as the entropy, mass and angular momentum.

It is worthwhile pointing out that the above results
in connection with the Lagrangian $\sqrt{-g}L_{(i)}$
can be naturally extended to the more generic one
\be
\sqrt{-g}L_{B}=
\sqrt{-g}L_{B}\big({g}^{\mu\nu},B_{\alpha_1\cdot\cdot\cdot\alpha_n},
\Box{B}_{\alpha_1\cdot\cdot\cdot\alpha_n},
\cdot\cdot\cdot,\Box^{m}B_{\alpha_1\cdot\cdot\cdot\alpha_n}\big)
\, , \label{LagBgen}
\ee
in which the rank-$n$ tensor
${B}_{\alpha_1\cdot\cdot\cdot\alpha_n}$ is assumed to
exhibit a more general form
\be
{B}_{\alpha_1\cdot\cdot\cdot\alpha_n}=
{B}_{\alpha_1\cdot\cdot\cdot\alpha_n}\left(g^{\mu\nu},
\Box^{j}{R}_{\mu\nu\rho\sigma}\right)
\, , \label{BinLagB}
\ee
with $j=0,1,2,\cdot\cdot\cdot$.
This Lagrangian can be also viewed as the generalization
of the one $\sqrt{-g}L_{\text{Riem}}$ under the transformation
${R}_{\mu\nu\rho\sigma}\rightarrow
{B}_{\alpha_1\cdot\cdot\cdot\alpha_n}$.
Within the framework of the Lagrangian (\ref{LagBgen}),
its variation can be expressed as the following form
\bea
\frac{\delta\left(\sqrt{-g}L_{B}\right)}{\sqrt{-g}}
&=&\left(\frac{\partial{L}_{B}}{\partial{g}^{\mu\nu}}
-\frac{1}{2}L_{B}g_{\mu\nu}\right)\delta{g}^{\mu\nu}
+\sum^m_{i=0}A^{\alpha_1\cdot\cdot\cdot\alpha_n}_{B(i)}
\delta\Box^{i}B_{\alpha_1\cdot\cdot\cdot\alpha_n}
\nn \\
&=&\sum^m_{i=1}\sum^i_{k=1}\nabla_\mu\Theta_{(i,k)}^\mu
-\frac{1}{2}\sum^m_{i=1}\sum^i_{k=1}
\left(2{X}^{\mu\nu}_{(i,k)}
+\nabla_\lambda{H}^{\nu\lambda\mu}_{(i,k)}\right)
\delta{g}_{\mu\nu} \nn \\
&&+\sum^m_{i=0}
\big(\Box^{i}A^{\alpha_1\cdot\cdot\cdot\alpha_n}_{B(i)}\big)
\delta{B}_{\alpha_1\cdot\cdot\cdot\alpha_n}
+\left(\frac{\partial{L}_{B}}{\partial{g}^{\mu\nu}}
-\frac{1}{2}L_{B}g_{\mu\nu}\right)\delta{g}^{\mu\nu}
\, , \label{VariLagBgen0}
\eea
with the rank-$n$ tensors
$A^{\alpha_1\cdot\cdot\cdot\alpha_n}_{B(i)}$s
$(i=0,1,\cdot\cdot\cdot,m)$ defined by
\be
A^{\alpha_1\cdot\cdot\cdot\alpha_n}_{B(i)}=
\frac{\partial{L_{B}}}{\partial
{\Box^{i}B_{\alpha_1\cdot\cdot\cdot\alpha_n}}}
\, . \label{AforLagBgen}
\ee
After dealing with the term
$\big(\Box^{i}A^{\alpha_1\cdot\cdot\cdot\alpha_n}_{B(i)}\big)
\delta{B}_{\alpha_1\cdot\cdot\cdot\alpha_n}$ in light of
Eq. (\ref{AdelBBoxRiem2}), equation
(\ref{AforLagBgen}) turns into the conventional form
\be
\delta\left(\sqrt{-g}L_{B}\right)=
\sqrt{-g}\left(-{E}^{\mu\nu}_{B}\delta
{g}_{\mu\nu}+\nabla_\mu\Theta^\mu_{B}\right)
\, . \label{VariLagBgen}
\ee
Within Eq. (\ref{VariLagBgen}),
the expression ${E}^{\mu\nu}_{B}$ for the field
equations takes the following form
\bea
{E}^{\mu\nu}_{B}&=&\frac{\partial{L}_{B}}{\partial{g}^{\mu\nu}}
-\frac{1}{2}L_{B}g_{\mu\nu}
+\sum^m_{i=0}{E}^{\mu\nu}_{\text{GenB}(i,j)}
+\frac{1}{2}\sum^m_{i=1}\sum^i_{k=1}
\left(2{X}^{\mu\nu}_{(i,k)}
+\nabla_\lambda{H}^{\nu\lambda\mu}_{(i,k)}\right) \nn \\
&=&\frac{\partial{L}_{B}}{\partial{g}^{\mu\nu}}
-\frac{1}{2}L_{B}g_{\mu\nu}
+\sum^m_{i=0}{E}^{\mu\nu}_{\text{GenB}(i,j)}
+\frac{1}{2}\sum^m_{i=1}\sum^i_{k=1}
\left(2{X}^{(\mu\nu)}_{(i,k)}
+\nabla_\lambda{H}^{(\mu|\lambda|\nu)}_{(i,k)}\right)
\, , \label{EomLagBgen0}
\eea
with the second-rank symmetric tensor
${E}^{\mu\nu}_{\text{GenB}(i,j)}$ given by
Eq. (\ref{EgenBij}), and the surface term
$\Theta^\mu_{B}$ is presented by
\be
\Theta^\mu_{B}=
\sum^m_{i=0}{\Theta}^\mu_{\text{GenB}(i,j)}
+\sum^m_{i=1}\sum^i_{k=1}\Theta_{(i,k)}^\mu
\, , \label{ThetaLagB}
\ee
where ${\Theta}^\mu_{\text{GenB}(i,j)}$ can be found in
Eq. (\ref{GenThetij}). Furthermore, with the help of the tensor
${X}^{\mu\nu}_{\text{GenB}(i,j)}$ appearing in Eq. (\ref{XGenBij}),
the expression ${E}^{\mu\nu}_{B}$ for the field equations is
able to be alternatively written as an economic and simple
form that is irrelevant to
the $\partial{L}_{B}/\partial{g}^{\mu\nu}$
and $\partial{B}_{\alpha_1\cdot\cdot\cdot\alpha_n}
/\partial{g}^{\mu\nu}$ terms, that is,
\be
{E}^{\mu\nu}_{B}=\sum^m_{i=0}{X}^{\mu\nu}_{\text{GenB}(i,j)}
+\sum^m_{i=1}\sum^i_{k=1}{X}^{\mu\nu}_{(i,k)}
-\frac{1}{2}L_{B}{g}^{\mu\nu}
\, . \label{EomLagBgen}
\ee

As a consequence of the comparison between
Eqs. (\ref{EomLagBgen0}) and (\ref{EomLagBgen}),
an identity related to the derivative of the Lagrangian
density $L_B$ with respect to the metric reads
\bea
\frac{\partial{L}_{B}}{\partial{g}^{\rho\sigma}}
{g}^{\mu\rho}{g}^{\nu\sigma}
&=&\sum^m_{i=0}\left({X}^{\mu\nu}_{\text{GenB}(i,j)}
-{E}^{\mu\nu}_{\text{GenB}(i,j)}\right)
-\frac{1}{2}\sum^m_{i=1}
\sum^i_{k=1}\nabla_\lambda{H}^{\nu\lambda\mu}_{(i,k)}
\, . \label{IdofDLagBg}
\eea
Substituting Eq. (\ref{XGenBij2}) into the above identity
results in the following form
\bea
\frac{\partial{L}_{B}}{\partial{g}^{\rho\sigma}}
{g}^{\mu\rho}{g}^{\nu\sigma}
+\sum^m_{i=0}P^{\mu\nu}_{\text{AB}(i)}
&=&2\sum^m_{i=0}Q^{\mu\lambda\rho\sigma}_{(i,j)}
\Box^{j}R^{\nu}_{~\lambda\rho\sigma}
-\frac{1}{2}\sum^m_{i=1}
\sum^i_{k=1}\nabla_\lambda{H}^{\nu\lambda\mu}_{(i,k)}
\, , \label{IdofDLagBg2}
\eea
in which the rank-2 symmetric tensor $P^{\mu\nu}_{\text{AB}(i)}$
and the fourth-rank one $Q^{\alpha\beta\rho\sigma}_{(i,j)}$
are given respectively by Eqs. (\ref{Qi24def}) and (\ref{Qil4def})
with the substitution
$A^{\alpha_1\cdot\cdot\cdot\alpha_n}\rightarrow
{A}^{\alpha_1\cdot\cdot\cdot\alpha_n}_{B(i)}$ in them.
Apart from this, the symmetry for ${E}^{\mu\nu}_{B}$
gives rise to another identity
\be
\sum^m_{i=0}{X}^{[\mu\nu]}_{\text{GenB}(i,j)}
=-\frac{1}{2}\sum^m_{i=1}
\sum^i_{k=1}\nabla_\lambda{H}^{[\mu|\lambda|\nu]}_{(i,k)}
=-\sum^m_{i=1}\sum^i_{k=1}{X}^{[\mu\nu]}_{(i,k)}
\, , \label{IdXLagBsymm}
\ee
which can equivalently be written as
\be
\sum^m_{i=0}\left(\Box^{j}R^{[\mu}_{~~\lambda\rho\sigma}\right)
Q^{\nu]\lambda\rho\sigma}_{(i,j)}
=\frac{1}{4}\sum^m_{i=1}
\sum^i_{k=1}\nabla_\lambda{H}^{[\mu|\lambda|\nu]}_{(i,k)}
\, . \label{IdXLagBsymm2}
\ee

Like before, the Noether potential ${K}^{\mu\nu}_{B}$
associated to the Lagrangian (\ref{LagBgen}) can be derived
out of the surface term $\Theta^\mu_{B}$ under the
transformation $\delta\rightarrow\mathcal{L}_\zeta$,
which is read off as
\be
{K}^{\mu\nu}_{B}=
\sum^m_{i=0}{K}^{\mu\nu}_{\text{GenB}(i,j)}
+\sum^m_{i=1}\sum^i_{k=1}{K}^{\mu\nu}_{(i,k)}
\, , \label{NoePotLagB}
\ee
with the second-rank anti-symmetric tensor
${K}^{\mu\nu}_{\text{GenB}(i,j)}$
given by Eq. (\ref{KGenBij}). In fact, according to
Eqs. (\ref{TheikdelLie}) and (\ref{TheXgenBijRel}),
it is easy to verify that the surface term
$\Theta^\mu_{B}(\delta\rightarrow\mathcal{L}_\zeta)$
establishes the connection between ${E}^{\mu\nu}_{B}$
in Eq. (\ref{EomLagBgen}) and ${K}^{\mu\nu}_{B}$ through
\be
\Theta^\mu_{B}(\delta\rightarrow\mathcal{L}_\zeta)
=2\zeta_\nu\left({E}^{\mu\nu}_{B}
+\frac{1}{2}L_{B}{g}^{\mu\nu}\right)
-\nabla_\nu{K}^{\mu\nu}_{B}
\, , \label{ThetBdelLie}
\ee
which has the same structure as the one in Eq. (\ref{TheABiLie}).
The off-shell Noether current $J^{\mu}_{B}$
corresponding to the Noether potential ${K}^{\mu\nu}_{B}$
takes the following form
\be
J^{\mu}_{B}=\nabla_\nu{K}^{\mu\nu}_{B}
=2\zeta_\nu{E}^{\mu\nu}_{B}+\zeta^\mu{L}_{B}
-\Theta^\mu_{B}(\delta\rightarrow\mathcal{L}_\zeta)
\, . \label{ConCurLagB}
\ee

Strictly speaking, within Eq. (\ref{VariLagBgen0}) and
all the above equations from (\ref{VariLagBgen}) to
(\ref{ConCurLagB}), the tensor
$A^{\alpha_1\cdot\cdot\cdot\alpha_n}$ in all the quantities
${X}^{\mu\nu}_{\text{GenB}(i,j)}$, ${X}^{\mu\nu}_{(i,k)}$,
${\Theta}^\mu_{\text{GenB}(i,j)}$, $\Theta_{(i,k)}^\mu$,
${E}^{\mu\nu}_{\text{GenB}(i,j)}$, ${H}^{\nu\lambda\mu}_{(i,k)}$,
${K}^{\mu\nu}_{\text{GenB}(i,j)}$,
and ${K}^{\mu\nu}_{(i,k)}$ is replaced with the
rank-$n$ one $A^{\alpha_1\cdot\cdot\cdot\alpha_n}_{B(i)}$.

When the tensor ${B}_{\alpha_1\cdot\cdot\cdot\alpha_n}$ in
the Lagrangian (\ref{LagBgen}) takes a more general form
\be
{B}_{\alpha_1\cdot\cdot\cdot\alpha_n}=
{B}_{\alpha_1\cdot\cdot\cdot\alpha_n}\left(g^{\mu\nu},
{R}_{\mu\nu\rho\sigma},\Box{R}_{\mu\nu\rho\sigma},
\cdot\cdot\cdot,
\Box^{\hat{m}}{R}_{\mu\nu\rho\sigma}\right)
\, , \label{GenBinLagB}
\ee
where $\hat{m}$ denotes an arbitrary nonnegative integer,
the aforementioned results related to the Lagrangian
(\ref{LagBgen}) can be directly extended to
such a situation. Correspondingly, the surface term
(\ref{ThetaLagB}) behaves like
\be
\Theta^\mu_{B}\rightarrow
\sum^{\hat{m}}_{j=0}\sum^m_{i=0}
{\Theta}^\mu_{\text{GenB}(i,j)}
+\sum^m_{i=1}\sum^i_{k=1}\Theta_{(i,k)}^\mu
\, , \label{ThetaLaggenB}
\ee
the expression (\ref{EomLagBgen}) for equations of motion
possesses an alternative form through
\be
{E}^{\mu\nu}_{B}\rightarrow
\sum^{\hat{m}}_{j=0}\sum^m_{i=0}{X}^{\mu\nu}_{\text{GenB}(i,j)}
+\sum^m_{i=1}\sum^i_{k=1}{X}^{\mu\nu}_{(i,k)}
-\frac{1}{2}L_{B}{g}^{\mu\nu}
\, , \label{EomLaggenB}
\ee
and the Noether potential ${K}^{\mu\nu}_{B}$ is transformed
into another form in the way
\be
{K}^{\mu\nu}_{B}\rightarrow
\sum^{\hat{m}}_{j=0}\sum^m_{i=0}{K}^{\mu\nu}_{\text{GenB}(i,j)}
+\sum^m_{i=1}\sum^i_{k=1}{K}^{\mu\nu}_{(i,k)}
\, . \label{NoePotLaggenB}
\ee

It has been proved that the expression ${E}^{\mu\nu}_{B}$
for equations of motion is divergenceless
in Appendix \ref{appendC}. The Noether potential ${K}^{\mu\nu}_{B}$
can be directly used to define the Wald entropy associated to the
Lagrangian $\sqrt{-g}L_{B}$. Particularly, when $j=0$ and
the tensor $B_{\alpha_1\cdot\cdot\cdot\alpha_n}$ are specific to
the values $\big(R,{R}_{\mu\nu},{R}_{\mu\nu\rho\sigma}\big)$ so that
$L_B=\big(L_R,L_{\text{Ric}},L_{\text{Riem}}\big)$, the three tensors
$A^{\alpha_1\cdot\cdot\cdot\alpha_n}_{B(i)}$,
$P^{\mu\nu}_{\text{AB}(i)}$, and
$Q^{\mu\nu\rho\sigma}_{(i,0)}$ accordingly behave like
\bea
&&A^{\alpha_1\cdot\cdot\cdot\alpha_n}_{B(i)}\rightarrow
\left({F}_{(i)},{P}^{\mu\nu}_{(i)},
{P}^{\mu\nu\rho\sigma}_{(i)}\right)\, ,\nn \\
&&P^{\mu\nu}_{\text{AB}(i)}\rightarrow
\left(2R^{\mu\nu}\Box^i{F}_{(i)},
R^{\mu\rho\nu\sigma}
\Box^i{P}_{(i)\rho\sigma},0\right)\, ,\nn \\
&&Q^{\mu\nu\rho\sigma}_{(i,0)}\rightarrow
\left(g^{\mu[\rho}g^{\sigma]\nu}\Box^i{F}_{(i)},
g^{[\mu|[\rho}\Box^i{P}^{\sigma]|\nu]}_{(i)},
\Box^i{P}^{\mu\nu\rho\sigma}_{(i)}\right)
\, , \label{Qi0ofRRicRiem}
\eea
respectively. Then substituting them into the
expression ${E}^{\mu\nu}_{B}$ for equations of motion and
the Noether potential ${K}^{\mu\nu}_{B}$ leads to the results
coinciding with those associated with the Lagrangians
$\sqrt{-g}L_R$, $\sqrt{-g}L_{\text{Ric}}$,
and $\sqrt{-g}L_{\text{Riem}}$, respectively. To this point,
one may conclude that the Lagrangian $\sqrt{-g}L_{B}$ provides
a unified perspective for all the mentioned three ones.
Apart from this, the combination
of ${K}^{\mu\nu}_{B}$ with the surface term $\Theta^\mu_{B}$
is able to produce the Iyer-Wald potential for the definition of
conserved charges corresponding to an arbitrary Killing vector
$\xi^\mu$, which is read off as \cite{LeeWald,IyWald,WalZo}
\be
Q^{\mu\nu}_{B}=\frac{1}{\sqrt{-g}}
\delta\left(\sqrt{-g}{K}^{\mu\nu}_{B}(\zeta\rightarrow\xi)\right)
-\xi^{[\mu}\Theta^{\nu]}_{B}
\, . \label{IWpotofLagB}
\ee
Here the potential $Q^{\mu\nu}_{B}$ can be adopted to define
the mass and the angular momentum, as well as to investigate
the black hole thermodynamics, for the theories of gravity
admitting the Lagrangian $\sqrt{-g}L_{B}$.

%%%%%%%%%%%%%%%%%%%%%%%%%%%%%%%%%%%%%%%%%%%%%%%%%%%%
\subsection{The equations of motion and Noether potentials
for the Lagrangians $\sqrt{-g}A(R)\Box^{i}B(R)$,
$\sqrt{-g}A(R)\Box^{-i}B(R)$ and
$\sqrt{-g}f_{(i,j)}(\Box^{i}B,\Box^{-j}D)$} \label{Five3}
%%%%%%%%%%%%%%%%%%%%%%%%%%%%%%%%%%%%%%%%%%%

Within the present subsection, as specific applications
to elucidate the generic results obtained in Subsection
\ref{Five1}, we perform investigation on the equations
of motion and Noether potentials associated to the
Lagrangians that consist of two functionals,
unlike in the previous situations where the Lagrangian
denstities, such as $L_R$, $L_{\text{Ric}}$,
$L_{\text{Riem}}$, $L_i$, and $L_B$, merely involve
one functional.

We first concentrate on the derivation for the field
equations and the Noether potential corresponding
to the Lagrangian
\be
\sqrt{-g}f_{(i)}=\sqrt{-g}A(R)\Box^{i}B(R)
\, , \label{LagfiABiB}
\ee
where the scalars $A(R)$ and $B(R)$ represent two analytic
functions merely depending on the Ricci scalar $R$. The
Lagrangian (\ref{LagfiABiB}) can be treated as the fundamental
element that constitutes the Lagrangians for nonlocal gravity
models \cite{ALS97,BGKM12,Mods12,BCKM14,KKSrev23,CBnonloc22}.
Varying the Lagrangian (\ref{LagfiABiB}) with respect to
the metric and the Ricci scalar, we obtain
\be
\delta\big(\sqrt{-g}f_{(i)}\big)=
\sqrt{-g}\left[\frac{1}{2}f_{(i)}g^{\mu\nu}\delta{g}_{\mu\nu}
+\big(\Box^{i}B\big)\delta{A}
+{A}\big(\delta\Box^{i}B\big)\right]
\, . \label{VaLagfiABiB}
\ee
With the help of Eq. (\ref{SumOmegik2}), equation (\ref{VaLagfiABiB})
can be further recast into
\bea
\frac{\delta\big(\sqrt{-g}f_{(i)}\big)}{\sqrt{-g}}
&=&A_{(i)}\delta{R}
+\frac{1}{2}f_{(i)}g^{\mu\nu}\delta{g}_{\mu\nu}
-\frac{1}{2}g^{\mu\nu}\sum^i_{k=1}\nabla_\lambda
\left[\left(\Box^{k-1}A\right)
\nabla^{\lambda}\Box^{i-k}B\right]\delta g_{\mu\nu}\nn \\
&&+\sum^i_{k=1}
\left(\nabla^{(\mu}\Box^{k-1}A\right)
\left(\nabla^{\nu)}\Box^{i-k}B\right)\delta g_{\mu\nu}
+\sum^i_{k=1}\nabla_\mu\Theta_{\text{SB}(i,k)}^\mu
\, . \label{VaLagfiABiB2}
\eea
Within the above equation, the scalar $A_{(i)}$ is defined
through
\be
A_{(i)}=\frac{dA}{dR}\big(\Box^i{B}\big)
+\big(\Box^i{A}\big)\frac{dB}{dR}
\, , \label{Aidef}
\ee
and the vector $\Theta_{\text{SB}(i,k)}^\mu$ can be derived out
of Eq. (\ref{Thegenik2}), which is read off as
\bea
\Theta_{\text{SB}(i,k)}^\mu&=&
\left(\Box^{k-1}A\right)
\left(\delta\nabla^{\mu}\Box^{i-k}B\right)
-\left(\nabla^{\mu}\Box^{k-1}A\right)
\left(\delta\Box^{i-k}B\right) \nn \\
&&+\frac{1}{2}g^{\rho\sigma}
\left(\Box^{k-1}A\right)\left(\nabla^{\mu}\Box^{i-k}B\right)
\delta{g}_{\rho\sigma}
\, . \label{TheSBidef}
\eea
Furthermore, by expressing $A_{(i)}\delta{R}$ as the sum
of terms proportional to the variation of the metric
and a divergence term,
\be
A_{(i)}\delta{R}=\big(\nabla^\mu\nabla^\nu{A}_{(i)}
-A_{(i)}{R}^{\mu\nu}
-{g}^{\mu\nu}\Box{A}_{(i)}\big)\delta{g}_{\mu\nu}
+\nabla_\mu\Theta_{\text{SA}(i)}^\mu
\, , \label{AidelRdef}
\ee
where the surface term $\Theta_{\text{SA}(i)}^\mu$ is given by
\be
\Theta_{\text{SA}(i)}^\mu=2A_{(i)}g^{\rho[\mu}
\nabla^{\nu]}\delta g_{\rho\nu}
-2g^{\rho[\mu}\big(\nabla^{\nu]}A_{(i)}\big)\delta g_{\rho\nu}
\, , \label{TheSAi}
\ee
the variation of the Lagrangian $\sqrt{-g}f_{(i)}$ is finally
written as
\be
\delta\big(\sqrt{-g}f_{(i)}\big)
=\sqrt{-g}\Big(-E^{\mu\nu}_{\text{AB}(i)}\delta{g}_{\mu\nu}
+\nabla_\mu\Theta_{\text{AB}(i)}^\mu\Big)
\, . \label{VaLagfiABiB3}
\ee
Within the above equation, the second-rank symmetric tensor
$E^{\mu\nu}_{\text{AB}(i)}$ is the desired expression for
equations of motion, being of the form
\bea
E^{\mu\nu}_{\text{AB}(i)}&=&
\frac{1}{2}g^{\mu\nu}\sum^i_{k=1}\nabla_\lambda
\left[\left(\Box^{k-1}A\right)
\nabla^{\lambda}\Box^{i-k}B\right]
-\sum^i_{k=1}
\left(\nabla^{(\mu}\Box^{k-1}A\right)
\nabla^{\nu)}\Box^{i-k}B \nn \\
&&+{g}^{\mu\nu}\Box{A}_{(i)}+A_{(i)}{R}^{\mu\nu}
-\nabla^\mu\nabla^\nu{A}_{(i)}
-\frac{1}{2}g^{\mu\nu}A\Box^{i}B
\, , \label{EoMABoxiB}
\eea
while the surface term $\Theta_{\text{AB}(i)}^\mu$
is presented by
\be
\Theta_{\text{AB}(i)}^\mu=\Theta_{\text{SA}(i)}^\mu
+\sum^i_{k=1}\Theta_{\text{SB}(i,k)}^\mu
\, . \label{TheSABi}
\ee
Through direct computations, one can prove that the
expression $E^{\mu\nu}_{\text{AB}(i)}$
for field equations is indeed divergence-free. Actually,
\bea
\nabla_\nu{E}^{\mu\nu}_{\text{AB}(i)}&=&
\frac{1}{2}A\nabla^\mu\Box^iB
-\frac{1}{2}\big(\Box^iA\big)\nabla^\mu{B}
-\frac{1}{2}\nabla^\mu\big(A\Box^{i}B\big)
+\frac{1}{2}{A}_{(i)}\nabla^\mu{R} \nn \\
&&+{R}^{\mu\nu}\nabla_\nu{A}_{(i)}+\nabla^\mu\Box{A}_{(i)}
-\nabla^\nu\nabla^\mu\nabla_\nu{A}_{(i)}\nn \\
&=&0
\, . \label{DivESABi}
\eea
In order to get the second equality in the above equation,
we have used the identity
$2\nabla^{[\mu}\nabla^{\nu]}\nabla_\nu{A}_{(i)}
=-{R}^{\mu\nu}\nabla_\nu{A}_{(i)}$.

Next, with the purpose to check the field equations through the
straightforward variation of the Lagrangian, as well as to
produce the Noether potential, we are going to pay our attention to
follow the method based on the conserved current to reproduce
Eq. (\ref{EoMABoxiB}) in an alternative way. To achieve such a
purpose, according to this method, it is sufficient to merely
compute the surface term $\Theta_{\text{AB}(i)}^\mu$ with the
variation $\delta$ in it replaced with the Lie derivative
along an arbitrary vector field.

According to Eq. (\ref{TheikdelLie}), when the variation
$\delta$ in $\Theta_{\text{SB}(i,k)}^\mu$ is transformed into
the Lie derivative $\mathcal{L}_\zeta$ with respect to
an arbitrary vector $\zeta^\mu$, this quantity turns into
\be
\Theta_{\text{SB}(i,k)}^\mu(\delta\rightarrow\mathcal{L}_\zeta)
=2\zeta_\nu{X}^{\mu\nu}_{\text{SB}(i,k)}
-\nabla_\nu{K}^{\mu\nu}_{\text{SB}(i,k)}
\, , \label{TheSBikdelLie}
\ee
in which the second-rank symmetric tensor
${X}^{\mu\nu}_{\text{SB}(i,k)}$ has the form
\be
{X}^{\mu\nu}_{\text{SB}(i,k)}=\frac{1}{2}g^{\mu\nu}
\nabla_\lambda\left[\left(\Box^{k-1}A\right)
\nabla^{\lambda}\Box^{i-k}B\right]
-\left(\nabla^{(\mu}\Box^{k-1}A\right)
\nabla^{\nu)}\Box^{i-k}B
\, , \label{XSBik}
\ee
and the anti-symmetric tensor ${K}^{\mu\nu}_{\text{SB}(i,k)}$
is given by
\be
{K}^{\mu\nu}_{\text{SB}(i,k)}=
2\zeta^{[\mu}\Big(\nabla^{\nu]}\Box^{i-k}B\Big)
\Big(\Box^{k-1}A\Big)
\, . \label{KSBik}
\ee
Additionally, by the aid of the second-rank symmetric tensor
\be
{X}^{\mu\nu}_{\text{SA}(i)}={g}^{\mu\nu}\Box{A}_{(i)}
+A_{(i)}{R}^{\mu\nu}
-\nabla^\mu\nabla^\nu{A}_{(i)}
\, , \label{XSAik}
\ee
together with the second-rank anti-symmetric tensor
${K}^{\mu\nu}_{\text{SA}(i)}$ defined through
\be
{K}^{\mu\nu}_{\text{SA}(i)}
=2{A}_{(i)}\nabla^{[\mu}\zeta^{\nu]}
+4\zeta^{[\mu}\nabla^{\nu]}{A}_{(i)}
\, , \label{KmnSAi}
\ee
the substitution of $\delta$ in $\Theta_{\text{SA}(i)}^\mu$
by $\mathcal{L}_\zeta$ leads to
\be
\Theta_{\text{SA}(i)}^\mu(\delta\rightarrow\mathcal{L}_\zeta)
=2\zeta_\nu{X}^{\mu\nu}_{\text{SA}(i)}
-\nabla_\nu{K}^{\mu\nu}_{\text{SA}(i)}
\, . \label{TheSAidelLie}
\ee
As a consequence of the combination for Eqs. (\ref{TheSBikdelLie})
and (\ref{TheSAidelLie}), the surface term
$\Theta_{\text{AB}(i)}^\mu(\delta\rightarrow\mathcal{L}_\zeta)$
is read off as
\be
\Theta_{\text{AB}(i)}^\mu(\delta\rightarrow\mathcal{L}_\zeta)
=2\zeta_\nu\left(E^{\mu\nu}_{\text{AB}(i)}
+\frac{1}{2}g^{\mu\nu}{f}_{(i)}\right)
-\nabla_\nu{K}^{\mu\nu}_{\text{AB}(i)}
\, . \label{TheSABidelLie}
\ee
According to Eqs. (\ref{TheLiegen}) and (\ref{EoMgen}),
the rank-2 tensor $E^{\mu\nu}_{\text{AB}(i)}$ in
Eq. (\ref{TheSABidelLie}) is just
the expression for field equations associated to the Lagrangian
$\sqrt{-g}{f}_{(i)}$, while the Noether potential for
this Lagrangian is the second-rank anti-symmetric tensor
${K}^{\mu\nu}_{\text{AB}(i)}$, given by
\be
{K}^{\mu\nu}_{\text{AB}(i)}
=2{A}_{(i)}\nabla^{[\mu}\zeta^{\nu]}
+4\zeta^{[\mu}\nabla^{\nu]}{A}_{(i)}
+2\sum^i_{k=1}\zeta^{[\mu}\Big(\nabla^{\nu]}\Box^{i-k}B\Big)
\big(\Box^{k-1}A\big)
\, . \label{KmnSABi}
\ee
In light of Eq. (\ref{TheSABidelLie}), the Noether current
${J}^{\mu}_{\text{AB}(i)}$ corresponding to the potential
${K}^{\mu\nu}_{\text{AB}(i)}$ is written as
\be
{J}^{\mu}_{\text{AB}(i)}=
\nabla_\nu{K}^{\mu\nu}_{\text{AB}(i)}
=2\zeta_\nu{E}^{\mu\nu}_{\text{AB}(i)}+{f}_{(i)}\zeta^\mu
-\Theta_{\text{AB}(i)}^\mu(\delta\rightarrow\mathcal{L}_\zeta)
\, . \label{CCofSABi}
\ee
Particularly, if $A=R^m$, $B=R$ and $i=n$, the Lagrangian
(\ref{LagfiABiB}) becomes the one $\sqrt{-g}L_{R1}=R^m\Box^n{R}$ in
Eq. (\ref{LagRmBoxRn}). In such a case, it can be confirmed that
${E}^{\mu\nu}_{\text{AB}(i)}$ coincides with the expression
${E}^{\mu\nu}_{R1}$ for field equations given by
Eq. (\ref{EoMofLagRm}) and the Noether potential
${K}^{\mu\nu}_{\text{AB}(i)}$ becomes
${K}^{\mu\nu}_{R1}$ in Eq. (\ref{KmnR1def}).

As an extension for the situation with respect to the Lagrangian
(\ref{LagfiABiB}), we switch to consider the Lagrangian
\be
\sqrt{-g}\tilde{f}_{(i)}=\sqrt{-g}A(R)\Box^{-i}B(R)
\, . \label{LagtilfiABiB}
\ee
In such a situation, introducing two scalars
\be
\tilde{A}=\Box^{-i}A \, , \quad
\tilde{B}=\Box^{-i}B
\, , \label{tildABdef}
\ee
which satisfy obviously $\Box^{i}\tilde{A}=A$ and
$\Box^{i}\tilde{B}=B$, respectively, we are able to
perform parallel analysis like in the case of the Lagrangian
$\sqrt{-g}{f}_{(i)}$. The variation of the Lagrangian
$\sqrt{-g}\tilde{f}_{(i)}$ is of the form
\bea
\delta\big(\sqrt{-g}\tilde{f}_{(i)}\big)&=&
\sqrt{-g}\left[\frac{1}{2}\tilde{f}_{(i)}g^{\mu\nu}\delta{g}_{\mu\nu}
+\big(\Box^{-i}B\big)\delta{A}
+{A}\big(\delta\Box^{-i}B\big)\right] \nn \\
&=&\sqrt{-g}\left[\frac{1}{2}\tilde{f}_{(i)}g^{\mu\nu}\delta{g}_{\mu\nu}
+\tilde{B}\delta{A}
+\big(\Box^i\tilde{A}\big)\delta\tilde{B}\right]
\, . \label{VaLagtildfiAB}
\eea
By making use of Eq. (\ref{AidelRdef}), together with
an identity for both the quantities $\tilde{A}$ and $\tilde{B}$
derived in terms of Eq. (\ref{SumOmegik2}),
\bea
\tilde{A}\big(\delta\Box^i\tilde{B}\big)
&=&\big(\Box^i\tilde{A}\big)\delta\tilde{B}
-\frac{1}{2}g^{\mu\nu}\sum^i_{k=1}
\nabla_\lambda\big[\big(\Box^{k-1}\tilde{A}\big)
\nabla^{\lambda}\Box^{i-k}\tilde{B}\big]\delta g_{\mu\nu}\nn \\
&&+\sum^i_{k=1}
\Big(\nabla^{(\mu}\Box^{k-1}\tilde{A}\Big)
\Big(\nabla^{\nu)}\Box^{i-k}\tilde{B}\Big)\delta g_{\mu\nu}
+\sum^i_{k=1}\nabla_\mu\tilde{\Theta}_{\text{SB}(i,k)}^\mu
\, , \label{TilfiABmiB}
\eea
where the surface term $\tilde{\Theta}_{\text{SB}(i,k)}^\mu$
is defined by
\be
\tilde{\Theta}_{\text{SB}(i,k)}^\mu=
\Theta_{\text{SB}(i,k)}^\mu
\big(A\rightarrow\tilde{A},B\rightarrow\tilde{B}\big)
\, , \label{tildThetilAB}
\ee
Eq. (\ref{VaLagtildfiAB}) can be written as the form
\be
\delta\big(\sqrt{-g}\tilde{f}_{(i)}\big)
=\sqrt{-g}\Big(-\tilde{E}^{\mu\nu}_{\text{AB}(i)}\delta{g}_{\mu\nu}
+\nabla_\mu\tilde{\Theta}_{\text{AB}(i)}^\mu\Big)
\, . \label{VaLagtilfitilAB}
\ee
Within Eq. (\ref{VaLagtilfitilAB}), the expression
$\tilde{E}^{\mu\nu}_{\text{AB}(i)}$ for field equations
is read off as
\bea
\tilde{E}^{\mu\nu}_{\text{AB}(i)}&=&
\sum^i_{k=1} \left(\nabla^{(\mu}\Box^{k-1}\tilde{A}\right)
\nabla^{\nu)}\Box^{i-k}\tilde{B}
-\frac{1}{2}g^{\mu\nu}\sum^i_{k=1}\nabla_\lambda
\left[\left(\Box^{k-1}\tilde{A}\right)
\nabla^{\lambda}\Box^{i-k}\tilde{B}\right]
\nn \\
&&+{g}^{\mu\nu}\Box\tilde{A}_{(i)}+\tilde{A}_{(i)}{R}^{\mu\nu}
-\nabla^\mu\nabla^\nu\tilde{A}_{(i)}
-\frac{1}{2}g^{\mu\nu}A\Box^{-i}B
\, , \label{EoMtildAB}
\eea
with the scalar $\tilde{A}_{(i)}$ defined through
\be
\tilde{A}_{(i)}=\frac{dA}{dR}\big(\Box^{-i}{B}\big)
+\big(\Box^{-i}{A}\big)\frac{dB}{dR}
\, , \label{TildAidef}
\ee
and the surface term $\tilde{\Theta}_{\text{AB}(i)}^\mu$ is
given by
\be
\tilde{\Theta}_{\text{AB}(i)}^\mu=
\Theta_{\text{SA}(i)}^\mu\big(A_{(i)}\rightarrow\tilde{A}_{(i)}\big)
-\sum^i_{k=1}\tilde{\Theta}_{\text{SB}(i,k)}^\mu
\, . \label{TildTheSABi}
\ee
In addition, repeating the same procedure adopted to derive
the Noether potential ${K}^{\mu\nu}_{\text{AB}(i)}$,
we obtain the Noether potential
$\tilde{K}^{\mu\nu}_{\text{AB}(i)}$ corresponding to
the Lagrangian $\sqrt{-g}\tilde{f}_{(i)}$, which is of the form
\be
\tilde{K}^{\mu\nu}_{\text{AB}(i)}
=2\tilde{A}_{(i)}\nabla^{[\mu}\zeta^{\nu]}
+4\zeta^{[\mu}\nabla^{\nu]}\tilde{A}_{(i)}
-2\sum^i_{k=1}\zeta^{[\mu}\Big(\nabla^{\nu]}\Box^{i-k}\tilde{B}\Big)
\big(\Box^{k-1}\tilde{A}\big)
\, . \label{TildKmnSABi}
\ee
Meanwhile, we can also reproduce the expression
$\tilde{E}^{\mu\nu}_{\text{AB}(i)}$ for equations of motion
out of the surface term
$\tilde{\Theta}_{\text{AB}(i)}^\mu
(\delta\rightarrow\mathcal{L}_\zeta)$.

Like in the works \cite{DDRS18,DDRS22,CCL23}, both the expressions
${E}^{\mu\nu}_{\text{AB}(i)}$ and
$\tilde{E}^{\mu\nu}_{\text{AB}(i)}$ for equations of motion can
be directly adopted to derive the field equations
for nonlocal gravities. For instance, if both of them are adapted to
the same nonlocal gravity models considered in \cite{DDRS18,DDRS22},
one is able to reproduce the corresponding field equations within
these works, which were obtained via the variation of the actions.

In the remainder of this subsection, as a natural generalization
for the combination of the Lagrangians (\ref{LagfiABiB})
and (\ref{LagtilfiABiB}), we consider the one
\be
\sqrt{-g}f_{(i,j)}=\sqrt{-g}f_{(i,j)}
\big(\Box^{i}B,\Box^{-j}D\big)
\, , \label{LagfijBD}
\ee
in which both the scalars $B$ and $D$ are supposed to
take the forms
\be
B=B(g^{\mu\nu},R_{\alpha\beta\rho\sigma})\, , \qquad
D=D(g^{\mu\nu},R_{\alpha\beta\rho\sigma})
\, , \label{BDijform}
\ee
respectively. Here the Lagrangian (\ref{LagfijBD}) can
be directly extended to polynomial-derivative theories
of gravity \cite{ALS97}, as well as to infinite
derivative theories of gravity \cite{BGKM12,Mods12}.
For convenience, let us introduce two scalars
\be
A_{(i,j)}=\frac{\partial{f}_{(i,j)}}{\partial\Box^iB}
\, , \qquad
C_{(i,j)}=\frac{\partial{f}_{(i,j)}}{\partial\Box^{-j}D}
\, , \label{ACijdef}
\ee
two vectors
\bea
\check{\Theta}^{\mu}_{\text{AB}(i,j)}&=&\sum^i_{k=1}
{\Theta}^{\mu}_{\text{SB}(i,k)}\big(A\rightarrow{A}_{(i,j)}\big)
\, , \nn \\
\check{\Theta}^{\mu}_{\text{CD}(i,j)}&=&\sum^j_{k=1}
{\Theta}^{\mu}_{\text{SB}(j,k)}
\big(A\rightarrow\Box^{-j}{C}_{(i,j)},
B\rightarrow\Box^{-j}{D}\big)
\, , \label{ThetABCDij}
\eea
and two second-rank tensors
\bea
\check{X}^{\mu\nu}_{\text{AB}(i,j)}
&=&\frac{1}{2}g^{\mu\nu}\sum^i_{k=1}
\nabla_\lambda\left[\left(\Box^{k-1}{A}_{(i,j)}\right)
\nabla^{\lambda}\Box^{i-k}B\right]
-\sum^i_{k=1}\left(\nabla^{(\mu}\Box^{k-1}{A}_{(i,j)}\right)
\nabla^{\nu)}\Box^{i-k}B\nn \\
&=&\sum^i_{k=1}
{X}^{\mu\nu}_{\text{SB}(i,k)}\big(A\rightarrow{A}_{(i,j)}\big)
\, , \nn \\
\check{X}^{\mu\nu}_{\text{CD}(i,j)}
&=&\frac{1}{2}g^{\mu\nu}\sum^j_{k=1}
\nabla_\lambda\left[\left(\Box^{k-j-1}{C}_{(i,j)}\right)
\nabla^{\lambda}\Box^{-k}D\right]
-\sum^j_{k=1}\left(\nabla^{(\mu}\Box^{k-j-1}{C}_{(i,j)}\right)
\nabla^{\nu)}\Box^{-k}D\nn \\
&=&\sum^j_{k=1}
{X}^{\mu\nu}_{\text{SB}(j,k)}
\big(A\rightarrow\Box^{-j}{C}_{(i,j)},
B\rightarrow\Box^{-j}{D}\big)
\, . \label{XABCDij}
\eea
Within Eqs. (\ref{ThetABCDij}) and (\ref{XABCDij}),
the quantities ${\Theta}^{\mu}_{\text{SB}(i,k)}$
and ${X}^{\mu\nu}_{\text{SB}(i,k)}$ are given by
Eqs. (\ref{TheSBidef}) and (\ref{XSBik}), respectively.
According to the identity
\bea
A_{(i,j)}\big(\delta\Box^i{B}\big)
&=&\big(\Box^iA_{(i,j)}\big)\delta{B}
-\check{X}^{\mu\nu}_{\text{AB}(i,j)}\delta g_{\mu\nu}
+\nabla_\mu\check{\Theta}^{\mu}_{\text{AB}(i,j)}
\, , \label{AijDelBoiB}
\eea
together with the one
\bea
C_{(i,j)}\big(\delta\Box^{-j}{D}\big)
&=&\big(\Box^{-j}C_{(i,j)}\big)\delta{D}
+\check{X}^{\mu\nu}_{\text{CD}(i,j)}\delta g_{\mu\nu}
-\nabla_\mu\check{\Theta}^{\mu}_{\text{CD}(i,j)}
\, , \label{CijDelBojD}
\eea
the variation of the Lagrangian (\ref{LagfijBD}) can be
expressed as
\be
\delta\big(\sqrt{-g}{f}_{(i,j)}\big)
=\sqrt{-g}\left(g_{\mu\rho}g_{\nu\sigma}
{E}^{\mu\nu}_{\text{BD}(i,j)}\delta{g}^{\rho\sigma}
+\nabla_\mu\Theta_{\text{BD}(i,j)}^\mu\right)
\, . \label{VaLagfijBD}
\ee
In the above equation, the expression
${E}^{\mu\nu}_{\text{BD}(i,j)}$ for field equations has the form
\bea
{E}^{\mu\nu}_{\text{BD}(i,j)}&=&
\check{X}^{\mu\nu}_{\text{AB}(i,j)}
-\check{X}^{\mu\nu}_{\text{CD}(i,j)}
-\frac{1}{2}{f}_{(i,j)}{g}^{\mu\nu}
+P^{\mu\nu}_{\text{BD}(i,j)} \nn \\
&&-P^{\mu\lambda\rho\sigma}_{\text{BD}(i,j)}
R^\nu_{~\lambda\rho\sigma}-2\nabla_\rho\nabla_\sigma
P^{\rho\mu\nu\sigma}_{\text{BD}(i,j)}
\, , \label{EBDijdef}
\eea
where both the tensors $P^{\mu\nu}_{\text{BD}(i,j)}$
and $P^{\mu\nu\rho\sigma}_{\text{BD}(i,j)}$ are defined
respectively by
\bea
P^{\mu\nu}_{\text{BD}(i,j)}
&=&g^{\mu\rho}g^{\nu\sigma}
\left[\frac{\partial{B}}{\partial{g}^{\rho\sigma}}
\big(\Box^i{A}_{(i,j)}\big)
+\big(\Box^{-j}{C}_{(i,j)}\big)
\frac{\partial{D}}{\partial{g}^{\rho\sigma}}\right] \, , \nn \\
P^{\mu\nu\rho\sigma}_{\text{BD}(i,j)}
&=&\frac{\partial{B}}{\partial{R}_{\mu\nu\rho\sigma}}
\big(\Box^i{A}_{(i,j)}\big)
+\big(\Box^{-j}{C}_{(i,j)}\big)
\frac{\partial{D}}{\partial{R}_{\mu\nu\rho\sigma}}
\, , \label{PBD24def}
\eea
while the surface term $\Theta_{\text{BD}(i,j)}^\mu$
is given by
\bea
\Theta_{\text{BD}(i,j)}^\mu&=&
\check{\Theta}^{\mu}_{\text{AB}(i,j)}
-\check{\Theta}^{\mu}_{\text{CD}(i,j)}
+2P^{\mu\nu\rho\sigma}_{\text{BD}(i,j)}
\nabla_\sigma\delta g_{\rho\nu}
-2(\delta g_{\nu\rho})
\nabla_\sigma{P}^{\mu\nu\rho\sigma}_{\text{BD}(i,j)}
\, . \label{ThetBDij}
\eea
More specifically, the surface term
$\Theta_{\text{BD}(i,j)}^\mu$ can be expressed as
\bea
\Theta_{\text{BD}(i,j)}^\mu&=&
-\sum^j_{k=1}\left[\big(\Box^{k-j-1}{C}_{(i,j)}\big)
\delta\nabla^{\mu}\Box^{-k}D
-\big(\nabla^{\mu}\Box^{k-j-1}{C}_{(i,j)}\big)
\delta\Box^{-k}D \right.\nn \\
&&\left.+\frac{1}{2}(g^{\rho\sigma}\delta{g}_{\rho\sigma})
\big(\Box^{k-j-1}{C}_{(i,j)}\big)
\nabla^{\mu}\Box^{-k}D\right]
-2(\delta g_{\nu\rho})
\nabla_\sigma{P}^{\mu\nu\rho\sigma}_{\text{BD}(i,j)}\nn \\
&&+\sum^i_{k=1}\left[\big(\Box^{k-1}{A}_{(i,j)}\big)
\delta\nabla^{\mu}\Box^{i-k}B
-\big(\nabla^{\mu}\Box^{k-1}{A}_{(i,j)}\big)
\delta\Box^{i-k}B \right. \nn \\
&&\left.+\frac{1}{2}(g^{\rho\sigma}\delta{g}_{\rho\sigma})
\big(\Box^{k-1}{A}_{(i,j)}\big)
\nabla^{\mu}\Box^{i-k}B\right]
+2P^{\mu\nu\rho\sigma}_{\text{BD}(i,j)}
\nabla_\sigma\delta g_{\rho\nu}
\, . \label{ThetBDij2}
\eea

Moreover, computing $\Theta_{\text{BD}(i,j)}^\mu
(\delta\rightarrow\mathcal{L}_\zeta)$ and putting it
into the form
\be
\Theta_{\text{BD}(i,j)}^\mu
(\delta\rightarrow\mathcal{L}_\zeta)=
2\zeta_\nu{X}^{\mu\nu}_{\text{BD}(i,j)}
-\nabla_\nu{K}^{\mu\nu}_{\text{BD}(i,j)}
\, , \label{RelThEBDijLie}
\ee
in which the rank-2 symmetric tensor ${X}^{\mu\nu}_{\text{BD}(i,j)}$
is read off as
\be
{X}^{\mu\nu}_{\text{BD}(i,j)}=
\check{X}^{\mu\nu}_{\text{AB}(i,j)}
-\check{X}^{\mu\nu}_{\text{CD}(i,j)}
+P^{\mu\lambda\rho\sigma}_{\text{BD}(i,j)}
R^\nu_{~\lambda\rho\sigma}-2\nabla_\rho\nabla_\sigma
P^{\rho\mu\nu\sigma}_{\text{BD}(i,j)}
\, , \label{XBDijdef}
\ee
while the anti-symmetric tensor
${K}^{\mu\nu}_{\text{BD}(i,j)}$, representing the Noether
potential associated to the Lagrangian (\ref{LagfijBD}),
takes the form
\bea
{K}^{\mu\nu}_{\text{BD}(i,j)}&=&
2\sum^i_{k=1}\zeta^{[\mu}\Big(\nabla^{\nu]}\Box^{i-k}{B}\Big)
\Big(\Box^{k-1}{A}_{(i,j)}\Big)
-2\sum^j_{k=1}\zeta^{[\mu}\Big(\nabla^{\nu]}\Box^{-k}{D}\Big)
\Big(\Box^{k-j-1}{C}_{(i,j)}\Big) \nn \\
&&+2P^{\mu\nu\rho\sigma}_{\text{BD}(i,j)}\nabla_{\rho}\zeta_{\sigma}
+4\zeta_\rho\nabla_\sigma{P}^{\mu\nu\rho\sigma}_{\text{BD}(i,j)}
-6P^{\mu[\nu\rho\sigma]}_{\text{BD}(i,j)}\nabla_\rho\zeta_\sigma
\, , \label{KmnBDij}
\eea
one obtains an alternative form for ${E}^{\mu\nu}_{\text{BD}(i,j)}$,
written as
\be
{E}^{\mu\nu}_{\text{BD}(i,j)}=
{X}^{\mu\nu}_{\text{BD}(i,j)}
-\frac{1}{2}{f}_{(i,j)}{g}^{\mu\nu}
\, . \label{EBDijaltf}
\ee
With the help of Eq. (\ref{XABCDij}), the above equation
is explicitly expressed as
\bea
{E}^{\mu\nu}_{\text{BD}(i,j)}&=&
\frac{1}{2}g^{\mu\nu}\sum^i_{k=1}\nabla_\lambda
\left[\big(\Box^{k-1}{A}_{(i,j)}\big)
\nabla^{\lambda}\Box^{i-k}{B}\right]
-\frac{1}{2}g^{\mu\nu}\sum^j_{k=1}\nabla_\lambda
\left[\big(\Box^{k-j-1}{C}_{(i,j)}\big)
\nabla^{\lambda}\Box^{-k}{D}\right]
 \nn \\
&&+\sum^j_{k=1} \left(\nabla^{(\mu}
\Box^{k-j-1}{C}_{(i,j)}\right)
\nabla^{\nu)}\Box^{-k}{D}
-\sum^i_{k=1} \left(\nabla^{(\mu}\Box^{k-1}{A}_{(i,j)}\right)
\nabla^{\nu)}\Box^{i-k}{B} \nn \\
&&+P^{\mu\lambda\rho\sigma}_{\text{BD}(i,j)}
R^\nu_{~\lambda\rho\sigma}-2\nabla_\rho\nabla_\sigma
P^{\rho\mu\nu\sigma}_{\text{BD}(i,j)}
-\frac{1}{2}{f}_{(i,j)}{g}^{\mu\nu}
\, . \label{EBDijaltf2}
\eea
It has been proven in Appendix \ref{appendC} that
the expression ${E}^{\mu\nu}_{\text{BD}(i,j)}$ fulfills
the Bianchi-type identity
$\nabla_\mu{E}^{\mu\nu}_{\text{BD}(i,j)}=0$.
On the basis of the surface term (\ref{ThetBDij}) and
the Noether potential (\ref{KmnBDij}), the off-shell
conserved current $J_{\text{BD}(i,j)}^\mu$ is defined by
\be
J_{\text{BD}(i,j)}^\mu=
2\zeta_\nu{X}^{\mu\nu}_{\text{BD}(i,j)}
-\Theta_{\text{BD}(i,j)}^\mu
(\delta\rightarrow\mathcal{L}_\zeta)
=\nabla_\nu{K}^{\mu\nu}_{\text{BD}(i,j)}
\, , \label{CCurJBDij}
\ee
and the Iyer-Wald potential associated to the
Lagrangian (\ref{LagfijBD}) possesses the form
\be
{Q}^{\mu\nu}_{\text{BD}(i,j)}=\frac{1}{\sqrt{-g}}
\delta\left(\sqrt{-g}{K}^{\mu\nu}_{\text{BD}(i,j)}
(\zeta\rightarrow\xi)\right)
-\xi^{[\mu}{\Theta}^{\nu]}_{\text{BD}(i,j)}
\, . \label{IWptLagfij}
\ee
Finally, it is worthwhile to point out that all the
aforementioned analysis related to the Lagrangian
(\ref{LagfijBD}) can be naturally extended to the one
$\sqrt{-g}\tilde{f}_{(i,j)}\big(\Box^{i}B,\Box^{j}D\big)$.
Besides, the expression (\ref{EBDijaltf2}) can be adopted to
reproduce the field equations for the nonlocal gravity
theories given by \cite{BCKM14} in a much simpler manner.
We shall achieve this in the future work.

%%%%%%%%%%%%%%%%%%%%%%%%%%%%%%%%%%%%%%%%%%%%%%%%%%%%
\subsection{Applications in the Lagrangian densities
$C(g^{\mu\nu},R_{\mu\nu\rho\sigma})
\Box^{i}D(g^{\mu\nu},R_{\mu\nu\rho\sigma})$,
$\hat{C}^{\alpha_1\cdot\cdot\cdot\alpha_n}
\Box^{i}\hat{D}_{\alpha_1\cdot\cdot\cdot\alpha_n}$
and
$\big(\Box^{i}\hat{C}^{\alpha_1\cdot\cdot\cdot\alpha_n}\big)
\Box^{j}\hat{D}_{\alpha_1\cdot\cdot\cdot\alpha_n}$} \label{Five4}
%%%%%%%%%%%%%%%%%%%%%%%%%%%%%%%%%%%%%%%%%%%

In the present subsection, in order to see the above results
related to the Lagrangian (\ref{LagfiABiB}) from a more generic
perspective, as well as to provide more generic examples to
demonstrate the results within Subsection \ref{Five1},
we pay attention to their applications in three
types of more generic Lagrangians than the one (\ref{LagfiABiB}),
which still comprise two functionals.
First, we take into consideration of the field equations and
the Noether potential associated to the following Lagrangian:
\be
\sqrt{-g}h_{(i)}=\sqrt{-g}C(g^{\mu\nu},R_{\mu\nu\rho\sigma})
\Box^{i}D(g^{\mu\nu},R_{\mu\nu\rho\sigma})
\, , \label{LagCBoxiD}
\ee
in which both the scalars $C(g^{\mu\nu},R_{\mu\nu\rho\sigma})$
and $D(g^{\mu\nu},R_{\mu\nu\rho\sigma})$ are restricted to
rely on the variables for both the inverse metric
$g^{\mu\nu}$ and the Riemann curvature tensor
$R_{\mu\nu\rho\sigma}$ for simplicity. Here the Lagrangian
(\ref{LagCBoxiD}) can be regarded as a special case of
the one (\ref{LagfijBD}) with $j=0$, $D=C$ and $B=D$.
By analogy with the
situation of the Lagrangian (\ref{LagfiABiB}), the variation
for the Lagrangian (\ref{LagCBoxiD}) with regard to
$g^{\mu\nu}$ and $R_{\mu\nu\rho\sigma}$ is expressed as
\bea
\delta\big(\sqrt{-g}h_{(i)}\big)
&=&\sqrt{-g}\left(\frac{1}{2}h_{(i)}g^{\mu\nu}\delta{g}_{\mu\nu}
-\bar{X}^{\mu\nu}_{\text{CD}(i)}\delta{g}_{\mu\nu}
 +\nabla_\mu\bar{\Theta}_{\text{CD}(i)}^\mu\right)\nn \\
&&+\sqrt{-g}\left[\big(\Box^{i}D\big)\delta{C}
+\big(\Box^{i}C\big)\delta{D}\right] \nn \\
&=&\sqrt{-g}\Big(-E^{\mu\nu}_{\text{CD}(i)}\delta{g}_{\mu\nu}
+\nabla_\mu\Theta_{\text{CD}(i)}^\mu\Big)
\, . \label{VaLagCBoxiD}
\eea
By the aid of a rank-2 symmetric tensor $P^{\mu\nu}_{\text{CD}(i)}$
and a rank-4 one $P^{\mu\nu\rho\sigma}_{\text{CD}(i)}$ that inherits
the algebraic symmetries from the Riemann tensor, defined via
\bea
P^{\mu\nu}_{\text{CD}(i)}
&=&g^{\mu\rho}g^{\nu\sigma}
\left[\frac{\partial{C}}{\partial{g}^{\rho\sigma}}
\big(\Box^i{D}\big)
+\big(\Box^i{C}\big)
\frac{\partial{D}}{\partial{g}^{\rho\sigma}}\right] \, , \nn \\
P^{\mu\nu\rho\sigma}_{\text{CD}(i)}
&=&\frac{\partial{C}}{\partial{R}_{\mu\nu\rho\sigma}}
\big(\Box^i{D}\big)
+\big(\Box^i{C}\big)
\frac{\partial{D}}{\partial{R}_{\mu\nu\rho\sigma}}
\, , \label{PCD24def}
\eea
respectively, the expression for field equations
$E^{\mu\nu}_{\text{CD}(i)}$ within Eq. (\ref{VaLagCBoxiD})
is written as
\bea
E^{\mu\nu}_{\text{CD}(i)}&=&
\bar{X}^{\mu\nu}_{\text{CD}(i)}+P^{\mu\nu}_{\text{CD}(i)}
-P^{\mu\lambda\rho\sigma}_{\text{CD}(i)}
R^\nu_{~\lambda\rho\sigma}-2\nabla_\rho\nabla_\sigma
P^{\rho\mu\nu\sigma}_{\text{CD}(i)}
-\frac{1}{2}g^{\mu\nu}C\Box^{i}D
\, , \label{EoMCBoxiD}
\eea
while the surface term $\Theta_{\text{CD}(i)}^\mu$ is presented
by
\be
\Theta_{\text{CD}(i)}^\mu
=\bar{\Theta}^{\mu}_{\text{CD}(i)}
+2P^{\mu\nu\rho\sigma}_{\text{CD}(i)}
\nabla_\sigma\delta g_{\rho\nu}
-2(\delta g_{\nu\rho})
\nabla_\sigma{P}^{\mu\nu\rho\sigma}_{\text{CD}(i)}
\, . \label{TheCBoxiD}
\ee
In addition, within Eqs. (\ref{VaLagCBoxiD}) and
(\ref{EoMCBoxiD}), the second-rank symmetric tensor
$\bar{X}^{\mu\nu}_{\text{CD}(i)}$ is of the form
\bea
\bar{X}^{\mu\nu}_{\text{CD}(i)}&=&
\frac{1}{2}g^{\mu\nu}\sum^i_{k=1}\nabla_\lambda
\left[\left(\Box^{k-1}C\right)
\nabla^{\lambda}\Box^{i-k}D\right]
-\sum^i_{k=1}
\left(\nabla^{(\mu}\Box^{k-1}C\right)
\nabla^{\nu)}\Box^{i-k}D
\, , \label{EbarCDi}
\eea
and the vector $\bar{\Theta}^{\mu}_{\text{CD}(i)}$
in Eqs. (\ref{VaLagCBoxiD}) and (\ref{TheCBoxiD})
is defined through
\bea
\bar{\Theta}^{\mu\nu}_{\text{CD}(i)}&=&
\sum^i_{k=1}\Theta_{\text{SB}(i,k)}^\mu
(A\rightarrow{C},B\rightarrow{D})
\nn \\
&=&\sum^i_{k=1}\left[\big(\Box^{k-1}C\big)
\delta\nabla^{\mu}\Box^{i-k}D
-\big(\nabla^{\mu}\Box^{k-1}C\big)
\delta\Box^{i-k}D \right.\nn \\
&&\left.+\frac{1}{2}g^{\rho\sigma}
\big(\Box^{k-1}C\big)\big(\nabla^{\mu}\Box^{i-k}D\big)
\delta{g}_{\rho\sigma}\right]
\, . \label{ThebarCDi}
\eea
Under the transformation $\delta\rightarrow\mathcal{L}_\zeta$,
the surface term $\bar{\Theta}^{\mu}_{\text{CD}(i)}$ is related to
the symmetric tensor $\bar{X}^{\mu\nu}_{\text{CD}(i)}$
in the following way
\be
\bar{\Theta}^\mu_{\text{CD}(i)}
(\delta\rightarrow\mathcal{L}_\zeta)=
2\zeta_\nu\bar{X}^{\mu\nu}_{\text{CD}(i)}
-\nabla_\nu\bar{K}^{\mu\nu}_{\text{CD}(i)}
\, , \label{RelBaThECDi}
\ee
where the anti-symmetric tensor $\bar{K}^{\mu\nu}_{\text{CD}(i)}$
has the form
\be
\bar{K}^{\mu\nu}_{\text{CD}(i)}=
2\sum^i_{k=1}\zeta^{[\mu}\Big(\nabla^{\nu]}\Box^{i-k}{D}\Big)
\Big(\Box^{k-1}{C}\Big)
\, . \label{KbarmnCDi}
\ee
Like before, by means of the computation on the surface term
$\Theta^\mu_{\text{CD}(i)}(\delta\rightarrow\mathcal{L}_\zeta)$,
we obtain the Noether potential
\be
K^{\mu\nu}_{\text{CD}(i)}=\bar{K}^{\mu\nu}_{\text{CD}(i)}
+2P^{\mu\nu\rho\sigma}_{\text{CD}(i)}\nabla_{\rho}\zeta_{\sigma}
+4\zeta_\rho\nabla_\sigma{P}^{\mu\nu\rho\sigma}_{\text{CD}(i)}
-6P^{\mu[\nu\rho\sigma]}_{\text{CD}(i)}\nabla_\rho\zeta_\sigma
\, , \label{KmnCBoxiD}
\ee
together with the following identity
\be
P^{\mu\nu}_{\text{CD}(i)}=
2P^{\mu\lambda\rho\sigma}_{\text{CD}(i)}
R^\nu_{~\lambda\rho\sigma}
\, , \label{PCD24Rel}
\ee
and the two ones
\be
P^{[\mu|\lambda\rho\sigma|}_{\text{CD}(i)}
R^{\nu]}_{~~\lambda\rho\sigma}=0 \, , \quad
\nabla_\rho\nabla_\sigma
P^{\rho[\mu\nu]\sigma}_{\text{CD}(i)}=0
\, . \label{PCD24Rel2}
\ee
Here the three identities take the same structures as those
corresponding to the Lagrangian
$\sqrt{-g}L(g^{\mu\nu},R_{\mu\nu\rho\sigma})$ given by the works
\cite{JJP2306,Pady}. As a matter of fact, when the integer $i=0$,
the scalar $C=1$ and $D=L(g^{\mu\nu},R_{\mu\nu\rho\sigma})$, the
Lagrangian (\ref{LagCBoxiD}) returns to the one
$\sqrt{-g}L(g^{\mu\nu},R_{\mu\nu\rho\sigma})$. Consequently,
both $E^{\mu\nu}_{\text{CD}(i)}$ and $K^{\mu\nu}_{\text{CD}(i)}$
become respectively to the expression for the field equations
and the Noether potential associated to the Lagrangian
$\sqrt{-g}L(g^{\mu\nu},R_{\mu\nu\rho\sigma})$.

Moreover, substituting the first equation within Eq. (\ref{PCD24Rel})
into Eq. (\ref{EoMCBoxiD}) to eliminate the rank-2 symmetric tensor
$P^{\mu\nu}_{\text{CD}(i)}$ in the latter, we get a simpler
expression for field equations without the term comprising
the derivative of the Lagrangian density with respect to
the metric, namely,
\bea
E^{\mu\nu}_{\text{CD}(i)}&=&
P^{\mu\lambda\rho\sigma}_{\text{CD}(i)}
R^\nu_{~\lambda\rho\sigma}-2\nabla_\rho\nabla_\sigma
P^{\rho\mu\nu\sigma}_{\text{CD}(i)}
-\sum^i_{k=1}\left(\nabla^{(\mu}\Box^{k-1}C\right)
\nabla^{\nu)}\Box^{i-k}D \nn \\
&&+\frac{1}{2}g^{\mu\nu}\sum^i_{k=1}\nabla_\lambda
\left[\left(\Box^{k-1}C\right)
\nabla^{\lambda}\Box^{i-k}D\right]
-\frac{1}{2}g^{\mu\nu}C\Box^{i}D
\, . \label{EoMCBoxiD2}
\eea
In particular, when $C=A$ and $D=B$, leading
to that the Lagrangian density $h_{(i)}$ coincides with
$f_{(i)}$, together with that $P^{\mu\nu\rho\sigma}_{\text{CD}(i)}
=g^{\mu[\rho}g^{\sigma]\nu}A_{(i)}$ and
$P^{\mu\nu}_{\text{CD}(i)}=2A_{(i)}R^{\mu\nu}$, Eqs. (\ref{EoMCBoxiD})
and (\ref{KmnCBoxiD}) with the rank-4 tensor
$P^{\mu\nu\rho\sigma}_{\text{CD}(i)}$ replaced by
the one $g^{\mu[\rho}g^{\sigma]\nu}A_{(i)}$ turn into
the expression $E^{\mu\nu}_{\text{AB}(i)}$ for equations
of motion and the Noether potential
$K^{\mu\nu}_{\text{AB}(i)}$ associated to the Lagrangian
$\sqrt{-g}f_{(i)}$, respectively.

Let us point out that all the above results related to the Lagrangian
(\ref{LagCBoxiD}) can be naturally extended to the Lagrangians
within which the two scalars $C$ and $D$ are allowed to depend
upon $\Box^m{R}_{\mu\nu\rho\sigma}$s $(m=1,2,\cdot\cdot\cdot)$
in addition to both the metric $g^{\mu\nu}$ and the Riemann tensor
${R}_{\mu\nu\rho\sigma}$. This is to be explicitly demonstrated
below. To avoid confusion, the Lagrangian (\ref{LagCBoxiD}) is
alternatively denoted by
\be
\sqrt{-g}\tilde{h}_{(i)}
=\sqrt{-g}C\Box^{i}D
\, . \label{TildLaghi}
\ee
However, here both the scalars $C$ and $D$ are supposed to have
the forms
\bea
C&=&C\left(g^{\mu\nu},
R_{\mu\nu\rho\sigma},\Box^m{R}_{\mu\nu\rho\sigma}\right)
\, , \nn \\
D&=&D\left(g^{\mu\nu},
R_{\mu\nu\rho\sigma},\Box^n{R}_{\mu\nu\rho\sigma}\right)
\, , \label{CDgendef}
\eea
respectively. For convenience, apart from the two tensors
$P^{\mu\nu}_{\text{CD}(i)}$
and $P^{\mu\nu\rho\sigma}_{\text{CD}(i)}$ in Eq. (\ref{PCD24def}),
we introduce two additional fourth-rank ones
${F}^{\mu\nu\rho\sigma}_{\text{CD}(i,m)}$ and
${Q}^{\mu\nu\rho\sigma}_{\text{CD}(i,n)}$, which
are defined respectively through
\be
{F}^{\mu\nu\rho\sigma}_{\text{CD}(i,m)}
=\frac{\partial{C}}{\partial\Box^m{R}_{\mu\nu\rho\sigma}}
\big(\Box^i{D}\big)\, , \qquad
{Q}^{\mu\nu\rho\sigma}_{\text{CD}(i,n)}
=\big(\Box^i{C}\big)
\frac{\partial{D}}{\partial\Box^n{R}_{\mu\nu\rho\sigma}}
\, . \label{FCDimdef}
\ee
Especially, both ${F}^{\mu\nu\rho\sigma}_{\text{CD}(i,m)}$
and ${Q}^{\mu\nu\rho\sigma}_{\text{CD}(i,n)}$ with
$m=0=n$ are utilized to represent
\be
{F}^{\mu\nu\rho\sigma}_{\text{CD}(i,0)}
=\frac{\partial{C}}{\partial{R}_{\mu\nu\rho\sigma}}
\big(\Box^i{D}\big)\, , \qquad
{Q}^{\mu\nu\rho\sigma}_{\text{CD}(i,0)}
=\big(\Box^i{C}\big)
\frac{\partial{D}}{\partial{R}_{\mu\nu\rho\sigma}}
\, , \label{FCDi0def}
\ee
respectively. In terms of them,
equation (\ref{SumOmegik2}) enables us to move $\Box^i$
off $\delta\Box^i{D}$ and then to write down
the variation of the Lagrangian (\ref{TildLaghi}) as
\bea
\frac{\delta\left(\sqrt{-g}\tilde{h}_{(i)}\right)}{\sqrt{-g}}
&=&\left(\frac{1}{2}\tilde{h}_{(i)}g^{\mu\nu}
-P^{\mu\nu}_{\text{CD}(i)}
-\bar{X}^{\mu\nu}_{\text{CD}(i)}\right)\delta{g}_{\mu\nu}
+P^{\mu\nu\rho\sigma}_{\text{CD}(i)}
\delta{R}_{\mu\nu\rho\sigma}\nn \\
&&+{F}^{\mu\nu\rho\sigma}_{\text{CD}(i,m)}
\delta\Box^m{R}_{\mu\nu\rho\sigma}
+{Q}^{\mu\nu\rho\sigma}_{\text{CD}(i,n)}
\delta\Box^n{R}_{\mu\nu\rho\sigma} \nn \\
&&+\nabla_\mu\bar{\Theta}^{\mu}_{\text{CD}(i)}
\, . \label{VaTildLagCDi}
\eea
By the aid of Eqs. (\ref{P4mndelRic}) and (\ref{QijBoxjRiem}),
the variation equation (\ref{VaTildLagCDi}) can be further
written as the conventional form
\be
\delta\left(\sqrt{-g}\tilde{h}_{(i)}\right)
=\sqrt{-g}\left(-\tilde{E}^{\mu\nu}_{\text{CD}(i,m,n)}\delta{g}_{\mu\nu}
+\nabla_\mu\tilde{\Theta}_{\text{CD}(i,m,n)}^\mu\right)
\, . \label{VaTildLagCDi2}
\ee
Here the surface term
$\tilde{\Theta}_{\text{CD}(i,m,n)}^\mu$ is given by
\bea
\tilde{\Theta}_{\text{CD}(i,m,n)}^\mu&=&
2P^{\mu\nu\rho\sigma}_{\text{CD}(i,m,n)}
\nabla_\sigma\delta g_{\rho\nu}
-2(\delta g_{\nu\rho})
\nabla_\sigma{P}^{\mu\nu\rho\sigma}_{\text{CD}(i,m,n)}
+\bar{\Theta}^{\mu}_{\text{CD}(i)} \nn \\
&&+\sum^m_{k=1}\Theta_{(m,k)}^\mu
\left(A\rightarrow{F}_{\text{CD}(i,m)},B\rightarrow{R}\right)
\nn \\
&&+\sum^n_{k=1}\Theta_{(n,k)}^\mu
\left(A\rightarrow{Q}_{\text{CD}(i,n)},B\rightarrow{R}\right)
\, , \label{TildTheCDi}
\eea
with $\Theta^\mu_{(m,k)}=\Theta^\mu_{(i,k)}\big|_{i=m}$,
$\Theta^\mu_{(n,k)}=\Theta^\mu_{(i,k)}\big|_{i=n}$, and
the fourth-rank tensor
${P}^{\mu\nu\rho\sigma}_{\text{CD}(i,m,n)}$
being of the form
\bea
{P}^{\mu\nu\rho\sigma}_{\text{CD}(i,m,n)}
&=&{F}^{\mu\nu\rho\sigma}_{\text{CD}(i,0)}
+{Q}^{\mu\nu\rho\sigma}_{\text{CD}(i,0)}
+\Box^m{F}^{\mu\nu\rho\sigma}_{\text{CD}(i,m)}
+\Box^n{Q}^{\mu\nu\rho\sigma}_{\text{CD}(i,n)}\nn \\
&=&P^{\mu\nu\rho\sigma}_{\text{CD}(i)}
+\Box^m{F}^{\mu\nu\rho\sigma}_{\text{CD}(i,m)}
+\Box^n{Q}^{\mu\nu\rho\sigma}_{\text{CD}(i,n)}
\, . \label{tildPCDimn}
\eea
Within Eq. (\ref{VaTildLagCDi2}), the expression for
field equations $\tilde{E}^{\mu\nu}_{\text{CD}(i,m,n)}$
is written as
\bea
\tilde{E}^{\mu\nu}_{\text{CD}(i,m,n)}&=&
\bar{X}^{\mu\nu}_{\text{CD}(i)}
+P^{\mu\nu}_{\text{CD}(i)}
-P^{\mu\lambda\rho\sigma}_{\text{CD}(i,m,n)}
R^\nu_{~\lambda\rho\sigma}-2\nabla_\rho\nabla_\sigma
P^{\rho\mu\nu\sigma}_{\text{CD}(i,m,n)}
-\frac{1}{2}g^{\mu\nu}\tilde{h}_{(i)} \nn \\
&&+{X}^{\mu\nu}_{\text{CD}(i,m,n)}
+2R^{\nu}_{~\lambda\rho\sigma}
\Box^{m}{F}^{\mu\lambda\rho\sigma}_{\text{CD}(i,m)}
-2{F}^{\mu\lambda\rho\sigma}_{\text{CD}(i,m)}
\Box^{m}R^{\nu}_{~\lambda\rho\sigma} \nn \\
&&+2R^{\nu}_{~\lambda\rho\sigma}
\Box^{n}{Q}^{\mu\lambda\rho\sigma}_{\text{CD}(i,n)}
-2{Q}^{\mu\lambda\rho\sigma}_{\text{CD}(i,n)}
\Box^{n}R^{\nu}_{~\lambda\rho\sigma}
\, , \label{TildECDimn}
\eea
with ${X}^{\mu\nu}_{\text{CD}(i,m,n)}$ presented by
\bea
{X}^{\mu\nu}_{\text{CD}(i,m,n)}&=&
\sum^m_{k=1}{X}^{\mu\nu}_{(m,k)}
\left(A\rightarrow{F}_{\text{CD}(i,m)},B\rightarrow{R}\right) \nn \\
&&+\sum^n_{k=1}{X}^{\mu\nu}_{(n,k)}
\left(A\rightarrow{Q}_{\text{CD}(i,n)},B\rightarrow{R}\right)
\, , \label{XCDimn}
\eea
in which ${X}^{\mu\nu}_{(m,k)}=
{X}^{\mu\nu}_{(i,k)}\big|_{i=m}$ and
${X}^{\mu\nu}_{(n,k)}=
{X}^{\mu\nu}_{(i,k)}\big|_{i=n}$, with
${X}^{\mu\nu}_{(i,k)}$ given by Eq. (\ref{Xmngenik})
or (\ref{Xmngenik2}). In particular, when $m=0=n$,
${X}^{\mu\nu}_{\text{CD}(i,m,n)}=0$. It is worth
mentioning that the tensor
${X}^{\mu\nu}_{\text{CD}(i,m,n)}$ can be also
determined by ${X}^{\mu\nu}_{\text{Riem}(i,k)}$ in
Eq. (\ref{XmnRiemik2}), that is,
\bea
{X}^{\mu\nu}_{\text{CD}(i,m,n)}&=&
\sum^m_{k=1}{X}^{\mu\nu}_{\text{Riem}(m,k)}
\left({P}_{(m)}
\rightarrow
{F}_{\text{CD}(i,m)}\right) \nn \\
&&+\sum^n_{k=1}{X}^{\mu\nu}_{\text{Riem}(n,k)}
\left({P}_{(n)}
\rightarrow
{Q}_{\text{CD}(i,n)}\right)
\, . \label{XCDimn2}
\eea
Employing Eq. (\ref{SumXikdef}) to compute
${X}^{[\mu\nu]}_{\text{CD}(i,m,n)}$ gives rise to
\bea
{X}^{[\mu\nu]}_{\text{CD}(i,m,n)}&=&
2R^{[\mu}_{~~\lambda\rho\sigma}
\Box^{m}{F}^{\nu]\lambda\rho\sigma}_{\text{CD}(i,m)}
-2\left(\Box^{m}R^{[\mu}_{~~\lambda\rho\sigma}\right)
{F}^{\nu]\lambda\rho\sigma}_{\text{CD}(i,m)} \nn \\
&&+2R^{[\mu}_{~~\lambda\rho\sigma}
\Box^{n}{Q}^{\nu]\lambda\rho\sigma}_{\text{CD}(i,n)}
-2\left(\Box^{n}R^{[\mu}_{~~\lambda\rho\sigma}\right)
{Q}^{\nu]\lambda\rho\sigma}_{\text{CD}(i,n)}
\, . \label{AnSyXCDimn}
\eea
Furthermore, according to Eqs. (\ref{ThetRiem0Lie}),
(\ref{TheikdelLie}), and (\ref{RelBaThECDi}), after the variation
operator $\delta$ in $\tilde{\Theta}_{\text{CD}(i,m,n)}^\mu$ is replaced
with the Lie derivative $\mathcal{L}_\zeta$, one obtains
\be
\tilde{\Theta}_{\text{CD}(i,m,n)}^\mu
(\delta\rightarrow\mathcal{L}_\zeta)=
2\zeta_\nu\left(\tilde{E}^{\mu\nu}_{\text{CD}(i,m,n)}
+\frac{1}{2}g^{\mu\nu}\tilde{h}_{(i)}\right)
-\nabla_\nu\tilde{K}^{\mu\nu}_{\text{CD}(i,m,n)}
\, . \label{TildTheCDiLie}
\ee
In the above equation, the expression
$\tilde{E}^{\mu\nu}_{\text{CD}(i,m,n)}$ for equations of
motion is alternatively written as a much simpler
form
\bea
\tilde{E}^{\mu\nu}_{\text{CD}(i,m,n)}&=&
P^{\mu\lambda\rho\sigma}_{\text{CD}(i,m,n)}
R^\nu_{~\lambda\rho\sigma}-2\nabla_\rho\nabla_\sigma
P^{\rho\mu\nu\sigma}_{\text{CD}(i,m,n)} \nn \\
&&+\bar{X}^{\mu\nu}_{\text{CD}(i)}
+{X}^{\mu\nu}_{\text{CD}(i,m,n)}
-\frac{1}{2}g^{\mu\nu}\tilde{h}_{(i)}
\, , \label{TildECDimn2}
\eea
with $\nabla_\mu\tilde{E}^{\mu\nu}_{\text{CD}(i,m,n)}=0$
proved in Appendix \ref{appendC},
and the anti-symmetric tensor
$\tilde{K}^{\mu\nu}_{\text{CD}(i,m,n)}$ represents the Noether
potential associated to the Lagrangian (\ref{TildLaghi}),
which is expressed as
\bea
\tilde{K}^{\mu\nu}_{\text{CD}(i,m,n)}
&=&
2P^{\mu\nu\rho\sigma}_{\text{CD}(i,m,n)}\nabla_{\rho}\zeta_{\sigma}
+4\zeta_\rho\nabla_\sigma{P}^{\mu\nu\rho\sigma}_{\text{CD}(i,m,n)}
-6P^{\mu[\nu\rho\sigma]}_{\text{CD}(i,m,n)}\nabla_\rho\zeta_\sigma
\nn \\
&&+\bar{K}^{\mu\nu}_{\text{CD}(i)}
+\sum^m_{k=1}{K}^{\mu\nu}_{\text{Riem}(m,k)}
\left({P}_{(m)}
\rightarrow
{F}_{\text{CD}(i,m)}\right) \nn \\
&&+\sum^n_{k=1}{K}^{\mu\nu}_{\text{Riem}(n,k)}
\left({P}_{(n)}
\rightarrow
{Q}_{\text{CD}(i,n)}\right)
\, . \label{KtildmnCDi}
\eea
Here
${K}^{\mu\nu}_{\text{Riem}(m,k)}=
{K}^{\mu\nu}_{\text{Riem}(i,k)}\big|_{i=m}$ and
${K}^{\mu\nu}_{\text{Riem}(n,k)}=
{K}^{\mu\nu}_{\text{Riem}(i,k)}\big|_{i=n}$ with
the anti-symmetric tensor ${K}^{\mu\nu}_{\text{Riem}(i,k)}$
given by Eq. (\ref{KmnRiemik2}). In accordance with
Eq. (\ref{TildTheCDiLie}), the conserved current corresponding to
an arbitrary vector reads
\be
\tilde{J}^\mu_{\text{CD}(i,m,n)}=
2\zeta_\nu\tilde{E}^{\mu\nu}_{\text{CD}(i,m,n)}
+\zeta^{\mu}\tilde{h}_{(i)}
-\tilde{\Theta}_{\text{CD}(i,m,n)}^\mu
(\delta\rightarrow\mathcal{L}_\zeta)
=\nabla_\nu\tilde{K}^{\mu\nu}_{\text{CD}(i,m,n)}
\, . \label{CCurrtildhi}
\ee
By means of the comparison between Eqs. (\ref{TildECDimn})
and (\ref{TildECDimn2}), one obtains an identity
for the second-rank symmetric tensor
$P^{\mu\nu}_{\text{CD}(i)}=P^{(\mu\nu)}_{\text{CD}(i)}$,
\be
P^{\mu\nu}_{\text{CD}(i)}=
2P^{\mu\lambda\rho\sigma}_{\text{CD}(i)}
R^\nu_{~\lambda\rho\sigma}
+2{F}^{\mu\lambda\rho\sigma}_{\text{CD}(i,m)}
\Box^{m}R^{\nu}_{~\lambda\rho\sigma}
+2{Q}^{\mu\lambda\rho\sigma}_{\text{CD}(i,n)}
\Box^{n}R^{\nu}_{~\lambda\rho\sigma}
\, . \label{IdPQmnCDi}
\ee
Apart from this, there exists another identity
${X}^{[\mu\nu]}_{\text{CD}(i,m,n)}=
-2P^{[\mu|\lambda\rho\sigma|}_{\text{CD}(i,m,n)}
R^{\nu]}_{~~\lambda\rho\sigma}$, or equivalently,
\be
P^{[\mu|\lambda\rho\sigma|}_{\text{CD}(i)}
R^{\nu]}_{~~\lambda\rho\sigma}=
-{F}^{\mu|\lambda\rho\sigma|}_{\text{CD}(i,m)}
\Box^{m}R^{\nu]}_{~~\lambda\rho\sigma}
-{Q}^{[\mu|\lambda\rho\sigma|}_{\text{CD}(i,n)}
\Box^{n}R^{\nu]}_{~~\lambda\rho\sigma}
\, , \label{IDfromAnSP}
\ee
arising from $P^{[\mu\nu]}_{\text{CD}(i)}=0$.

As a simple example to check the expression
$\tilde{E}^{\mu\nu}_{\text{CD}(i,m,n)}$
for field equations, we consider the situation in which
$C=\text{Const}$ and $i\neq0$. In such a case,
within Eq. (\ref{TildECDimn2}), except for that
$\bar{X}^{\mu\nu}_{\text{CD}(i)}=g^{\mu\nu}\tilde{h}_{(i)}/2$,
both the quantities
$P^{\mu\nu\rho\sigma}_{\text{CD}(i,m,n)}$ and
${X}^{\mu\nu}_{\text{CD}(i,m,n)}$ disappear. As a result,
$\tilde{E}^{\mu\nu}_{\text{CD}(i,m,n)}=0$. This is in accordance
with the fact that the Lagrangian density
$\tilde{h}_{(i)}=\nabla_\mu\big(C\nabla^\mu\Box^{i-1}D\big)$
is a total derivative term. 
In addition, when both the scalars $C$ and $D$ have
the following forms
\be
C=C\left(g^{\mu\nu},\Box^m{R}_{\mu\nu\rho\sigma}\right)
\, , \qquad
D=D\left(g^{\mu\nu},\Box^n{R}_{\mu\nu\rho\sigma}\right)
\, , \label{CDgendef2}
\ee
respectively,
the fourth-rank tensor $P^{\mu\nu\rho\sigma}_{\text{CD}(i,m,n)}$
in the surface term $\tilde{\Theta}_{\text{CD}(i,m,n)}^\mu$,
the expression for field equations $\tilde{E}^{\mu\nu}_{\text{CD}(i,m,n)}$
and the Noether potential $\tilde{K}^{\mu\nu}_{\text{CD}(i,m,n)}$,
given by Eqs. (\ref{TildTheCDi}), (\ref{TildECDimn2}) and
(\ref{KtildmnCDi}), respectively, has to be replaced with the one
\be
\mathcal{P}^{\mu\nu\rho\sigma}_{\text{CD}(i,m,n)}=
\Box^m{F}^{\mu\nu\rho\sigma}_{\text{CD}(i,m)}
+\Box^n{Q}^{\mu\nu\rho\sigma}_{\text{CD}(i,n)}
\, . \label{MatPCDimn}
\ee

Next, as a generalization of the Lagrangian (\ref{LagCBoxiD}),
we proceed to take into account the situation involving
the Lagrangian that consists of two rank-$n$ tensors instead
of two scalars, which takes the following form
\be
\sqrt{-g}\hat{h}_{(i)}=\sqrt{-g}
\hat{C}^{\alpha_1\cdot\cdot\cdot\alpha_n}
\Box^{i}\hat{D}_{\alpha_1\cdot\cdot\cdot\alpha_n}
\, , \label{Laghathi}
\ee
in which both the tensors
$\hat{C}^{\alpha_1\cdot\cdot\cdot\alpha_n}$
and
$\hat{D}_{\alpha_1\cdot\cdot\cdot\alpha_n}$
are restricted to
\bea
\hat{C}^{\alpha_1\cdot\cdot\cdot\alpha_n}&=&
\hat{C}^{\alpha_1\cdot\cdot\cdot\alpha_n}
\big(g^{\mu\nu},\Box^a{R}_{\mu\nu\rho\sigma}\big)
\, , \nn \\
\hat{D}_{\alpha_1\cdot\cdot\cdot\alpha_n}&=&
\hat{D}_{\alpha_1\cdot\cdot\cdot\alpha_n}
\big(g^{\mu\nu},\Box^b{R}_{\mu\nu\rho\sigma}\big)
\, , \label{hatCDvariab}
\eea
with $a$ and $b$ standing for two arbitrary
nonnegative integers. Particularly, when $a=0=b$,
it indicates that both the tensors
$\hat{C}^{\alpha_1\cdot\cdot\cdot\alpha_n}$ and
$\hat{D}_{\alpha_1\cdot\cdot\cdot\alpha_n}$ merely
depend on the metric and the Riemann tensor.
Varying the Lagrangian
(\ref{Laghathi}) gives rise to
\be
\delta\big(\sqrt{-g}\hat{h}_{(i)}\big)
=\sqrt{-g}\Big({g}_{\mu\rho}{g}_{\nu\sigma}
\hat{E}^{\mu\nu}_{\text{CD}(i,a,b)}\delta{g}^{\rho\sigma}
+\nabla_\mu\hat{\Theta}_{\text{CD}(i,a,b)}^\mu\Big)
\, . \label{VaLahatCBiD}
\ee
Here the expression $\hat{E}^{\mu\nu}_{\text{CD}(i,a,b)}$
for field equations will be determined below.
According to Eq. (\ref{ThetaABi}), the surface term
$\hat{\Theta}_{\text{CD}(i,a,b)}^\mu$ is given by
\bea
\hat{\Theta}_{\text{CD}(i,a,b)}^\mu&=&
2\hat{P}^{\mu\nu\rho\sigma}_{\text{CD}(i,a,b)}
\nabla_\sigma\delta g_{\rho\nu}
-2(\delta g_{\nu\rho})\nabla_\sigma
\hat{P}^{\mu\nu\rho\sigma}_{\text{CD}(i,a,b)}\nn \\
&&+\hat{\bar{\Theta}}_{\text{CD}(i)}^\mu
+\hat{\Theta}_{\text{CDF}(i,a)}^\mu
+\hat{\Theta}_{\text{CDQ}(i,b)}^\mu
\, , \label{hatThetaCD}
\eea
where the three surface terms
$\hat{\bar{\Theta}}_{\text{CD}(i)}^\mu$,
$\hat{\Theta}_{\text{CDF}(i,a)}^\mu$,
and $\hat{\Theta}_{\text{CDQ}(i,b)}^\mu$ can be defined
in terms of $\Theta_{(i,k)}^\mu$ presented by
Eq. (\ref{Thegenik2}). Specifically, the surface term
$\hat{\bar{\Theta}}_{\text{CD}(i)}^\mu$ is written as
\be
\hat{\bar{\Theta}}_{\text{CD}(i)}^\mu=
\sum^i_{k=1}\Theta_{(i,k)}^\mu\left(A\rightarrow\hat{C},
B\rightarrow\hat{D}\right)
\, , \label{HatBarThCDi}
\ee
the one $\hat{\Theta}_{\text{CDF}(i,a)}^\mu$ is given by
\bea
\hat{\Theta}^\mu_{\text{CDF}(i,a)}&=&
\sum^a_{k=1}\Theta_{(a,k)}^\mu
\left(A^{\alpha_1\cdot\cdot\cdot\alpha_n}\rightarrow
\hat{F}^{\gamma\lambda\rho\sigma}_{\text{CD}(i,a)},
B_{\alpha_1\cdot\cdot\cdot\alpha_n}
\rightarrow{R}_{\gamma\lambda\rho\sigma}\right)\nn \\
&=&\sum^a_{k=1}\Theta^\mu_{\text{Riem}(a,k)}
\left({P}_{(a)}\rightarrow
\hat{F}_{\text{CD}(i,a)}\right)
\, , \label{HatTheCDFia}
\eea
and $\hat{\Theta}_{\text{CDQ}(i,b)}^\mu$ takes the similar
form as $\hat{\Theta}_{\text{CDF}(i,a)}^\mu$, namely,
\bea
\hat{\Theta}^\mu_{\text{CDQ}(i,b)}&=&
\sum^b_{k=1}\Theta_{(b,k)}^\mu
\left(A^{\alpha_1\cdot\cdot\cdot\alpha_n}\rightarrow
\hat{Q}^{\gamma\lambda\rho\sigma}_{\text{CD}(i,b)},
B_{\alpha_1\cdot\cdot\cdot\alpha_n}
\rightarrow{R}_{\gamma\lambda\rho\sigma}\right) \nn \\
&=&\sum^b_{k=1}\Theta^\mu_{\text{Riem}(b,k)}
\left({P}_{(b)}\rightarrow
\hat{Q}_{\text{CD}(i,b)}\right)
\, . \label{HatTheCDQib}
\eea
In Eqs. (\ref{HatTheCDFia}) and (\ref{HatTheCDQib}),
the surface term $\Theta^\mu_{\text{Riem}(i,k)}$
is presented by Eq. (\ref{TheRiemik}).
The fourth-rank tensor
$\hat{P}^{\mu\nu\rho\sigma}_{\text{CD}(i,a,b)}$
in Eq. (\ref{hatThetaCD}) is defined through
\be
\hat{P}^{\mu\nu\rho\sigma}_{\text{CD}(i,a,b)}=
\Box^a\hat{F}^{\mu\nu\rho\sigma}_{\text{CD}(i,a)}
+\Box^b\hat{Q}^{\mu\nu\rho\sigma}_{\text{CD}(i,b)}
\, , \label{hatPCD4def}
\ee
with $\hat{F}^{\mu\nu\rho\sigma}_{\text{CD}(i,a)}$ and
$\hat{Q}^{\mu\nu\rho\sigma}_{\text{CD}(i,b)}$ given
respectively by
\bea
\hat{F}^{\mu\nu\rho\sigma}_{\text{CD}(i,a)}
&=&\frac{\partial\hat{C}^{\alpha_1\cdot\cdot\cdot\alpha_n}}{
\partial\Box^a{R}_{\mu\nu\rho\sigma}}
\big(\Box^i\hat{D}_{\alpha_1\cdot\cdot\cdot\alpha_n}\big)
=\frac{\partial\hat{C}_{\alpha_1\cdot\cdot\cdot\alpha_n}}{
\partial\Box^a{R}_{\mu\nu\rho\sigma}}
\big(\Box^i\hat{D}^{\alpha_1\cdot\cdot\cdot\alpha_n}\big)
\, , \nn \\
\hat{Q}^{\mu\nu\rho\sigma}_{\text{CD}(i,b)}
&=&\frac{\partial\hat{D}_{\alpha_1\cdot\cdot\cdot\alpha_n}}{
\partial\Box^b{R}_{\mu\nu\rho\sigma}}
\big(\Box^i\hat{C}^{\alpha_1\cdot\cdot\cdot\alpha_n}\big)
=\frac{\partial\hat{D}^{\alpha_1\cdot\cdot\cdot\alpha_n}}{
\partial\Box^b{R}_{\mu\nu\rho\sigma}}
\big(\Box^i\hat{C}_{\alpha_1\cdot\cdot\cdot\alpha_n}\big)
\, . \label{haFQCDimdef}
\eea

With the surface term (\ref{hatThetaCD}) in hand,
we switch to figure out the expression
$\hat{E}^{\mu\nu}_{\text{CD}(i,a,b)}$ for field equations and the
Noether potential associated to the Lagrangian (\ref{Laghathi}).
In order to achieve this, by the aid of Eq. (\ref{TheABiLie}),
we follow the method based
on the conserved current to deal with the surface term
$\hat{\Theta}_{\text{CD}(i,a,b)}^\mu$ under the condition that the
variation operator $\delta$ in it is transformed into the
Lie derivative $\mathcal{L}_\zeta$ along an arbitrary vector
$\zeta^\mu$. After some manipulations to the three quantities
$\hat{\bar{\Theta}}_{\text{CD}(i)}^\mu$,
$\hat{\Theta}_{\text{CDF}(i,a)}^\mu$ and
$\hat{\Theta}_{\text{CDQ}(i,b)}^\mu$, we obtain
\bea
\hat{\bar{\Theta}}_{\text{CD}(i)}^\mu
(\delta\rightarrow\mathcal{L}_\zeta)&=&
2\zeta_\nu\hat{\bar{X}}^{\mu\nu}_{\text{CD}(i)}
-\nabla_\nu\hat{\bar{K}}^{\mu\nu}_{\text{CD}(i)}
\, , \label{HatTheCDLie}
\eea
together with
\bea
\hat{\Theta}_{\text{CDF}(i,a)}^\mu
(\delta\rightarrow\mathcal{L}_\zeta)&=&
2\zeta_\nu\hat{X}^{\mu\nu}_{\text{CDF}(i,a)}
-\nabla_\nu\hat{K}^{\mu\nu}_{\text{CDF}(i,a)} \, , \nn \\
\hat{\Theta}_{\text{CDQ}(i,b)}^\mu
(\delta\rightarrow\mathcal{L}_\zeta)&=&
2\zeta_\nu\hat{X}^{\mu\nu}_{\text{CDQ}(i,b)}
-\nabla_\nu\hat{K}^{\mu\nu}_{\text{CDQ}(i,b)}
\, . \label{HatTheCDLie2}
\eea
In the above equations, the tensor
$\hat{\bar{X}}^{\mu\nu}_{\text{CD}(i)}$
is expressed as
\be
\hat{\bar{X}}^{\mu\nu}_{\text{CD}(i)}=
\sum^i_{k=1}{X}^{\mu\nu}_{(i,k)}
\left(A\rightarrow\hat{C},B\rightarrow\hat{D}\right)
\, , \label{HatBarXCDi}
\ee
while both the tensors $\hat{X}^{\mu\nu}_{\text{CDF}(i,a)}$
and $\hat{X}^{\mu\nu}_{\text{CDQ}(i,b)}$ are defined respectively
through
\bea
\hat{X}^{\mu\nu}_{\text{CDF}(i,a)}&=&
\sum^a_{k=1}{X}^{\mu\nu}_{(a,k)}
\left(A^{\alpha_1\cdot\cdot\cdot\alpha_n}\rightarrow
\hat{F}^{\gamma\lambda\rho\sigma}_{\text{CD}(i,a)},
B_{\alpha_1\cdot\cdot\cdot\alpha_n}
\rightarrow{R}_{\gamma\lambda\rho\sigma}\right) \nn \\
&=&
\sum^a_{k=1}{X}^{\mu\nu}_{\text{Riem}(a,k)}
\left({P}_{(a)}\rightarrow
\hat{F}_{\text{CD}(i,a)}\right) \, , \nn \\
\hat{X}^{\mu\nu}_{\text{CDQ}(i,b)}
&=&\sum^b_{k=1}{X}^{\mu\nu}_{(b,k)}
\left(A^{\alpha_1\cdot\cdot\cdot\alpha_n}\rightarrow
\hat{Q}^{\gamma\lambda\rho\sigma}_{\text{CD}(i,b)},
B_{\alpha_1\cdot\cdot\cdot\alpha_n}
\rightarrow{R}_{\gamma\lambda\rho\sigma}\right) \nn \\
&=&\sum^b_{k=1}{X}^{\mu\nu}_{\text{Riem}(b,k)}
\left({P}_{(b)}\rightarrow
\hat{Q}_{\text{CD}(i,b)}\right)
\, . \label{hatXCDimn}
\eea
Within Eq. (\ref{hatXCDimn}), the tensors 
${X}^{\mu\nu}_{\text{Riem}(a,k)}$ and
${X}^{\mu\nu}_{\text{Riem}(b,k)}$ are given by
Eq. (\ref{XmnRiemik2}).
For convenience to compute the field
equations, by the aid of Eqs. (\ref{Xmnikgensym})
and (\ref{Xmnikgenansym}), the tensor
$\hat{\bar{X}}^{\mu\nu}_{\text{CD}(i)}$ is explicitly
expressed as
\bea
\hat{\bar{X}}^{\mu\nu}_{\text{CD}(i)}&=&
\frac{1}{2}g^{\mu\nu}\sum^i_{k=1}\nabla_\lambda
\left[\big(\Box^{k-1}
\hat{C}^{\alpha_1\cdot\cdot\cdot\alpha_n}\big)
\nabla^{\lambda}\Box^{i-k}
\hat{D}_{\alpha_1\cdot\cdot\cdot\alpha_n}\right] \nn \\
&&-\sum^i_{k=1}\left(\nabla^{(\mu}\Box^{k-1}
\hat{C}^{|\alpha_1\cdot\cdot\cdot\alpha_n|}\right)
\nabla^{\nu)}\Box^{i-k}
\hat{D}_{\alpha_1\cdot\cdot\cdot\alpha_n}\nn \\
&&+\frac{1}{2}\sum^i_{k=1}\nabla_\lambda\left(
\hat{H}^{(\mu\nu)\lambda}_{\text{CD}(i,k)}
-\hat{H}^{\lambda(\mu\nu)}_{\text{CD}(i,k)}
+\hat{H}^{[\mu|\lambda|\nu]}_{\text{CD}(i,k)}\right)
\, , \label{HatBarXCDi2}
\eea
where the third-rank tensor
$\hat{H}^{\lambda\mu\nu}_{\text{CD}(i,k)}
={H}^{\lambda\mu\nu}_{(i,k)}\big|_{
A\rightarrow\hat{C},B\rightarrow\hat{D}}$, 
with ${H}^{\lambda\mu\nu}_{(i,k)}$ given by
Eq. (\ref{Hikdef}),
takes the concrete form
\bea
\hat{H}^{\lambda\mu\nu}_{\text{CD}(i,k)}
&=&g^{\lambda\rho}
\sum^n_{j=1}\left(\nabla^{\mu}\Box^{k-1}
\hat{C}^{\alpha_1\cdot\cdot\cdot\alpha_{j-1}\nu
\alpha_{j+1}\cdot\cdot\cdot\alpha_n}\right)
\left(\Box^{i-k}
\hat{D}_{\alpha_1\cdot\cdot\cdot\alpha_{j-1}\rho
\alpha_{j+1}\cdot\cdot\cdot\alpha_n}\right) \nn \\
&&-g^{\lambda\rho}
\sum^n_{j=1}\left(\Box^{k-1}
\hat{C}^{\alpha_1\cdot\cdot\cdot\alpha_{j-1}\nu
\alpha_{j+1}\cdot\cdot\cdot\alpha_n}\right)
\nabla^{\mu}\Box^{i-k}
\hat{D}_{\alpha_1\cdot\cdot\cdot\alpha_{j-1}\rho
\alpha_{j+1}\cdot\cdot\cdot\alpha_n}
\, . \label{HikCDdef}
\eea
In particular, when both
$\hat{C}^{\alpha_1\cdot\cdot\cdot\alpha_n}$ and
$\hat{D}_{\alpha_1\cdot\cdot\cdot\alpha_n}$ are
scalars, the tensor
$\hat{H}^{\lambda\mu\nu}_{\text{CD}(i,k)}$
disappears. As a consequence of Eq. (\ref{HatBarXCDi2}), 
we have
\bea
\hat{X}^{\mu\nu}_{\text{CDF}(i,a)}&=&
\hat{\bar{X}}^{\mu\nu}_{\text{CD}(a)}
\left(\hat{C}^{\alpha_1\cdot\cdot\cdot\alpha_n}\rightarrow
\hat{F}^{\gamma\lambda\rho\sigma}_{\text{CD}(i,a)},
\hat{D}_{\alpha_1\cdot\cdot\cdot\alpha_n}
\rightarrow{R}_{\gamma\lambda\rho\sigma}\right) \, ,\nn \\
\hat{X}^{\mu\nu}_{\text{CDQ}(i,b)}
&=&\hat{\bar{X}}^{\mu\nu}_{\text{CD}(b)}
\left(\hat{C}^{\alpha_1\cdot\cdot\cdot\alpha_n}\rightarrow
\hat{Q}^{\gamma\lambda\rho\sigma}_{\text{CD}(i,b)},
\hat{D}_{\alpha_1\cdot\cdot\cdot\alpha_n}
\rightarrow{R}_{\gamma\lambda\rho\sigma}\right)
\, . \label{hatXCDimn2}
\eea
Additionally, within Eq. (\ref{HatTheCDLie}),
the anti-symmetric tensor $\hat{\bar{K}}^{\mu\nu}_{\text{CD}(i)}$
is of the form
\be
\hat{\bar{K}}^{\mu\nu}_{\text{CD}(i)}=
\sum^i_{k=1}{K}^{\mu\nu}_{(i,k)}
\left(A\rightarrow\hat{C},B\rightarrow\hat{D}\right)
\, , \label{HatBarKmmCDi}
\ee
with ${K}^{\mu\nu}_{(i,k)}$ given by Eq. (\ref{Kmngenik})
or (\ref{Kmngenik2}), and the other two anti-symmetric tensors
$\hat{K}^{\mu\nu}_{\text{CDF}(i,a)}$ and
$\hat{K}^{\mu\nu}_{\text{CDQ}(i,b)}$ are expressed respectively
as
\bea
\hat{K}^{\mu\nu}_{\text{CDF}(i,a)}&=&
\sum^a_{k=1}{K}^{\mu\nu}_{\text{Riem}(a,k)}
\left({P}_{(a)}\rightarrow
\hat{F}_{\text{CD}(i,a)}\right) \, , \nn \\
\hat{K}^{\mu\nu}_{\text{CDQ}(i,b)}&=&
\sum^b_{k=1}{K}^{\mu\nu}_{\text{Riem}(b,k)}
\left({P}_{(b)}\rightarrow
\hat{Q}_{\text{CD}(i,b)}\right)
\, , \label{HatKCDFQ}
\eea
where the two anti-symmetric tensors 
${K}^{\mu\nu}_{\text{Riem}(a,k)}$ and
${K}^{\mu\nu}_{\text{Riem}(b,k)}$ are
presented by Eq. (\ref{KmnRiemik2}). 

As a consequence of
Eqs. (\ref{HatTheCDLie}) and (\ref{HatTheCDLie2}), the expression
$\hat{E}^{\mu\nu}_{\text{CD}(i,a,b)}$ for equations
of motion in Eq. (\ref{VaLahatCBiD}) is read off as
\bea
\hat{E}^{\mu\nu}_{\text{CD}(i,a,b)}&=&
\hat{P}^{\mu\lambda\rho\sigma}_{\text{CD}(i,a,b)}
R^\nu_{~\lambda\rho\sigma}-2\nabla_\rho\nabla_\sigma
\hat{P}^{\rho\mu\nu\sigma}_{\text{CD}(i,a,b)}
-\frac{1}{2}{g}^{\mu\nu}
\hat{C}^{\alpha_1\cdot\cdot\cdot\alpha_n}
\Box^{i}\hat{D}_{\alpha_1\cdot\cdot\cdot\alpha_n} \nn \\
&&+\hat{\bar{X}}^{\mu\nu}_{\text{CD}(i)}
+\hat{X}^{\mu\nu}_{\text{CDF}(i,a)}
+\hat{X}^{\mu\nu}_{\text{CDQ}(i,b)}
\, . \label{EomhatCDhi}
\eea
Specially, when $a=0=b$, $\hat{X}^{\mu\nu}_{\text{CDF}(i,0)}
=0=\hat{X}^{\mu\nu}_{\text{CDQ}(i,0)}$ and the expression
$\hat{E}^{\mu\nu}_{\text{CD}(i,a,b)}$ turns into
\bea
\hat{E}^{\mu\nu}_{\text{CD}(i,0,0)}&=&
\hat{F}^{\mu\lambda\rho\sigma}_{\text{CD}(i,0)}
R^\nu_{~\lambda\rho\sigma}
-2\nabla_\rho\nabla_\sigma
\hat{F}^{\rho\mu\nu\sigma}_{\text{CD}(i,0)}
+\hat{Q}^{\mu\lambda\rho\sigma}_{\text{CD}(i,0)}
R^\nu_{~\lambda\rho\sigma}
-2\nabla_\rho\nabla_\sigma
\hat{Q}^{\rho\mu\nu\sigma}_{\text{CD}(i,0)} \nn \\
&&+\hat{\bar{X}}^{\mu\nu}_{\text{CD}(i)}
-\frac{1}{2}{g}^{\mu\nu}
\hat{C}^{\alpha_1\cdot\cdot\cdot\alpha_n}
\Box^{i}\hat{D}_{\alpha_1\cdot\cdot\cdot\alpha_n}
\, . \label{EomhatCDhi0ab}
\eea
Here the expression $\hat{E}^{\mu\nu}_{\text{CD}(i,0,0)}$
can be utilized to provide a practical way to verify equations of motion
for the nonlocal theories of gravity appearing in \cite{BCKM14}.
In the mean time, on the basis of Eq. (\ref{TheABiLie}), the Noether
potential $\hat{K}^{\mu\nu}_{\text{CD}(i,a,b)}$ corresponding to any
smooth vector $\zeta^\mu$ for the Lagrangian (\ref{Laghathi})
is presented by
\bea
\hat{K}^{\mu\nu}_{\text{CD}(i,a,b)}&=&
2\hat{P}^{\mu\nu\rho\sigma}_{\text{CD}(i,a,b)}\nabla_{\rho}\zeta_{\sigma}
+4\zeta_\rho\nabla_\sigma\hat{P}^{\mu\nu\rho\sigma}_{\text{CD}(i,a,b)}
-6\hat{P}^{\mu[\nu\rho\sigma]}_{\text{CD}(i,a,b)}\nabla_\rho\zeta_\sigma
\nn \\
&&+\hat{\bar{K}}^{\mu\nu}_{\text{CD}(i)}
+\hat{K}^{\mu\nu}_{\text{CDF}(i,a)}
+\hat{K}^{\mu\nu}_{\text{CDQ}(i,b)}
\, . \label{hatKmmCDi}
\eea
From the above equation, the off-shell Noether current
$\hat{J}^{\mu}_{\text{CD}(i,a,b)}$ corresponding to the
Noether potential
$\hat{K}^{\mu\nu}_{\text{CD}(i,a,b)}$ is read off as
\bea
\hat{J}^{\mu}_{\text{CD}(i,a,b)}
&=&2\zeta_\nu\hat{E}^{\mu\nu}_{\text{CD}(i,a,b)}
+\zeta^\mu\hat{h}_{(i)}
-\hat{\Theta}_{\text{CD}(i,a,b)}^\mu
(\delta\rightarrow\mathcal{L}_\zeta) \nn \\
&=&\nabla_\nu\hat{K}^{\mu\nu}_{\text{CD}(i,a,b)}
\, . \label{CCurLaghathi}
\eea
What is more, according to $\hat{E}^{\mu\nu}_{\text{CD}(i,a,b)}
=\hat{E}^{\nu\mu}_{\text{CD}(i,a,b)}$, one gets the following identity
\be
2\hat{P}^{[\mu|\lambda\rho\sigma|}_{\text{CD}(i,a,b)}
R^{\nu]}_{~~\lambda\rho\sigma}
=-\hat{\bar{X}}^{[\mu\nu]}_{\text{CD}(i)}
-\hat{X}^{[\mu\nu]}_{\text{CDF}(i,a)}
-\hat{X}^{[\mu\nu]}_{\text{CDQ}(i,b)}
\, . \label{IdASymHaECD}
\ee
After utilizing Eq. (\ref{SumXikdef}) to simplify
the identity (\ref{IdASymHaECD}), one arrives at
\be
\hat{\bar{X}}^{[\mu\nu]}_{\text{CD}(i)}=
2\left(\Box^a{R}^{[\mu}_{~~\lambda\rho\sigma}\right)
\hat{F}^{\nu]\lambda\rho\sigma}_{\text{CD}(i,a)}
+2\left(\Box^b{R}^{[\mu}_{~~\lambda\rho\sigma}\right)
\hat{Q}^{\nu]\lambda\rho\sigma}_{\text{CD}(i,b)}
=\frac{1}{2}\sum^i_{k=1}\nabla_\lambda
\hat{H}^{[\mu|\lambda|\nu]}_{\text{CD}(i,k)}
\, . \label{IdASymHaECD2}
\ee

In particular, when both the tensors
$\big(\hat{C}^{\alpha_1\cdot\cdot\cdot\alpha_n},
\hat{D}_{\alpha_1\cdot\cdot\cdot\alpha_n}\big)$ take the values
$\big(R^{\mu\nu\rho\sigma},R_{\mu\nu\rho\sigma}\big)$ and the
integer $i=n$, the Lagrangian (\ref{Laghathi}) reduces to the one
$\sqrt{-g}L_{\text{Riem1}}$ given by Eq. (\ref{LagRiemBnRiem}),
one is able to verify that both the quantities
$\hat{K}^{\mu\nu}_{\text{CD}(i,a,b)}$ and
$\hat{E}^{\mu\nu}_{\text{CD}(i,a,b)}$ are in agreement with the Noether
potential $K^{\mu\nu}_{\text{Riem1}}$ in Eq. (\ref{KmnRieBnRiem})
and the expression $E^{\mu\nu}_{\text{Riem1}}$ for equations of motion
in Eq. (\ref{EoMLagRemBnRem}), respectively. Furthermore, by
utilizing Eqs. (\ref{IdeRiemHik2}) and (\ref{DivtildXBi}), the
divergence for $\hat{E}^{\mu\nu}_{\text{CD}(i,a,b)}$ reads
\bea
\nabla_\mu\hat{E}^{\mu\nu}_{\text{CD}(i,a,b)}&=&
\frac{1}{2}\hat{C}^{\alpha_1\cdot\cdot\cdot\alpha_n}
\nabla^{\nu}\Box^{i}\hat{D}_{\alpha_1\cdot\cdot\cdot\alpha_n}
-\frac{1}{2}
\left(\Box^{i}\hat{C}^{\alpha_1\cdot\cdot\cdot\alpha_n}\right)
\nabla^{\nu}\hat{D}_{\alpha_1\cdot\cdot\cdot\alpha_n}
\nn \\
&&+\frac{1}{2}\hat{P}^{\alpha\beta\rho\sigma}_{\text{CD}(i,a,b)}
\nabla^{\nu}R_{\alpha\beta\rho\sigma}-\frac{1}{2}
\nabla^{\nu}\left(\hat{C}^{\alpha_1\cdot\cdot\cdot\alpha_n}
\Box^{i}\hat{D}_{\alpha_1\cdot\cdot\cdot\alpha_n}\right) \nn \\
&&+\frac{1}{2}\hat{F}^{\alpha\beta\rho\sigma}_{\text{CD}(i,a)}
\nabla^\nu\Box^a{R}_{\alpha\beta\rho\sigma}
-\frac{1}{2}\left(\Box^a
\hat{F}^{\alpha\beta\rho\sigma}_{\text{CD}(i,a)}\right)
\nabla^\nu{R}_{\alpha\beta\rho\sigma} \nn \\
&&+\frac{1}{2}\hat{Q}^{\alpha\beta\rho\sigma}_{\text{CD}(i,b)}
\nabla^\nu\Box^b{R}_{\alpha\beta\rho\sigma}
-\frac{1}{2}\left(\Box^b
\hat{Q}^{\alpha\beta\rho\sigma}_{\text{CD}(i,b)}\right)
\nabla^\nu{R}_{\alpha\beta\rho\sigma} \nn \\
&=&0
\, . \label{divhatEomCD}
\eea
Hence one gets the generalized Bianchi identity associated
to $\hat{E}^{\mu\nu}_{\text{CD}(i,a,b)}$. Particularly,
when $a=0=b$, the identity (\ref{divhatEomCD}) can be adopted to
give a direct proof for the Bianchi-type identity
in \cite{BCKM14}.

At the end, due to the fact that the scalar
$\big(\Box^{i}\hat{C}^{\alpha_1\cdot\cdot\cdot\alpha_n}\big)
\Box^{j}\hat{D}_{\alpha_1\cdot\cdot\cdot\alpha_n}$ can be
expressed as the form
\be
\big(\Box^{i}\hat{C}^{\alpha_1\cdot\cdot\cdot\alpha_n}\big)
\Box^{j}\hat{D}_{\alpha_1\cdot\cdot\cdot\alpha_n}
=\hat{C}^{\alpha_1\cdot\cdot\cdot\alpha_n}
\Box^{i+j}\hat{D}_{\alpha_1\cdot\cdot\cdot\alpha_n}
+\nabla_\mu \mathcal{B}^\mu
\, , \label{BhatCBhatD1}
\ee
with the vector $\mathcal{B}^\mu$ given by
\be
\mathcal{B}^\mu
=\sum^i_{k=1}\big[
\big(\nabla^\mu\Box^{i-k}
\hat{C}^{\alpha_1\cdot\cdot\cdot\alpha_n}\big)
\Box^{j+k-1}\hat{D}_{\alpha_1\cdot\cdot\cdot\alpha_n}
-\big(\Box^{i-k}
\hat{C}^{\alpha_1\cdot\cdot\cdot\alpha_n}\big)
\nabla^\mu\Box^{j+k-1}
\hat{D}_{\alpha_1\cdot\cdot\cdot\alpha_n}\big]
\, , \label{CalvecBdef}
\ee
the expression (\ref{EomhatCDhi}) for equations of
motion is able to be straightforwardly extended to
the following Lagrangian
\be
\sqrt{-g}\hat{h}_{(i,j)}=\sqrt{-g}
\left(\Box^{i}\hat{C}^{\alpha_1\cdot\cdot\cdot\alpha_n}\right)
\Box^{j}\hat{D}_{\alpha_1\cdot\cdot\cdot\alpha_n}
\, , \label{LagCDhij}
\ee
yielding the expression for field equations
\be
\hat{E}^{\mu\nu}_{\text{CD}(i,j,a,b)}=
\hat{E}^{\mu\nu}_{\text{CD}(i+j,a,b)}
\, . \label{EomhCDij}
\ee
Besides, the surface term $\hat{\Theta}_{\text{CD}(i,j,a,b)}^\mu$
derived from the variation of the Lagrangian
$\sqrt{-g}\hat{h}_{(i,j)}$ is read off as
\be
\hat{\Theta}_{\text{CD}(i,j,a,b)}^\mu=
\hat{\Theta}_{\text{CD}(i+j,a,b)}^\mu
+\delta\mathcal{B}^\mu
+\frac{1}{2}\mathcal{B}^\mu
{g}^{\rho\sigma}\delta{g}_{\rho\sigma}
\, . \label{TheCDijdef}
\ee
As before, on the basis of Eq. (\ref{TheCDijdef}), through
the computations on the surface term $\hat{\Theta}_{\text{CD}(i,j,a,b)}^\mu$
with the variation operator in it substituted by
the Lie derivative along the arbitrary vector field
$\zeta^\mu$, one is able to reproduce the expression
$\hat{E}^{\mu\nu}_{\text{CD}(i,j,a,b)}$ for the field equations,
as well as to acquire the Noether potential
\be
\hat{K}^{\mu\nu}_{\text{CD}(i,j,a,b)}=
\hat{K}^{\mu\nu}_{\text{CD}(i+j,a,b)}
+2\zeta^{[\mu}\mathcal{B}^{\nu]}
\, , \label{NoPforhCDij}
\ee
which corresponds to the conserved current
\bea
\hat{J}^{\mu}_{\text{CD}(i,j,a,b)}&=&
2\zeta_\nu\hat{E}^{\mu\nu}_{\text{CD}(i,j,a,b)}
+\zeta^\mu\hat{h}_{(i,j)}
-\hat{\Theta}_{\text{CD}(i,j,a,b)}^\mu
(\delta\rightarrow\mathcal{L}_\zeta) \nn \\
&=&\nabla_\nu\hat{K}^{\mu\nu}_{\text{CD}(i,j,a,b)}
\, . \label{CCforTildCDi}
\eea
In light of the Noether potential
$\hat{K}^{\mu\nu}_{\text{CD}(i,j,a,b)}$, the Iyer-Wald potential
associated to the Lagrangian (\ref{LagCDhij}) takes
the form
\be
\hat{Q}^{\mu\nu}_{\text{CD}(i,j,a,b)}=\frac{1}{\sqrt{-g}}
\delta\left(\sqrt{-g}\hat{K}^{\mu\nu}_{\text{CD}(i,j,a,b)}
(\zeta\rightarrow\xi)\right)
-\xi^{[\mu}\hat{\Theta}^{\nu]}_{\text{CD}(i,j,a,b)}
\, . \label{IWpotofLahij}
\ee
The Lagrangian density $\hat{h}_{(0,i)}=\hat{h}_{(i)}$,
while $\hat{h}_{(i)}$ includes ${h}_{(i)}$ and
${f}_{(i)}$ as its special cases. As a consequence, here
the Iyer-Wald potential $\hat{Q}^{\mu\nu}_{\text{CD}(i,j,a,b)}$
is applicable to the Lagrangians (\ref{LagfiABiB}),
(\ref{LagCBoxiD}) and (\ref{Laghathi}).

%%%%%%%%%%%%%%%%%%%%%%%%%%%%%%%%%%%%%%%%%%%%%%%%%%%%%%%%%%%%%%%%%%%%%%%%
\section{Summary}\label{six}
%%%%%%%%%%%%%%%%%%%%%%%%%%%%%%%%%%%%%%%%%%%%%%%%%%%%%%%%%%%%%%%%%%%%%%%%

With the purpose to reveal how higher-order derivatives
of the Riemann curvature tensor make contributions to
equations of motion and conserved
quantities, we systematically investigate
the field equations and Noether potentials corresponding
to an arbitrary smooth vector field within the framework
of higher-order gravity theories armed with Lagrangians
involving the variables $\Box^i{R}$s,
$\Box^i{R}_{\mu\nu}$s, and $\Box^i{R}_{\mu\nu\rho\sigma}$s.
Firstly, we starting with a direct variation to the
Lagrangian (\ref{LagBoxR}) that depends on the Ricci
scalar $R$ and its higher-order derivatives $\Box^iR$s
to derive the expression (\ref{EoMforLagR})
for field equations. Then we follow the method based on
conserved current to reproduce such an expression, as well
as to gain the Noether potential (\ref{KmnRdef}).
Secondly, by analogy with the analysis to the Lagrangian
(\ref{LagBoxR}), we derive both the expressions
(\ref{EoMforLagRic}) and (\ref{EoMforLagRic3})
for field equations together with the Noether potential
(\ref{KmnRic}) associated to a more general Lagrangian
(\ref{LagBoxRic}), which includes the one (\ref{LagBoxR})
as a special case. It has been demonstrated that the identity
(\ref{Ident2Ric2}) for
$\partial{L}_{\text{Ric}}/\partial{g}^{\mu\nu}$ establishes
the equivalence relation between Eqs.
(\ref{EoMforLagRic}) and (\ref{EoMforLagRic3}). Thirdly,
in terms of all the results for the Lagrangians
(\ref{LagBoxR}) and (\ref{LagBoxRic}), we further generalize
them to the Lagrangian (\ref{LagBoxRiem}), which is
supposed to be dependent of the inverse metric $g^{\mu\nu}$,
the Riemann curvature tensor $R_{\mu\nu\rho\sigma}$ and the variables
generated through $\Box^i$ acting on $R_{\mu\nu\rho\sigma}$.
The expression $E^{\mu\nu}_{\text{Riem}}$ for field equations is given
by Eq. (\ref{EoMforLagRiem}) or (\ref{EoMforLagRiem2}), while the
Noether potential $K^{\mu\nu}_{\text{Riem}}$ is presented by
Eq. (\ref{KmnRiem}). By the aid of this potential, we derive
the off-shell Noether current (\ref{ConCurLRiim}) and the
Iyer-Wald potential (\ref{IWpotofLagRiem}). What is more, we
obtain two identities (\ref{Ident1Riem2}) and
(\ref{Ident2Riem2}) in connection with the expression
for equations of motion. The latter assists us to eliminate
the term composed of
$\partial{L}_{\text{Riem}}/\partial{g}^{\mu\nu}$ in
the field equations. As an application, it has been explicitly
illustrated that all the results for the Lagrangian
(\ref{LagBoxRiem}) cover those corresponding to the
Lagrangians (\ref{LagBoxR}) and (\ref{LagBoxRic}).

Within the situations for the three Lagrangians
(\ref{LagBoxR}), (\ref{LagBoxRic}) and (\ref{LagBoxRiem}), there
exist three relations given by Eqs. (\ref{SumPhik}), (\ref{SumPsik0})
and (\ref{SumUpsik0}), respectively. They have been utilized to
peel off the operators $\Box^i$s in the variation terms and play an
important role in figuring out the expressions for equations of
motion and the Noether potentials. Furthermore, these three
crucial relations are generalized to the one (\ref{SumOmegik})
for a scalar
$A^{\alpha_1\cdot\cdot\cdot\alpha_n}
(\delta\Box^{i}B_{\alpha_1\cdot\cdot\cdot\alpha_n})$,
where $A^{\alpha_1\cdot\cdot\cdot\alpha_n}$ and
$B_{\alpha_1\cdot\cdot\cdot\alpha_n}$ represent
two generic rank-$n$ tensors. On the basis of this relation,
we analyse in detail how the scalar
$A^{\alpha_1\cdot\cdot\cdot\alpha_n}
(\delta\Box^{i}B_{\alpha_1\cdot\cdot\cdot\alpha_n})$
makes contributions to the field equations and the Noether
potentials. Subsequently, the results are applied to the Lagrangian
(\ref{LagBgen}), in which the tensor
$B_{\alpha_1\cdot\cdot\cdot\alpha_n}$ is assumed to depend upon
the variables $g^{\mu\nu}$, $R_{\mu\nu\rho\sigma}$ and
$\Box^iR_{\mu\nu\rho\sigma}$s.
It has been illustrated that the Lagrangian
(\ref{LagBgen}) provides a platform to unify all
the results associated to the three Lagrangians
(\ref{LagBoxR}), (\ref{LagBoxRic}) and (\ref{LagBoxRiem}).
Besides, the results from the scalar
$A^{\alpha_1\cdot\cdot\cdot\alpha_n}
(\delta\Box^{i}B_{\alpha_1\cdot\cdot\cdot\alpha_n})$
are utilized to derive equations of motion and Noether
potentials for the Lagrangian (\ref{LagfijBD})
$\sqrt{-g}f_{(i,j)}$ and the ones belonging to the form
$\sqrt{-g}A^{\bullet}\Box^{i}B_{\bullet}$. Here
both $A^{\bullet}$ and $B_{\bullet}$ stand for two
generic tensors relying on $g_{\mu\nu}$ (or $g^{\mu\nu}$),
$R_{\mu\nu\rho\sigma}$, and $\Box^iR_{\mu\nu\rho\sigma}$s.
All the expressions for the field equations and the Noether
potentials are summarized in TABLE \ref{LagEomNP}
within Appendix \ref{appendD}. In terms of the Noether
potentials and the surface terms, the Noether currents
and the Iyer-Wald potentials are also presented. By making
use of the potentials and currents, one is able to further
define conserved quantities of these gravity theories.
In particular, we stress the potential applications
of our results in nonlocal theories of gravity.

In order to check all the expressions for field equations
obtained in the present work, we have proved that all of them
satisfy Bianchi-type identities by means of straightforward
computations on the divergences of these expressions.

\section*{Acknowledgments}

This work was supported by the National Natural Science
Foundation of China under Grant Nos. 11865006.

%%%%%%%%%%%%%%%%%%%%%%%%%%%%%%%%%%%%%%%%%%%%%%%%%%%%%%%%%%%
\appendix

%%%%%%%%%%%%%%%%%%%%%%%%%%%%%%%%%%%%%%%%%%%%%%%%%%%%%%%%
\section{The derivation of Eq. (\ref{PsikRel}) from
Eq. (\ref{UpsikRel})}\label{appendA}
%%%%%%%%%%%%%%%%%%%%%%%%%%%%%%%%%%%%%%%%%%%%%%%%%%%%

In this appendix, we demonstrate that Eq. (\ref{PsikRel})
can be also derived out of Eq. (\ref{UpsikRel}). To do this,
we compute $\Upsilon_{(i,k)}$ and $\Upsilon_{(i,k+1)}$
in light of Eq. (\ref{P4iP4forLRic}), giving rise to
\bea
\Upsilon_{(i,k)}\big|_{P\rightarrow\bar{P}}&=&
\Psi_{(i,k)}-\left(\Box^{k-1} P^{\rho\sigma}_{(i)}\right)
\left(\Box^{i-k+1}R_{\alpha\rho\beta\sigma}\right)
\delta{g}^{\alpha\beta} \, , \nn \\
\Upsilon_{(i,k+1)}\big|_{P\rightarrow\bar{P}}&=&
\Psi_{(i,k+1)}-\left(\Box^{k} P^{\rho\sigma}_{(i)}\right)
\left(\Box^{i-k}R_{\alpha\rho\beta\sigma}\right)
\delta{g}^{\alpha\beta}
\, . \label{Psikk1Ric}
\eea
Besides, doing some calculations for the divergence of
the vector ${U}^{\mu}_{(i,k)}$ on the basis of
Eq. (\ref{UVLNrel}) leads to
\bea
\left(\nabla_\mu{U}^{\mu}_{(i,k)}\right)\Big|_{P\rightarrow\bar{P}}
&=&\nabla_\mu{L}^{\mu}_{(i,k)}
-\left(\Box^{k-1} P^{\rho\sigma}_{(i)}\right)
\left(\Box^{i-k+1}R_{\alpha\rho\beta\sigma}\right)
\delta{g}^{\alpha\beta}\nn \\
&&-\left(\Box^{k-1} P^{\rho\sigma}_{(i)}\right)
\left(\nabla_{\mu}\Box^{i-k}R_{\alpha\rho\beta\sigma}\right)
\left(\nabla^\mu\delta{g}^{\alpha\beta}\right) \nn \\
&&+\left(\Box^{k} P^{\rho\sigma}_{(i)}\right)
\left(\Box^{i-k}R_{\alpha\rho\beta\sigma}\right)
\delta{g}^{\alpha\beta}\nn\\
&&+\left(\nabla_{\mu}\Box^{k-1} P^{\rho\sigma}_{(i)}\right)
\left(\Box^{i-k}R_{\alpha\rho\beta\sigma}\right)
\left(\nabla^\mu\delta{g}^{\alpha\beta}\right)
\, . \label{DivUmRic}
\eea
What is more, by the aid of Eq. (\ref{WMikrel}), we obtain
\bea
\left({g}_{\gamma\lambda}W^{\lambda\mu\nu}_{(i,k)}
\delta \Gamma^\gamma_{\mu\nu}\right)
\Big|_{P\rightarrow\bar{P}}&=&
{g}_{\rho\sigma}M^{\sigma\mu\nu}_{(i,k)}
\delta \Gamma^\rho_{\mu\nu}
+\left(\Box^{k-1} P^{\rho\sigma}_{(i)}\right)
\left(\nabla_{\mu}\Box^{i-k}R_{\alpha\rho\beta\sigma}\right)
\left(\nabla^\mu\delta{g}^{\alpha\beta}\right)
 \nn \\
&&-\left(\nabla_{\mu}\Box^{k-1} P^{\rho\sigma}_{(i)}\right)
\left(\Box^{i-k}R_{\alpha\rho\beta\sigma}\right)
\left(\nabla^\mu\delta{g}^{\alpha\beta}\right)
\, . \label{WdelGamRic}
\eea
Consequently, substituting Eqs. (\ref{Psikk1Ric}),
(\ref{DivUmRic}) and (\ref{WdelGamRic}) into
Eq. (\ref{UpsikRel}), we acquire Eq. (\ref{PsikRel})
corresponding to the Lagrangian
$\sqrt{-g}L_{\text{Ric}}$.

%%%%%%%%%%%%%%%%%%%%%%%%%%%%%%%%%%%%%%%%%%%%%%%%%%%%%%%%
\section{Two examples for computing
$(\Box^{i}A^{\alpha_1\cdot\cdot\cdot\alpha_n})
\delta{B}_{\alpha_1\cdot\cdot\cdot\alpha_n}$}\label{appendB}
%%%%%%%%%%%%%%%%%%%%%%%%%%%%%%%%%%%%%%%%%%%%%%%%%%%%

The present appendix is devoted to providing a more specific
example to illustrate Eqs. (\ref{BoxiAdelB}) and
(\ref{TheBidelLie}). For the sake of doing this,
we take into consideration of the scalar
$(\Box^{i}A^{\alpha_1\cdot\cdot\cdot\alpha_n})
\delta{B}_{\alpha_1\cdot\cdot\cdot\alpha_n}$
$(i=0,1,2,\cdot\cdot\cdot)$ in the situation where
the generic tensor ${B}_{\alpha_1\cdot\cdot\cdot\alpha_n}$ is restricted to
depend upon the metric $g_{\mu\nu}$ (or its inverse $g^{\mu\nu}$) together
with the Riemann curvature tensor $R_{\alpha\beta\rho\sigma}$.
Accordingly, the scalar $(\Box^{i}A^{\alpha_1\cdot\cdot\cdot\alpha_n})
\delta{B}_{\alpha_1\cdot\cdot\cdot\alpha_n}$ is generally
given as the sum of two terms proportional to
$\delta{g}_{\rho\sigma}$ and
$\delta{R}_{\gamma\lambda\rho\sigma}$, respectively,
\be
(\Box^{i}A^{\alpha_1\cdot\cdot\cdot\alpha_n})
\delta{B}_{\alpha_1\cdot\cdot\cdot\alpha_n}=
-P^{\rho\sigma}_{\text{AB}(i)}\delta{g}_{\rho\sigma}
+Q^{\gamma\lambda\rho\sigma}_{(i)}
\delta{R}_{\gamma\lambda\rho\sigma}
\, , \label{AdelBRiem}
\ee
in which the second-rank symmetric tensor
$P^{\rho\sigma}_{\text{AB}(i)}$
and the rank-4 one $Q^{\gamma\lambda\rho\sigma}_{(i)}$
are defined respectively through
\bea
P^{\rho\sigma}_{\text{AB}(i)}
&=&{g}^{\rho\mu}{g}^{\sigma\nu}
\big(\Box^{i}A^{\alpha_1\cdot\cdot\cdot\alpha_n}\big)
\frac{\partial{B}_{\alpha_1\cdot\cdot\cdot\alpha_n}}{
\partial{g}^{\mu\nu}}
\, , \nn \\
Q^{\gamma\lambda\rho\sigma}_{(i)}&=&
\big(\Box^{i}A^{\alpha_1\cdot\cdot\cdot\alpha_n}\big)
\frac{\partial{B}_{\alpha_1\cdot\cdot\cdot\alpha_n}}{
\partial{R}_{\gamma\lambda\rho\sigma}}
\, . \label{Qi24def}
\eea
The definition for $Q^{\gamma\lambda\rho\sigma}_{(i)}$ implies
that it possesses the same algebraic symmetries as those for
the Riemann tensor $R^{\gamma\lambda\rho\sigma}$, namely,
\be
Q^{\gamma\lambda\rho\sigma}_{(i)}
=Q^{[\gamma\lambda][\rho\sigma]}_{(i)}
=Q^{[\rho\sigma][\gamma\lambda]}_{(i)}
=Q^{\rho\sigma\gamma\lambda}_{(i)}
\, . \label{Qiran4alsym}
\ee
By the aid of the Palatini identity for
$\delta{R}_{\gamma\lambda\rho\sigma}$, Eq. (\ref{AdelBRiem})
can be further expressed as the linear combination for
a term proportional to the variation of the metric together
with a divergence term,
\be
(\Box^{i}A^{\alpha_1\cdot\cdot\cdot\alpha_n})
\delta{B}_{\alpha_1\cdot\cdot\cdot\alpha_n}=
-\tilde{E}^{\mu\nu}_{B(i)}\delta{g}_{\mu\nu}+
\nabla_\mu \tilde{\Theta}^\mu_{B(i)}
\, . \label{BxiAdelBsC}
\ee
In the above equation, the rank-2 symmetric tensor
$\tilde{E}^{\mu\nu}_{B(i)}$ is
read off as
\be
\tilde{E}^{\mu\nu}_{B(i)}
={P}^{\mu\nu}_{\text{AB}(i)}-Q^{\mu\tau\rho\sigma}_{(i)}
R^\nu_{~\tau\rho\sigma}-2\nabla_\rho\nabla_\sigma
Q^{\rho\mu\nu\sigma}_{(i)}
\, , \label{TildEBidef}
\ee
and the surface term $\tilde{\Theta}^\mu_{B(i)}$ is given by
\be
\tilde{\Theta}^\mu_{B(i)}=2Q^{\mu\nu\rho\sigma}_{(i)}
\nabla_\sigma\delta g_{\rho\nu}
-2(\delta g_{\nu\rho})
\nabla_\sigma{Q}^{\mu\nu\rho\sigma}_{(i)}
\, . \label{TilThetAB}
\ee
Moreover, on the basis of Eq. (\ref{TilThetAB}), replacing the variation
operator $\delta$ in $\tilde{\Theta}^\mu_{B(i)}$ with the Lie
derivative $\mathcal{L}_\zeta$, we obtain
\be
\tilde{\Theta}^\mu_{B(i)}(\delta\rightarrow\mathcal{L}_\zeta)
=2\zeta_\nu\tilde{X}^{\mu\nu}_{B(i)}
-\nabla_\nu\tilde{K}^{\mu\nu}_{B(i)}
\, , \label{TilTheBidelLie}
\ee
where the second-rank tensor $\tilde{X}^{\mu\nu}_{B(i)}$
is written as
\be
\tilde{X}^{\mu\nu}_{B(i)}=Q^{\mu\tau\rho\sigma}_{(i)}
R^\nu_{~\tau\rho\sigma}-2\nabla_\rho\nabla_\sigma
Q^{\rho\mu\nu\sigma}_{(i)}
\, , \label{TildXBidef}
\ee
and the anti-symmetric tensor $\tilde{K}^{\mu\nu}_{B(i)}$ is
presented by
\be
\tilde{K}^{\mu\nu}_{B(i)}=
2Q^{\mu\nu\rho\sigma}_{(i)}\nabla_{\rho}\zeta_{\sigma}
+4\zeta_\rho\nabla_\sigma{Q}^{\mu\nu\rho\sigma}_{(i)}
-6Q^{\mu[\nu\rho\sigma]}_{(i)}\nabla_\rho\zeta_\sigma
\, . \label{KmnABidef}
\ee
By using Eq. (\ref{TildEBidef}) to eliminate the
$-2\nabla_\rho\nabla_\sigma{Q}^{\rho\mu\nu\sigma}_{(i)}$
term in Eq. (\ref{TildXBidef}), one is able to
gain an identity
\be
\tilde{X}^{[\mu\nu]}_{B(i)}=2Q^{[\mu|\tau\rho\sigma|}_{(i)}
R^{\nu]}_{~~\tau\rho\sigma}
\, . \label{TildXBiantiId}
\ee
According to Eqs. (\ref{BxiAdelBsC}) and (\ref{TilTheBidelLie}),
both of them can be regarded as the specific illustrations of
Eqs. (\ref{BoxiAdelB}) and (\ref{TheBidelLie}), respectively.

In addition, when the rank-$n$ tensor
${B}_{\alpha_1\cdot\cdot\cdot\alpha_n}$
is assumed to take a more generic form
\be
{B}_{\alpha_1\cdot\cdot\cdot\alpha_n}=
{B}_{\alpha_1\cdot\cdot\cdot\alpha_n}\left(g^{\mu\nu},
\Box^{j}{R}_{\mu\nu\rho\sigma}\right)
\, , \label{Bngenform}
\ee
where the nonnegative integer $j=0,1,2,\cdot\cdot\cdot$,
introducing a rank-4 tensor exhibiting the same algebraic
symmetries as the Riemann tensor
\be
Q^{\mu\nu\rho\sigma}_{(i,j)}=
\big(\Box^{i}A^{\alpha_1\cdot\cdot\cdot\alpha_n}\big)
\frac{\partial{B}_{\alpha_1\cdot\cdot\cdot\alpha_n}}{
\partial\Box^{j}{R}_{\mu\nu\rho\sigma}}
\, , \label{Qil4def}
\ee
which fulfills $Q^{\mu\nu\rho\sigma}_{(i,0)}
=Q^{\mu\nu\rho\sigma}_{(i)}$ within the $j=0$
situation, one is able to express the scalar
$(\Box^{i}A^{\alpha_1\cdot\cdot\cdot\alpha_n})
\delta{B}_{\alpha_1\cdot\cdot\cdot\alpha_n}$ as the following
form
\be
(\Box^{i}A^{\alpha_1\cdot\cdot\cdot\alpha_n})
\delta{B}_{\alpha_1\cdot\cdot\cdot\alpha_n}=
-P^{\mu\nu}_{\text{AB}(i)}\delta{g}_{\mu\nu}
+Q^{\mu\nu\rho\sigma}_{(i,j)}
\delta\Box^{j}{R}_{\mu\nu\rho\sigma}
\, . \label{AdelBBoxRiem}
\ee
By utilizing Eq. (\ref{SumOmegik2}), here the scalar
$Q^{\mu\nu\rho\sigma}_{(i,j)}
\delta\Box^{j}{R}_{\mu\nu\rho\sigma}$ can be further
transformed into the one proportional to the variation
of the Riemann tensor
$\big(\Box^{j}Q^{\mu\nu\rho\sigma}_{(i,j)}\big)
\delta{R}_{\mu\nu\rho\sigma}$ through
\bea
Q^{\mu\nu\rho\sigma}_{(i,j)}
\delta\Box^{j}{R}_{\mu\nu\rho\sigma}&=&
-\left(X^{\mu\nu}_{Q(i,j)}+2R^{\nu}_{~\lambda\rho\sigma}
\Box^{j}Q^{\mu\lambda\rho\sigma}_{(i,j)}
-2Q^{\mu\lambda\rho\sigma}_{(i,j)}
\Box^{j}R^{\nu}_{~\lambda\rho\sigma}\right)
\delta g_{\mu\nu} \nn \\
&&+\left(\Box^{j}Q^{\mu\nu\rho\sigma}_{(i,j)}\right)
\delta{R}_{\mu\nu\rho\sigma}
+\nabla_\mu\Theta_{Q(i,j)}^\mu
\, , \label{QijBoxjRiem}
\eea
in which the quantities $X^{\mu\nu}_{Q(i,j)}$ and
$\Theta_{Q(i,j)}^\mu$ are defined respectively as
\bea
X^{\mu\nu}_{Q(i,j)}&=&\sum^j_{k=1}X^{\mu\nu}_{(j,k)}
\left(A\rightarrow{Q}_{(i,j)},B\rightarrow{R}\right)
=\sum^j_{k=1}X^{\mu\nu}_{\text{Riem}(j,k)}
\left({P}_{(j)}\rightarrow{Q}_{(i,j)}\right) \, , \nn \\
\Theta^\mu_{Q(i,j)}&=&\sum^j_{k=1}\Theta_{(j,k)}^\mu
\left(A\rightarrow{Q}_{(i,j)},B\rightarrow{R}\right)
=\sum^j_{k=1}\Theta^\mu_{\text{Riem}(j,k)}
\left({P}_{(j)}\rightarrow{Q}_{(i,j)}\right)
\, . \label{XTheQij}
\eea
In the above equation, the surface term ${\Theta}^{\mu}_{(j,k)}=
{\Theta}^{\mu}_{(i,k)}\big|_{i=j}$ and the rank-2 tensor
${X}^{\mu\nu}_{(j,k)}={X}^{\mu\nu}_{(i,k)}\big|_{i=j}$, with
${\Theta}^{\mu}_{(i,k)}$ and ${X}^{\mu\nu}_{(i,k)}$ given
by Eqs. (\ref{Thegenik2}) and (\ref{Xmngenik}), respectively, while
$\Theta^\mu_{\text{Riem}(j,k)}$ and $X^{\mu\nu}_{\text{Riem}(j,k)}$
can be found in Eqs. (\ref{TheRiemik}) and (\ref{XmnRiemik2}),
respectively. According to the definitions,
$X^{\mu\nu}_{Q(i,0)}=0$ and $\Theta^\mu_{Q(i,0)}=0$. With the help
of Eq. (\ref{TheikdelLie}), the tensor $X^{\mu\nu}_{Q(i,j)}$
is in connection with $\Theta^\mu_{Q(i,j)}$ via
\be
\Theta^\mu_{Q(i,j)}(\delta\rightarrow\mathcal{L}_\zeta)
=2\zeta_\nu{X}^{\mu\nu}_{Q(i,j)}
-\nabla_\nu{K}^{\mu\nu}_{Q(i,j)}
\, , \label{TheXQijRel}
\ee
where the anti-symmetric tensor ${K}^{\mu\nu}_{Q(i,j)}$ is
read off as
\bea
{K}^{\mu\nu}_{Q(i,j)}&=&
\sum^j_{k=1}K^{\mu\nu}_{(j,k)}
\left(A\rightarrow{Q}_{(i,j)},B\rightarrow{R}\right) \nn \\
&=&\sum^j_{k=1}K^{\mu\nu}_{\text{Riem}(j,k)}
\left({P}_{(j)}\rightarrow{Q}_{(i,j)}\right)
\, ,
\eea
with $K^{\mu\nu}_{(j,k)}$ and $K^{\mu\nu}_{\text{Riem}(j,k)}$
given by Eqs. (\ref{Kmngenik}) and (\ref{KmnRiemik2}), respectively.
Subsequently, with the help of the expression
\bea
\left(\Box^{j}Q^{\mu\nu\rho\sigma}_{(i,j)}\right)
\delta{R}_{\mu\nu\rho\sigma}&=&
2\nabla_\mu\left[\left(\Box^{j}Q^{\mu\nu\rho\sigma}_{(i,j)}\right)
\nabla_\sigma\delta g_{\rho\nu}
-(\delta g_{\nu\rho})
\nabla_\sigma\Box^{j}Q^{\mu\nu\rho\sigma}_{(i,j)}\right]
 \nn \\
&&+\left(R^\nu_{~\lambda\rho\sigma}
\Box^{j}Q^{\mu\lambda\rho\sigma}_{(i,j)}
+2\nabla_\rho\nabla_\sigma
\Box^{j}Q^{\rho\mu\nu\sigma}_{(i,j)}\right)\delta{g}_{\mu\nu}
\, , \label{BoxjQdelRiem}
\eea
substituting Eq. (\ref{QijBoxjRiem})
into Eq. (\ref{AdelBBoxRiem}) eventually leads to
\bea
(\Box^{i}A^{\alpha_1\cdot\cdot\cdot\alpha_n})
\delta{B}_{\alpha_1\cdot\cdot\cdot\alpha_n}
&=&-{E}^{\mu\nu}_{\text{GenB}(i,j)}\delta{g}_{\mu\nu}
+\nabla_\mu\Theta_{\text{GenB}(i,j)}^\mu
\, , \label{AdelBBoxRiem2}
\eea
where the second-rank tensor
${E}^{\mu\nu}_{\text{GenB}(i,j)}$ is given by
\bea
{E}^{\mu\nu}_{\text{GenB}(i,j)}
&=&P^{\mu\nu}_{\text{AB}(i)}+
X^{\mu\nu}_{Q(i,j)}+R^{\nu}_{~\lambda\rho\sigma}
\Box^{j}Q^{\mu\lambda\rho\sigma}_{(i,j)}
-2Q^{\mu\lambda\rho\sigma}_{(i,j)}
\Box^{j}R^{\nu}_{~\lambda\rho\sigma} \nn \\
&&-2\nabla_\rho\nabla_\sigma
\Box^{j}Q^{\rho\mu\nu\sigma}_{(i,j)}
\, , \label{EgenBij}
\eea
and the surface term $\Theta_{\text{GenB}(i,j)}^\mu$ takes
the form
\be
\Theta_{\text{GenB}(i,j)}^\mu=
2\left(\Box^{j}Q^{\mu\nu\rho\sigma}_{(i,j)}\right)
\nabla_\sigma\delta g_{\rho\nu}
-2(\delta g_{\nu\rho})
\nabla_\sigma\Box^{j}Q^{\mu\nu\rho\sigma}_{(i,j)}
+\Theta_{Q(i,j)}^\mu
\, . \label{GenThetij}
\ee
As a matter of fact, the tensor ${E}^{\mu\nu}_{\text{GenB}(i,j)}$
in Eq. (\ref{EgenBij}) is symmetric with respect to the indices
$(\mu\nu)$, arising from that
\be
\left(\Box^{j}Q^{[\mu|\lambda\rho\sigma|}_{(i,j)}\right)
R^{\nu]}_{~~\lambda\rho\sigma}=-2\nabla_\rho\nabla_\sigma
\Box^{j}Q^{\rho[\mu\nu]\sigma}_{(i,j)}
\, , \label{QijRanSym}
\ee
together with the identity obtained via Eq. (\ref{SumXikdef}),
namely,
\be
X^{[\mu\nu]}_{Q(i,j)}=2R^{[\mu}_{~~\lambda\rho\sigma}
\Box^{j}Q^{\nu]\lambda\rho\sigma}_{(i,j)}
-2\left(\Box^{j}R^{[\mu}_{~~\lambda\rho\sigma}\right)
Q^{\nu]\lambda\rho\sigma}_{(i,j)}
\, . \label{XQijanSym}
\ee
Obviously, ${E}^{\mu\nu}_{\text{GenB}(i,0)}
=\tilde{E}^{\mu\nu}_{B(i)}$ and ${\Theta}^{\mu}_{\text{GenB}(i,0)}
=\tilde{\Theta}^{\mu}_{B(i)}$. Furthermore, by making use of
Eqs. (\ref{TilTheBidelLie}) and (\ref{TheXQijRel}), under
the transformation $\delta\rightarrow\mathcal{L}_\zeta$,
one obtains
\be
\Theta^\mu_{\text{GenB}(i,j)}(\delta\rightarrow\mathcal{L}_\zeta)
=2\zeta_\nu{X}^{\mu\nu}_{\text{GenB}(i,j)}
-\nabla_\nu{K}^{\mu\nu}_{\text{GenB}(i,j)}
\, , \label{TheXgenBijRel}
\ee
in which
\be
{X}^{\mu\nu}_{\text{GenB}(i,j)}=
\left(\Box^{j}Q^{\mu\lambda\rho\sigma}_{(i,j)}\right)
R^\nu_{~\lambda\rho\sigma}-2\nabla_\rho\nabla_\sigma
\Box^{j}Q^{\rho\mu\nu\sigma}_{(i,j)}
+{X}^{\mu\nu}_{Q(i,j)}
\, , \label{XGenBij}
\ee
and the anti-symmetric tensor ${K}^{\mu\nu}_{\text{GenB}(i,j)}$
is given by
\be
{K}^{\mu\nu}_{\text{GenB}(i,j)}=
2\left(\Box^{j}Q^{\mu\nu\rho\sigma}_{(i,j)}\right)
\nabla_{\rho}\zeta_{\sigma}
+4\zeta_\rho\nabla_\sigma
\Box^{j}Q^{\mu\nu\rho\sigma}_{(i,j)}
-6\left(\Box^{j}Q^{\mu[\nu\rho\sigma]}_{(i,j)}\right)
\nabla_\rho\zeta_\sigma
+{K}^{\mu\nu}_{Q(i,j)}
\, . \label{KGenBij}
\ee
Both the tensors ${X}^{\mu\nu}_{\text{GenB}(i,j)}$
and ${E}^{\mu\nu}_{\text{GenB}(i,j)}$ are in connection
with each other in the way
\bea
{X}^{\mu\nu}_{\text{GenB}(i,j)}
&=&{E}^{\mu\nu}_{\text{GenB}(i,j)}
-P^{\mu\nu}_{\text{AB}(i)}
+2Q^{\mu\lambda\rho\sigma}_{(i,j)}
\Box^{j}R^{\nu}_{~\lambda\rho\sigma}
\, . \label{XGenBij2}
\eea
From Eqs. (\ref{XGenBij}) and (\ref{KGenBij}), one observes that
${X}^{\mu\nu}_{\text{GenB}(i,0)}=\tilde{X}^{\mu\nu}_{B(i)}$ and
${K}^{\mu\nu}_{\text{GenB}(i,0)}=\tilde{K}^{\mu\nu}_{B(i)}$.
On the basis of Eqs. (\ref{AdelBBoxRiem2}) and
(\ref{TheXgenBijRel}), performing the same analysis
in Sec. \ref{four}, one is able to figure out all the
contributions from the scalar
$(\Box^{i}A^{\alpha_1\cdot\cdot\cdot\alpha_n})
\delta{B}_{\alpha_1\cdot\cdot\cdot\alpha_n}$
to equations of motion and the Noether potentials.
For a concrete example see the derivation for
the field equations and the Noether potential associated to
the Lagrangian (\ref{TildLaghi}) within Subsec. \ref{Five4}.

%%%%%%%%%%%%%%%%%%%%%%%%%%%%%%%%%%%%%%%%%%%%%%%%%%%%%%%%
\section{The proofs for $\nabla_\mu
{E}^{\mu\nu}_{\text{Riem}}=0$, $\nabla_\mu
{E}^{\mu\nu}_{B}=0$,
$\nabla_\mu{E}^{\mu\nu}_{\text{BD}(i,j)}=0$
and $\nabla_\mu \tilde{E}^{\mu\nu}_{\text{CD}(i,m,n)}=0$}
\label{appendC}
%%%%%%%%%%%%%%%%%%%%%%%%%%%%%%%%%%%%%%%%%%%%%%%%%%%%

In the present appendix, firstly, we utilize Eq. (\ref{IdeRiemHik2})
to straightforwardly prove that the expression
${E}^{\mu\nu}_{\text{Riem}}$ for equations of motion is
divergence-free, that is, $\nabla_\mu
{E}^{\mu\nu}_{\text{Riem}}=0$. Without loss of generality,
here we adopt the form of ${E}^{\mu\nu}_{\text{Riem}}$ given by
Eq. (\ref{EoMforLagRiem2}) rather than the one presented by
Eq. (\ref{EoMforLagRiem}), attributed to the fact that
the latter has a shortcoming of dealing with the divergence
for the derivative of the Lagrangian density with respect
to the metric tensor. Within the situation
for the Lagrangian $\sqrt{-g}L_{\text{Riem}}$, replacing
the tensors  $(A^{\alpha_1\cdot\cdot\cdot\alpha_n},
B_{\alpha_1\cdot\cdot\cdot\alpha_n})$ in Eq. (\ref{IdeRiemHik2})
with the ones $\big(P^{\alpha\beta\rho\sigma}_{(i)},
R_{\alpha\beta\rho\sigma}\big)$,
we have the sum of the divergence for the tensor
${X}^{\mu\nu}_{\text{Riem}(i,k)}$ over $k$ from 1 to $i$,
being of the form
\be
\sum^i_{k=1}\nabla_\mu{X}^{\mu\nu}_{\text{Riem}(i,k)}=
\frac{1}{2}P^{\alpha\beta\rho\sigma}_{(i)}
\nabla^{\nu}\Box^{i}R_{\alpha\beta\rho\sigma}
-\frac{1}{2}\Big(\Box^{i}P^{\alpha\beta\rho\sigma}_{(i)}\Big)
\nabla^{\nu}R_{\alpha\beta\rho\sigma}
\, . \label{SumDivXRiemik}
\ee
By making use of Eq. (\ref{SumDivXRiemik}), the divergence of
${E}^{\mu\nu}_{\text{Riem}}$ in Eq. (\ref{EoMforLagRiem2})
is written as
\bea
\nabla_\mu{E}^{\mu\nu}_{\text{Riem}}
&=&R^{\nu}_{~\lambda\rho\sigma}
\nabla_\mu{P}^{\mu\lambda\rho\sigma}
+{P}^{\mu\lambda\rho\sigma}\nabla_\mu{R}^{\nu}_{~\lambda\rho\sigma}
+2\nabla_{[\rho}\nabla_{\mu]}
\nabla_\sigma{P}^{\rho\mu\nu\sigma} \nn \\
&&-\frac{1}{2}\nabla^\nu{L}_{\text{Riem}}
+\frac{1}{2}\sum^m_{i=1}P^{\alpha\beta\rho\sigma}_{(i)}
\nabla^{\nu}\Box^{i}R_{\alpha\beta\rho\sigma} \nn \\
&&-\frac{1}{2}\sum^m_{i=1}
\Big(\Box^{i}P^{\alpha\beta\rho\sigma}_{(i)}\Big)
\nabla^{\nu}R_{\alpha\beta\rho\sigma}
\, . \label{DivEoMLaRiem}
\eea
Furthermore, substituting the divergence for the Lagrangian
density
\be
\nabla^\nu{L}_{\text{Riem}}=
P^{\alpha\beta\rho\sigma}_{(0)}
\nabla^{\nu}R_{\alpha\beta\rho\sigma}
+\sum^m_{i=1}
P^{\alpha\beta\rho\sigma}_{(i)}
\nabla^{\nu}\Box^{i}R_{\alpha\beta\rho\sigma}
\, , \label{DelnuLRiem}
\ee
together with the identity
\bea
{P}^{\mu\lambda\rho\sigma}\nabla_\mu{R}^{\nu}_{~\lambda\rho\sigma}
&=&\frac{1}{2}P^{\alpha\beta\rho\sigma}_{(0)}
\nabla^{\nu}R_{\alpha\beta\rho\sigma}
+\frac{1}{2}\sum^m_{i=1}
\Big(\Box^{i}P^{\alpha\beta\rho\sigma}_{(i)}\Big)
\nabla^{\nu}R_{\alpha\beta\rho\sigma}
\,  \label{IdePRiem}
\eea
and the one
\bea
\nabla_{[\rho}\nabla_{\mu]}
\nabla_\sigma{P}^{\rho\mu\nu\sigma}&=&
-\frac{1}{2}R^{\nu}_{~\lambda\rho\sigma}
\nabla_\mu{P}^{\mu\lambda\rho\sigma}
\, , \label{IdePRiem2}
\eea
into Eq. (\ref{DivEoMLaRiem}), we ultimately arrive at the
generalized Bianchi identity,
\be
\nabla_\mu{E}^{\mu\nu}_{\text{Riem}}\equiv0
\, . \label{DivEoMLaRiem2}
\ee
This apparently demonstrates that ${E}^{\mu\nu}_{\text{Riem}}$ is
indeed divergence-free. Additionally, due to the fact that the
field equation expressions ${E}^{\mu\nu}_{R}$ and
${E}^{\mu\nu}_{\text{Ric}}$ can be interpreted as two special
cases of ${E}^{\mu\nu}_{\text{Riem}}$, one has the conclusion
that the aforementioned proof also works for both of them.
Henece they are proved to be divergenceless as well.

Secondly, by analogy with the above proof for
${E}^{\mu\nu}_{\text{Riem}}$, we switch to prove that
the expression ${E}^{\mu\nu}_{B}$ for field equations
given by Eq. (\ref{EomLagBgen}) is divergence-free as well.
In the situation for the Lagrangian (\ref{LagBgen}),
Eq. (\ref{IdeRiemHik2}) is transformed into
\bea
\sum^i_{k=1}\nabla_\mu{X}^{\mu\nu}_{(i,k)}&=&
\frac{1}{2}A^{\alpha_1\cdot\cdot\cdot\alpha_n}_{B(i)}
\nabla^{\nu}\Box^{i}B_{\alpha_1\cdot\cdot\cdot\alpha_n}
-\frac{1}{2}
\left(\Box^{i}A^{\alpha_1\cdot\cdot\cdot\alpha_n}_{B(i)}\right)
\nabla^{\nu}B_{\alpha_1\cdot\cdot\cdot\alpha_n}
\, . \label{IdDivXikLagB}
\eea
The divergence of ${X}^{\mu\nu}_{\text{GenB}(i,j)}$
is given by
\bea
\nabla_\mu{X}^{\mu\nu}_{\text{GenB}(i,j)}&=&
\frac{1}{2}
\left(\Box^{j}{Q}^{\alpha\beta\rho\sigma}_{(i,j)}\right)
\nabla^{\nu}R_{\alpha\beta\rho\sigma}
+\nabla_\mu{X}^{\mu\nu}_{Q(i,j)}\, , \nn \\
\nabla_\mu{X}^{\mu\nu}_{Q(i,j)}&=&
\frac{1}{2}Q^{\alpha\beta\rho\sigma}_{(i,j)}
\nabla^{\nu}\Box^{j}R_{\alpha\beta\rho\sigma}
-\frac{1}{2}\Big(\Box^{j}Q^{\alpha\beta\rho\sigma}_{(i,j)}\Big)
\nabla^{\nu}R_{\alpha\beta\rho\sigma}
\, . \label{DivtildXBi}
\eea
And the divergence of the Lagrangian density $L_{B}$ is
read off as
\be
\nabla^\nu{L}_{B}=Q^{\alpha\beta\rho\sigma}_{(0,j)}
\nabla^{\nu}\Box^{j}R_{\alpha\beta\rho\sigma}
+\sum^m_{i=1}A^{\alpha_1\cdot\cdot\cdot\alpha_n}_{B(i)}
\nabla^{\nu}\Box^{i}B_{\alpha_1\cdot\cdot\cdot\alpha_n}
\, . \label{DelnuLagB}
\ee
By means of Eqs. (\ref{IdDivXikLagB}), (\ref{DivtildXBi})
and (\ref{DelnuLagB}), one is able to obtain the vanishing
divergence for the field equations,
\bea
\nabla_\mu{E}^{\mu\nu}_{B}&=&
\nabla_\mu{X}^{\mu\nu}_{\text{GenB}(0,j)}
+\sum^m_{i=1}\left(\nabla_\mu{X}^{\mu\nu}_{\text{GenB}(i,j)}
+\sum^i_{k=1}\nabla_\mu{X}^{\mu\nu}_{(i,k)}\right)
-\frac{1}{2}\nabla^\nu{L}_{B} \nn \\
&=&\frac{1}{2}\sum^m_{i=1}
\left[Q^{\alpha\beta\rho\sigma}_{(i,j)}
\nabla^{\nu}\Box^{j}R_{\alpha\beta\rho\sigma}
-\left(\Box^{i}A^{\alpha_1\cdot\cdot\cdot\alpha_n}_{B(i)}\right)
\nabla^{\nu}B_{\alpha_1\cdot\cdot\cdot\alpha_n}\right] \nn \\
&=&0
\, . \label{DivEomLagB}
\eea
Within Eqs. (\ref{DivtildXBi}), (\ref{DelnuLagB}) and
(\ref{DivEomLagB}), the rank-4 tensor
$Q^{\alpha\beta\rho\sigma}_{(i,j)}$ is given by
Eq. (\ref{Qil4def}) with
$A^{\alpha_1\cdot\cdot\cdot\alpha_n}$
substituted by
$A^{\alpha_1\cdot\cdot\cdot\alpha_n}_{B(i)}$.

Thirdly, we move on to prove that the expression
${E}^{\mu\nu}_{\text{BD}(i,j)}$ given by Eq. (\ref{EBDijaltf})
satisfies the Bianchi-type identity
$\nabla_\mu{E}^{\mu\nu}_{\text{BD}(i,j)}=0$.
By the aid of the two identities
\bea
\nabla_\mu\check{X}^{\mu\nu}_{\text{AB}(i,j)}&=&
\frac{1}{2}A_{(i,j)}\nabla^{\nu}\Box^{i}B
-\frac{1}{2}\big(\Box^{i}A_{(i,j)}\big)\nabla^{\nu}B
\, , \nn \\
\nabla_\mu\check{X}^{\mu\nu}_{\text{CD}(i,j)}&=&
\frac{1}{2}\big(\Box^{-j}C_{(i,j)}\big)\nabla^{\nu}D
-\frac{1}{2}C_{(i,j)}\nabla^{\nu}\Box^{-j}D
\, , \label{DivchXBDij}
\eea
together with the one
\be
\frac{1}{2}P^{\alpha\beta\rho\sigma}_{\text{BD}(i,j)}
\nabla^\nu{R}_{\alpha\beta\rho\sigma}=
\nabla_\mu\left(P^{\mu\lambda\rho\sigma}_{\text{BD}(i,j)}
R^\nu_{~\lambda\rho\sigma}-2\nabla_\rho\nabla_\sigma
P^{\rho\mu\nu\sigma}_{\text{BD}(i,j)}\right)
\, , \label{divPBDij}
\ee
the divergence for the expression
${E}^{\mu\nu}_{\text{BD}(i,j)}$ is read off as
\bea
\nabla_\mu{E}^{\mu\nu}_{\text{BD}(i,j)}&=&
\nabla_\mu\check{X}^{\mu\nu}_{\text{AB}(i,j)}
-\nabla_\mu\check{X}^{\mu\nu}_{\text{CD}(i,j)}
+\frac{1}{2}P^{\alpha\beta\rho\sigma}_{\text{BD}(i,j)}
\nabla^\nu{R}_{\alpha\beta\rho\sigma}\nn \\
&&-\frac{1}{2}A_{(i,j)}\nabla^{\nu}\Box^{i}B
-\frac{1}{2}C_{(i,j)}\nabla^{\nu}\Box^{-j}D \nn \\
&=&\frac{1}{2}P^{\alpha\beta\rho\sigma}_{\text{BD}(i,j)}
\nabla^\nu{R}_{\alpha\beta\rho\sigma}
-\frac{1}{2}\big(\Box^{i}A_{(i,j)}\big)\nabla^{\nu}B
-\frac{1}{2}\big(\Box^{-j}C_{(i,j)}\big)\nabla^{\nu}D\nn \\
&=&0
\, . \label{BianIdEBDij}
\eea
This is our desired Bianchi-type identity for
${E}^{\mu\nu}_{\text{BD}(i,j)}$.

Fourthly, we focus on proving in a similar fashion
that the expression $\tilde{E}^{\mu\nu}_{\text{CD}(i,m,n)}$
for equations of motion in Eq. (\ref{TildECDimn2})
is divergence-free. By making use of
\be
\frac{1}{2}P^{\alpha\beta\rho\sigma}_{\text{CD}(i,m,n)}
\nabla^\nu{R}_{\alpha\beta\rho\sigma}=
\nabla_\mu\left(P^{\mu\lambda\rho\sigma}_{\text{CD}(i,m,n)}
R^\nu_{~\lambda\rho\sigma}-2\nabla_\rho\nabla_\sigma
P^{\rho\mu\nu\sigma}_{\text{CD}(i,m,n)}\right)
\, , \label{divPCDimn}
\ee
after performing some computations, the divergence
of $\tilde{E}^{\mu\nu}_{\text{CD}(i,m,n)}$ is read off as
\bea
\nabla_\mu\tilde{E}^{\mu\nu}_{\text{CD}(i,m,n)}&=&
\frac{1}{2}P^{\alpha\beta\rho\sigma}_{\text{CD}(i,m,n)}
\nabla^\nu{R}_{\alpha\beta\rho\sigma}
+\nabla_\mu\bar{X}^{\mu\nu}_{\text{CD}(i)}
+\nabla_\mu{X}^{\mu\nu}_{\text{CD}(i,m,n)}
-\frac{1}{2}C\nabla^\nu\Box^iD\nn \\
&&-\frac{1}{2}{F}^{\mu\nu\rho\sigma}_{\text{CD}(i,0)}
\nabla^\nu{R}_{\alpha\beta\rho\sigma}
-\frac{1}{2}{F}^{\alpha\beta\rho\sigma}_{\text{CD}(i,m)}
\nabla^\nu\Box^m{R}_{\alpha\beta\rho\sigma}
\, . \label{DivTilECDi}
\eea
For the Lagrangian (\ref{TildLaghi}), utilizing
Eq. (\ref{IdeRiemHik2}), we have
\bea
\nabla_\mu\bar{X}^{\mu\nu}_{\text{CD}(i)}&=&
\frac{1}{2}C\nabla^\nu\Box^iD
-\frac{1}{2}\big(\Box^iC\big)\nabla^\nu{D} \nn \\
&=&\frac{1}{2}C\nabla^\nu\Box^iD
-\frac{1}{2}{Q}^{\alpha\beta\rho\sigma}_{\text{CD}(i,0)}
\nabla^\nu{R}_{\alpha\beta\rho\sigma} \nn \\
&&-\frac{1}{2}{Q}^{\alpha\beta\rho\sigma}_{\text{CD}(i,n)}
\nabla^\nu\Box^n{R}_{\alpha\beta\rho\sigma}
\, , \label{IdDivXbarCDi}
\eea
together with
\bea
\nabla_\mu{X}^{\mu\nu}_{\text{CD}(i,m,n)}&=&
\frac{1}{2}{F}^{\alpha\beta\rho\sigma}_{\text{CD}(i,m)}
\nabla^\nu\Box^m{R}_{\alpha\beta\rho\sigma}
-\frac{1}{2}\left(\Box^m
{F}^{\alpha\beta\rho\sigma}_{\text{CD}(i,m)}\right)
\nabla^\nu{R}_{\alpha\beta\rho\sigma} \nn \\
&&+\frac{1}{2}{Q}^{\alpha\beta\rho\sigma}_{\text{CD}(i,n)}
\nabla^\nu\Box^n{R}_{\alpha\beta\rho\sigma}
-\frac{1}{2}\left(\Box^n
{Q}^{\alpha\beta\rho\sigma}_{\text{CD}(i,n)}\right)
\nabla^\nu{R}_{\alpha\beta\rho\sigma}
\, . \label{IdDivXCDimn}
\eea
Substituting Eqs. (\ref{IdDivXbarCDi}) and
(\ref{IdDivXCDimn}) into Eq. (\ref{DivTilECDi}),
we further arrive at
\be
\nabla_\mu\tilde{E}^{\mu\nu}_{\text{CD}(i,m,n)}=0
\, . \label{DivTilECDi2}
\ee
The above equation can be regarded as the generalized
Bianchi-type identity associated to the field equations
$\tilde{E}^{\mu\nu}_{\text{CD}(i,m,n)}=0$.

From the above proofs, one observes that it
perfectly avoids performing computations on the divergence for
the term composed of the
derivative of the Lagrangian density with respect to the metric
to adopt the expression of field equations obtained
through the method based upon the conserved current instead
of the one derived out of the variation of the Lagrangian.
In fact, the
latter renders it of great difficulty to prove the
vanishing divergence for the field equations if there is a
lack of a remedy to eliminate the derivative of the
Lagrangian density with
respect to the metric involved in them.

%%%%%%%%%%%%%%%%%%%%%%%%%%%%%%%%%%%%%%%%%%%%%%%%%%%%%%%%
\section{Notations and a summary for the main
results}\label{appendD}
%%%%%%%%%%%%%%%%%%%%%%%%%%%%%%%%%%%%%%%%%%%%%%%%%%%%

The notations in the present paper are in accordance with
those in the textbook \cite{WaldGR}. Specifically, the
Levi-Civita connection $\Gamma^\rho_{~\mu\nu}$, formed
from the metric and its derivatives, takes the form
\be
\Gamma^\rho_{~\mu\nu}=\frac{1}{2}g^{\rho\sigma}
\left(\partial_\mu{g}_{\sigma\nu}
+\partial_\nu{g}_{\mu\sigma}
-\partial_\sigma{g}_{\mu\nu}\right)
\, . \label{LeCivconn}
\ee
The Riemann curvature tensor $R_{\mu\nu\rho\sigma}$
is defined through
\be
\left(\nabla_\mu\nabla_\nu-\nabla_\nu\nabla_\mu\right)
V_\rho=R_{\mu\nu\rho\sigma}V^\sigma
\, , \label{RiemTdef}
\ee
in which $V^\mu$ represents an arbitrary vector field.
On the basis of the Riemann curvature tensor
$R_{\mu\nu\rho\sigma}$,
the Ricci tensor $R_{\mu\nu}$ and its scalar curvature
$R$ are defined respectively as
\be
R_{\mu\nu}=g^{\rho\sigma}R_{\rho\mu\sigma\nu}
\, , \qquad
R=g^{\mu\nu}R_{\mu\nu}
\, . \label{RiccTSdef}
\ee
The Lie derivative of a rank-$(m,n)$ tensor
$T^{\alpha_1\cdot\cdot\cdot\alpha_m}_{
~~~~~~~~\beta_1\cdot\cdot\cdot\beta_n}$ along an arbitrary vector
$\zeta^\mu$ is defined by
\bea
\mathcal{L}_{\zeta}{T}^{\alpha_1\cdot\cdot\cdot\alpha_m}_{
~~~~~~~~\beta_1\cdot\cdot\cdot\beta_n}&=&
\zeta^\nu\nabla_\nu{T}^{\alpha_1\cdot\cdot\cdot\alpha_m}_{
~~~~~~~~\beta_1\cdot\cdot\cdot\beta_n} \nn \\
&&-\sum^m_{i=1}{T}^{\alpha_1\cdot\cdot\cdot\cdot\alpha_{i-1}
\nu\alpha_{i+1}\cdot\cdot\alpha_m}_{
~~~~~~~~~~~~~~~~~~~~~\beta_1\cdot\cdot\cdot\beta_n}
\nabla_\nu\zeta^{\alpha_{i}} \nn \\
&&+\sum^n_{i=1}{T}^{\alpha_1\cdot\cdot\cdot\alpha_m}_{
~~~~~~~~\beta_1\cdot\cdot\cdot\cdot\beta_{i-1}
\nu\beta_{i+1}\cdot\cdot\beta_n}
\nabla_{\beta_{i}}\zeta^\nu
\, . \label{LieTdef}
\eea

Within this paper, we have obtained the equations of motion,
the Noether potentials and the surface terms associated to
a range of Lagrangians that
involve the variables $\Box^i R$s, $\Box^i R_{\mu\nu}$s and
$\Box^i{R}_{\mu\nu\rho\sigma}$s, together with the ones
$\Box^i{B}_{\alpha_1\cdot\cdot\cdot\alpha_n}$s, where
${B}_{\alpha_1\cdot\cdot\cdot\alpha_n}$ stands for an arbitrary
rank-$n$ tensor depending upon the metric, the Riemann curvature
tensor and the variables via $\Box^i$ acting on the latter.
All of them are summarized in TABLE \ref{LagEomNP}.

\begin{table}[H]
\caption{Lagrangians, expressions for equations of motion (EEoM),
         Noether potentials and surface terms}
    \vspace{15pt}
    \centering
    \begin{tabular}{p{3.25cm}p{3.25cm}p{3.25cm}p{3.25cm}}
        \hline
        \hline
     Lagrangian   &EEoM
     & Noether potential  &surface term \\
        \hline
        $L_R$ (\ref{LagBoxR})
        &$E^{\mu\nu}_{R}$ (\ref{EoMforLagR})
        &$K^{\mu\nu}_{R}$  (\ref{KmnRdef})
        &$\Theta^\mu_R$    (\ref{TThetR}) \\
        $R^m\Box^nR$
        &$E^{\mu\nu}_{R1}$ (\ref{EoMofLagRm})
        &$K^{\mu\nu}_{R1}$  (\ref{KmnR1def})
        &$\Theta^{\mu}_{R1}$ (\ref{ThetR1})\\
        $\big(\Box^iR\big)\Box^jR$
        &$E^{\mu\nu}_{R2}$ (\ref{EoMofBoxRij})
        &$K^{\mu\nu}_{R2}$  (\ref{KmnBoxRij})
        &$\Theta^\mu_{R2}$ (\ref{ThetR2def})\\
        $L_{\text{Ric}}$ (\ref{LagBoxRic})
        &$E^{\mu\nu}_{\text{Ric}}$ (\ref{EoMforLagRic3})
        &$K^{\mu\nu}_{\text{Ric}}$  (\ref{KmnRic})
        &$\Theta^\mu_{\text{Ric}}$  (\ref{TThetRic})\\
        $R^{\mu\nu}\Box^nR_{\mu\nu}$
        &$E^{\mu\nu}_{\text{Ric1}}$ (\ref{EomLagRic1c})
        &$K^{\mu\nu}_{\text{Ric1}}$  (\ref{KmnRicc1})
        &$\Theta^\mu_{\text{Ric1}}$ (\ref{TheRic1def})\\
        $L_{\text{Riem}}$ (\ref{LagBoxRiem})
        &$E^{\mu\nu}_{\text{Riem}}$ (\ref{EoMforLagRiem2})
        &$K^{\mu\nu}_{\text{Riem}}$  (\ref{KmnRiem})
        &$\Theta^\mu_{\text{Riem}}$  (\ref{TThetRiem})\\
        $R^{\mu\nu\rho\sigma}\Box^n{R}_{\mu\nu\rho\sigma}$
        &$E^{\mu\nu}_{\text{Riem1}}$ (\ref{EoMLagRemBnRem})
        &$K^{\mu\nu}_{\text{Riem1}}$  (\ref{KmnRieBnRiem})
        &$\Theta^\mu_{\text{Riem1}}$ (\ref{TheRiem1def})\\
        $L_{(i)}$ (\ref{LagBoxiB})
        &${E}^{\mu\nu}_{(i)}$ (\ref{EomABLi})
        &${K}^{\mu\nu}_{(i)}$  (\ref{NoePABiLie})
        &$\tilde{\Theta}_{(i)}^\mu$ (\ref{ThetaLagi})\\
        $L_{B}$ (\ref{LagBgen})
        &${E}^{\mu\nu}_{B}$ (\ref{EomLagBgen})
        &${K}^{\mu\nu}_{B}$  (\ref{NoePotLagB})
        &$\Theta^\mu_{B}$   (\ref{ThetaLagB})\\
        $A(R)\Box^{i}B(R)$
        &$E^{\mu\nu}_{\text{AB}(i)}$ (\ref{EoMABoxiB})
        &$K^{\mu\nu}_{\text{AB}(i)}$  (\ref{KmnSABi})
        &$\Theta_{\text{AB}(i)}^\mu$ (\ref{TheSABi})\\
        $A(R)\Box^{-i}B(R)$
        &$\tilde{E}^{\mu\nu}_{\text{AB}(i)}$ (\ref{EoMtildAB})
        &$\tilde{K}^{\mu\nu}_{\text{AB}(i)}$  (\ref{TildKmnSABi})
        &$\tilde{\Theta}_{\text{AB}(i)}^\mu$ (\ref{TildTheSABi})\\
       $f_{(i,j)}$ (\ref{LagfijBD})
        &$E^{\mu\nu}_{\text{BD}(i,j)}$ (\ref{EBDijaltf2})
        &$K^{\mu\nu}_{\text{BD}(i,j)}$  (\ref{KmnBDij})
        &$\Theta_{\text{BD}(i,j)}^\mu$  (\ref{ThetBDij})\\
       $h_{(i)}$ (\ref{LagCBoxiD})
        &$E^{\mu\nu}_{\text{CD}(i)}$ (\ref{EoMCBoxiD2})
        &$K^{\mu\nu}_{\text{CD}(i)}$  (\ref{KmnCBoxiD})
        &$\Theta_{\text{CD}(i)}^\mu$  (\ref{TheCBoxiD})\\
        $\tilde{h}_{(i)}$ (\ref{TildLaghi})
        &$\tilde{E}^{\mu\nu}_{\text{CD}(i,m,n)}$ (\ref{TildECDimn2})
        &$\tilde{K}^{\mu\nu}_{\text{CD}(i,m,n)}$  (\ref{KtildmnCDi})
        &$\tilde{\Theta}_{\text{CD}(i,m,n)}^\mu$  (\ref{TildTheCDi})\\
        $\hat{h}_{(i)}$ (\ref{Laghathi})
        &$\hat{E}^{\mu\nu}_{\text{CD}(i,a,b)}$ (\ref{EomhatCDhi})
        &$\hat{K}^{\mu\nu}_{\text{CD}(i,a,b)}$  (\ref{hatKmmCDi})
        &$\hat{\Theta}_{\text{CD}(i,a,b)}^\mu$ (\ref{hatThetaCD})\\
        $\hat{h}_{(i,j)}$ (\ref{LagCDhij})
        &$\hat{E}^{\mu\nu}_{\text{CD}(i,j,a,b)}$ (\ref{EomhCDij})
        &$\hat{K}^{\mu\nu}_{\text{CD}(i,j,a,b)}$  (\ref{NoPforhCDij})
        &$\hat{\Theta}_{\text{CD}(i,j,a,b)}^\mu$ (\ref{TheCDijdef})\\

      \hline
      \hline
    \end{tabular}
    \label{LagEomNP}
\end{table}

%%%%%%%%%%%%%%%%%%%%%%%%%%%%%%%%%%%%%%%%%%%%%%%%%%%%%%%
%%%%%%%%%%%%%%%%%%%%%%%%%%%%%%%%%%%%%%%%%%%%%%%%%%%%%%%


\begin{thebibliography}{100}
\bibitem{JJP2306}
J.J. Peng, A note on field equations
in generalized theories of gravity,
Phys. Scr. \textbf{99}, 105229 (2024).
%arXiv:2306.11561 [gr-qc].

\bibitem{Pady}
T. Padmanabhan,
Some aspects of field equations in generalised theories of gravity,
Phys. Rev. D \textbf{84}, 124041 (2011).
%arXiv:1109.3846 [gr-qc]
%DOI: 10.1103/PhysRevD.84.124041

%%%%%%%%%%%%%%%%%%%%%%%%%%%%%%%%%%%%%%%%Box R%%%%%%%%%%
\bibitem{HJSch90}
H.J. Schmidt,
Variational derivatives of arbitrarily high order
and multiinflation cosmological models,
Classical Quantum Gravity \textbf{7}, 1023 (1990).
%DOI: 10.1088/0264-9381/7/6/011

\bibitem{Wan93}
D. Wands,
Extended gravity theories and the Einstein-Hilbert action,
Classical Quantum Gravity \textbf{11}, 269 (1994).
%gr-qc/9307034 [gr-qc]

\bibitem{HOW96}
A. Hindawi, B.A. Ovrut and D. Waldram,
Nontrivial vacua in higher derivative gravitation,
Phys. Rev. D \textbf{53}, 5597 (1996).
%hep-th/9509147 [hep-th]

\bibitem{CdMP16}
R.R. Cuzinatto, C.A.M. de Melo, L.G. Medeiros and
P.J. Pompeia,
Scalar-multi-tensorial equivalence for higher order
$f(R,\nabla_\mu{R},\nabla_{\mu_1}\nabla_{\mu_2}{R},
...,\nabla_{\mu_1}...\nabla_{\mu_n}{R})$ theories of gravity,
Phys. Rev. D \textbf{93}, 124034 (2016)
[\emph{Erratum}: Phys. Rev. D \textbf{98}, 029901 (2018)].
%arXiv:1603.01563 [gr-qc]

%%%%%%%%%%%%%%%%%%%%%%%%%%%%%%%%%

\bibitem{PWG23}
J.J. Peng, Y. Wang and W.J. Guo,
Conserved quantities for asymptotically AdS spacetimes in quadratic
curvature gravity in terms of a rank-4 tensor,
Phys. Rev. D \textbf{108}, 104035 (2023).
%arXiv:2305.12611 [gr-qc].
%https://doi.org/10.1103/PhysRevD.108.104035

%%%%%%%%%%%%off-shell conserved current%%%%%%%%%%%%%%%%
\bibitem{TPad10}
T. Padmanabhan,
Thermodynamical aspects of gravity: new insights,
Rept. Prog. Phys. \textbf{73}, 046901 (2010).
%arXiv:0911.5004 [gr-qc]
%DOI: 10.1088/0034-4885/73/4/046901

\bibitem{RievLL} 	
T. Padmanabhan and D. Kothawala,
Lanczos-Lovelock models of gravity,
Phys. Rept. \textbf{531}, 115 (2013).
%arXiv:1302.2151 [gr-qc]
%DOI: 10.1016/j.physrep.2013.05.007

%%%%%%%%%%%%%%nonlocal gravity%%%%%%%%%%%%%%%%
\bibitem{ALS97}
M. Asorey, J.L. Lopez and I.L. Shapiro,
Some remarks on high derivative quantum gravity,
Int. J. Mod. Phys. A \textbf{12}, 5711 (1997).
%hep-th/9610006 [hep-th]

\bibitem{BGKM12}
T. Biswas, E. Gerwick, T. Koivisto and A. Mazumdar,
Towards singularity and ghost free theories of gravity,
Phys. Rev. Lett. \textbf{108}, 031101 (2012).
%arXiv:1110.5249 [gr-qc]

\bibitem{Mods12}
L. Modesto, Super-renormalizable quantum gravity,
Phys. Rev. D \textbf{86}, 044005 (2012).
%arXiv:1107.2403 [hep-th].

\bibitem{BCKM14}
T. Biswas, A. Conroy, A.S. Koshelev and A. Mazumdar,
Generalized ghost-free quadratic curvature gravity,
Classical Quantum Gravity \textbf{31}, 015022 (2014)
[\emph{Erratum}: Classical Quantum Gravity \textbf{31},
159501 (2014)].
%arXiv:1308.2319[hep-th]

\bibitem{KKSrev23}
A.S. Koshelev, K.S. Kumar and A.A. Starobinsky,
Cosmology in nonlocal gravity,
arXiv:2305.18716 [hep-th].

\bibitem{CBnonloc22}
S. Capozziello and F. Bajardi,
Nonlocal gravity cosmology: An overview,
Int. J. Mod. Phys. D \textbf{31}, 2230009 (2022).
%arXiv:2201.04512 [gr-qc]

\bibitem{DDRS18}
I. Dimitrijevic, B. Dragovich, Z. Rakic and J. Stankovic,
Variations of infinite derivative modified gravity,
Springer Proc. Mathematics $\&$ Statistics
\textbf{263}, 91-111 (2018).
%arXiv:1902.08820 [hep-th]
%DOI: https://doi.org/10.1007/978-981-13-2715-5_5

\bibitem{DDRS22}
I. Dimitrijevic, B. Dragovich, Z. Rakic and J. Stankovic,
Nonlocal de Sitter gravity and its exact cosmological solutions,
J. High Energy Phys. \textbf{2022}, 054 (2022).
%arXiv:2206.13515 [gr-qc]

\bibitem{CCL23}
S. Capozziello, M. Capriolo and G. Lambiase,
The energy-momentum complex in non-local gravity,
Int. J. Geom. Meth. Mod. Phys. \textbf{20}, 2350177 (2023).
%arXiv:2301.04023 [gr-qc]

%%%%%%Wald formalism%%%%%%%%%%%%%%%%%%%%%%%%%%%%%%%%%%%%%%%%%%%
\bibitem{LeeWald}
J. Lee and R. M. Wald,
Local symmetries and constraints,
J. Math. Phys. \textbf{31}, 725 (1990).

\bibitem{IyWald}
V. Iyer and R.M. Wald,
Some properties of the Noether charge and a proposal for dynamical
black hole entropy,
Phys. Rev. D \textbf{50}, 846 (1994).
%[arXiv:gr-qc/9403028].

\bibitem{WalZo}
R.M. Wald and A. Zoupas,
A general definition of `conserved quantities' in
general relativity and other theories of gravity,
Phys. Rev. D \textbf{61}, 084027 (2000).
%gr-qc/9911095
%%%%%%%%%%%%%%%%%%%%%%%%%%%%%%%%%%%

%%%%%%%%%%%%%%%%%%%%%%%%%%%%%%%%%%%%
\bibitem{WaldGR}
R. M. Wald, General Relativity (University of Chicago
Press, Chicago, 1984).
%%%%%%%%%%%%%%%%%%%%%%%%%%%%%%%%%%%%%%%%%%

\end{thebibliography}
\end{document}